\documentclass[a4paper,11pt]{article}
\usepackage{jinstpub} 
\usepackage{lineno}
\usepackage{tikz}

\title{\boldmath Modeling Light Signals Using Data from the First Pulsed Neutron Source Program at the DUNE Vertical Drift ColdBox Test Facility at the CERN Neutrino Platform}

\author[a,1]{A. Paudel,\note{Corresponding author.}}
\author[b,1]{W. Shi,}
\author[a]{P. Sala,}
\author[a]{F. Cavanna,}
\author[c]{W. Johnson,}
\author[c,1]{J. Wang,}
\author[a]{W. Ketchum,}
\author[d]{F. Resnati,}
\author[b]{A. Heindel,}
\author[e]{A. Ashkenazi,}
\author[e]{E. Bertholet,}
\author[f]{E. Bertolini,}
\author[c]{D. A. Martinez Caicedo,}
\author[q]{E. Calvo,}
\author[q]{A. Canto,}
\author[q]{S. Manthey Corchado,}
\author[q]{C. Cuesta,}
\author[g]{Z. Djurcic,}
\author[p]{M. Fani,}
\author[a]{A. Feld,}
\author[h]{S. Fogarty,}
\author[f,i]{F. Galizzi,}
\author[o]{S. Gollapinni,}
\author[j]{Y. Kermaïdic,}
\author[a]{A. Kish,}
\author[k]{F. Marinho,}
\author[c]{D. Torres Muñoz,}
\author[q]{A. Verdugo de Osa,}
\author[k]{L. Paulucci,}
\author[a]{W. Pellico,}
\author[e]{V. Popov,}
\author[c]{J. Rodriguez Rondon,}
\author[c]{D. Leon Silverio,}
\author[l]{S. Sacerdoti,}
\author[f]{H. Souza,}
\author[n]{R. C Svoboda,}
\author[h]{D. Totani,}
\author[m]{V. Trabattoni,}
\author[j]{L. Zambelli}

\affiliation[a]{Fermi National Accelerator Laboratory, Batavia, IL, 60510, USA}
\affiliation[b]{Stony Brook University, SUNY, Stony Brook, New York 11794, USA}
\affiliation[c]{South Dakota School of Mines and Technology, Rapid City, SD 57701, USA}
\affiliation[d]{CERN}
\affiliation[e]{Tel Aviv University, Tel Aviv, 69978, Israel}
\affiliation[f]{University of Milano-Bicocca
Department Physics, University of Milano-Bicocca, Milan, Italy}
\affiliation[g]{Argonne National Laboratory, Argonne, IL 60439, USA}
\affiliation[h]{Colorado State University, Limelight Ave, Castle Rock 80109, USA}
\affiliation[i]{INFN Sezione di Milano-Bicocca, Piazza della Scienza 3, Milan, Italy}
\affiliation[j]{Laboratoire d’Annecy-le-Vieux de Physique des Particules, Annecy-le-Vieux, France}
\affiliation[k]{Instituto Tecnologico de Aeronautica (ITA), São José dos Campos/SP, 12228, Brasil}
\affiliation[l]{AstroParticule et Cosmologie (APC)
10, rue Alice Domon et Léonie Duquet, 75013 Paris, France}
\affiliation[m]{Università degli Studi di Milano, Department of Physics, Via Giovanni Celoria 16, 20133 Milano, Italy}
\affiliation[n]{University of California Davis, Davis, CA 95616, USA}
\affiliation[o]{Los Alamos National Laboratory, Los Alamos, NM 87545, USA}
\affiliation[p]{University of Minnesota Twin Cities, Minneapolis, MN 55455, USA}
\affiliation[q]{Centro de Investigaciones Energéticas Medioambientales y Tecnológicas (CIEMAT), 28040 Madrid, Spain}

\emailAdd{apaudel@fnal.gov, wei.shi.1@stonybrook.edu, Jingbo.Wang@sdsmt.edu}

\abstract{In this paper, we present a first quantitative test of detected light signals produced in a pulsed neutron source run in a small vertical drift LArTPC at the CERN neutrino platform ColdBox test facility. The ColdBox cryostat, detectors, neutron sources, and particle interactions are modeled and simulated using Fluka. We demonstrate the ability to identify the contribution from neutron interactions using X-ARAPUCA photodetectors, and show first comparisons of data to simulation, which indicate reasonable agreement. A time constant is also fitted from the neutron-beam-off light signal spectrum and found consistent between data and simulation. Several important systematic effects are discussed and serve as guides for future runs at larger LArTPCs.}


\arxivnumber{2512.10790} 

\begin{document}

\begin{tikzpicture}[remember picture,overlay]
    \node[anchor=north east, align=right, font=\small] at (current page.north east) {FERMILAB-PUB-25-0856-LBNF};
\end{tikzpicture}

\maketitle
\flushbottom

\section{Motivation}
\label{sec:intro}

The Deep Underground Neutrino Experiment (DUNE)~\cite{DUNE_Physics_TDR} is a next-generation long-baseline neutrino oscillation experiment in the US. The far detector (FD) will consist of large modular liquid argon (LAr) time projection chambers (LArTPC) with a total active mass of roughly 40 kt. These modules are located 1500 meters underground at the Sanford Underground Research Facility (SURF), South Dakota, and about 1300 kilometers from the near detector complex at Fermilab, Chicago. The vertical drift (VD) FD module~\cite{DUNE_VD_TDR} will be the first of the four FD modules to be installed at SURF and is under construction. The VD LAr time projection chamber (TPC) features X-ARAPUCA (XA) photon detectors~\cite{Arapuca_2016, XA_2018} installed on the cathode plane and the cryostat membrane. The VD photon detection system (PDS) provides accurate timing of an event (referred to as $t_{0}$), background rejection, and calorimetric measurement capabilities. The R\&D and validation of the VD LArTPC detector technologies have been enabled by the Neutrino Platform at CERN through its ColdBox and ProtoDUNE programs. More details on ProtoDUNE, ColdBox, and future planned tests can be found in ref.~\cite{ProtoDUNE3}.

MeV energy scale calibration of the PDS is crucial for DUNE's low energy physics program. One promising candidate calibration technique utilizes the neutron capture on $^{40}$Ar. Once the neutron is captured, the gamma cascade from the excited $^{41}$Ar has a total energy release of 6.1 MeV, which may serve as a standard candle for PDS calorimetry calibration. By tagging neutron captures, it is possible to derive a reliable light yield map for the energy measurement of MeV neutrino events. This paper presents the first study of light signals produced by neutron interactions in LAr, including their capture on $^{40}$Ar, using a pulsed neutron source (PNS)~\cite{DUNE_SP_TDR}.

\section{Vertical Drift ColdBox Test Facility at the CERN Neutrino Platform}

\begin{figure}[h!]
  \centering  \includegraphics[width=0.9\columnwidth]{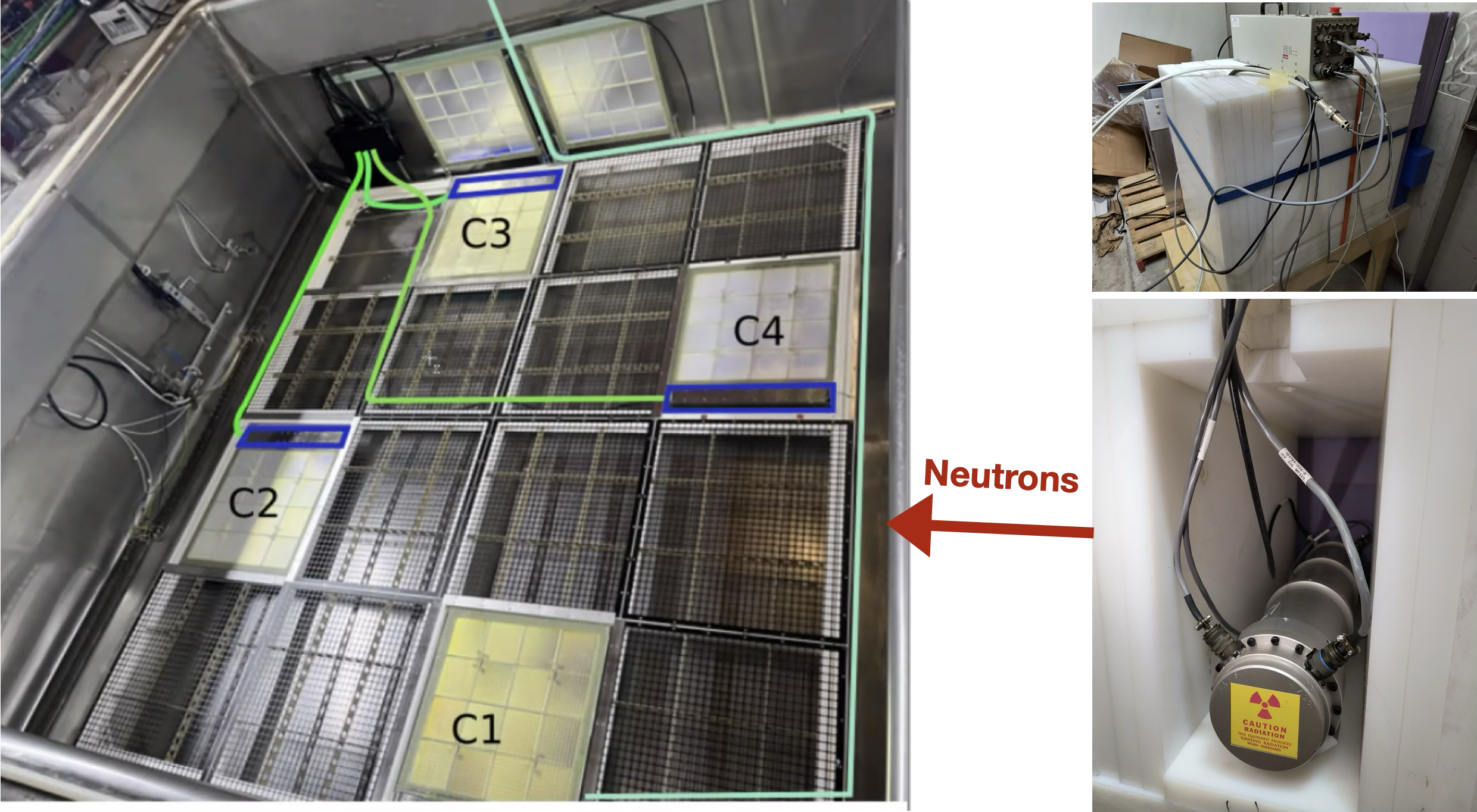}
  \caption{ColdBox detector and PNS geometry in April 2024 run. Left: top view of the cold box. Four XA photon detectors, C1-C4, are instrumented on the cathode. Two XA photon detectors are installed on the membrane. Right: Deuterium-Deuterium neutron generator (bottom) and its shielding (top). }
  \label{fig:CBgeo}
\end{figure}

The PNS program is performed at the CERN VD ColdBox (CB) test facility. The CB cryostat has an internal volume of 3.89 m (\textit{y}) × 3.91 m (\textit{z}) × 1 m (\textit{x}) and is filled with LAr to a height of $\sim$70 cm during normal operation. The drift direction is in \textit{+x}. A high voltage of -10 kV is applied to a cathode of size 3.37 m x 2.98 m in \textit{y-z} plane, establishing a uniform electric field of 454 V/cm in the drift region. The total drift time is 140 $\mu$s for a 21.5 cm drift distance. Four 60 cm $\times$ 60 cm-sized XA photon detectors, assembled at CIEMAT and NIU, are installed on the cathode surface facing the anode charge readout plane. The front-end cold readout electronics of these four XA detectors are powered using the Power-over-Fiber (PoF) technology~\cite{DUNE_PoF}, which enables safe and noise-free detector operation on the high-voltage cathode surface at the cryogenic temperature. The photodetector layout in the CB is shown in Fig.~\ref{fig:CBgeo} and implemented in Fluka geometry as shown in Fig.~\ref{fig:Flukageo}. A combination of different types of SiPMs is used in the four XA photon detectors. C1 is equipped with Fondazione Bruno Kessler (FBK) SiPMs, while C2, C3, and C4 are equipped with Hamamatsu (HPK) SiPMs. All SiPMs are operated at 5 V above their breakdown voltage. Two more XAs are also installed on one side of the membrane. However, they are not considered in this study. 

\section{Pulsed Neutron Source}
\label{sec:pns}

\begin{figure}[h!]
  \centering  \includegraphics[width=0.6\columnwidth]{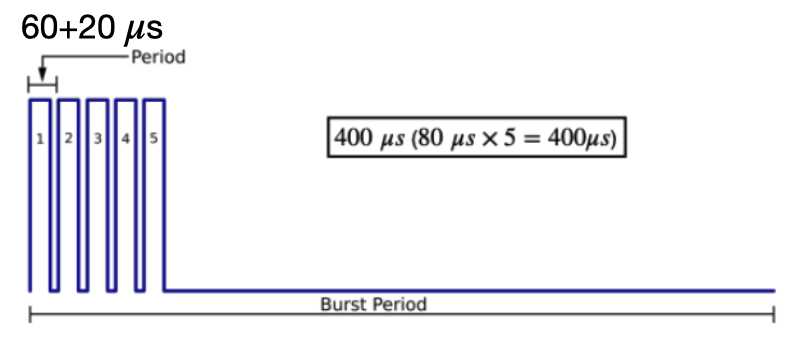}
  \caption{PNS burst mode in the ColdBox run in April 2024. The burst repeats at a frequency of 80 Hz. The data acquisition rate is 4 Hz, and the DAQ system records every 20th neutron burst generated by the neutron source.}
  \label{fig:burstmode}
\end{figure}

The PNS used in the run is a commercial Thermo Scientific MP 320 Deuterium-Deuterium generator (DDG). It is deployed on one side of the CB at the mid-height of the drift region. Custom polyethylene blocks are used as neutron shielding for radiation safety. Lead blocks and borated polyethylene are placed in front of the DDG to absorb gammas produced from the polyethylene. The DDG produces $\sim$1 million mono-energetic 2.5 MeV neutrons per second. However, in this run at the CB, the DDG is operated in a special mode (Fig.~\ref{fig:burstmode}), where five bursts of neutrons are generated every 12.5 ms (80 Hz). In this special mode, the actual emission intensity is unknown.  Each burst lasts 60 $\mu$s followed by a 20 $\mu$s long downtime. The data acquisition rate is 4 Hz, and the DAQ system records every 20th neutron burst generated by the neutron source. The produced neutrons penetrate the CB wall and the non-instrumented LAr region before reaching the bulk LAr with active charge readout. Neutron interaction products reaching the dead LAr region can produce scintillation light and can be detected by XAs. 

\section{Samples}
\label{sec:samples}

In this section, we describe the samples that are used in the later calibration analysis presented in Sec.~\ref{sec:lightana} and the result in Sec.~\ref{sec:mainresult}. The cosmic and PNS trigger and data taking are detailed in Sec.~\ref{sec:data}. The simulation of PNS and the ColdBox geometry, the neutron interactions with all relevant materials, and the light produced from these interactions are described in Sec.~\ref{sec:mc}.

\subsection{Trigger and Data Acquisition}
\label{sec:data}

PNS data were acquired with both the TPC and PDS read out using an external hardware Transistor-Transistor Logic (TTL) trigger provided by the DDG synchronous with the neutron beam. Cosmic ray triggers are taken with a software-issued random trigger. For both cosmic and PNS data taking, the trigger is unbiased. No external scintillator paddles are used. No signal amplitude threshold is applied. 

For the analysis, only runs with substantial statistics were considered, resulting in a total of over 160,000 PNS triggers, each recorded with a 1-ms readout window. More than 250,000 cosmic ray triggers were collected during the same period, using a 4-ms readout window. These cosmic datasets were used for calibration and background estimation. Test runs with limited statistics or used for debugging purposes were excluded from the study. These are summarized in Tab.~\ref{tab:runinfo}.

\begin{table}[h!]
\centering
\begin{tabular}{c|c|c}
\hline\hline
Trigger Type    & Triggers & Readout Window (ms)  \\
    \hline
PNS     & 162908   & 1           \\
Cosmic  & 253734   & 4           \\
\hline\hline
\end{tabular}
\caption{Data runs information.}
\label{tab:runinfo}
\end{table}

\subsection{Monte Carlo}
\label{sec:mc}

\begin{figure}[h!]
  \centering  \includegraphics[width=0.53\columnwidth]{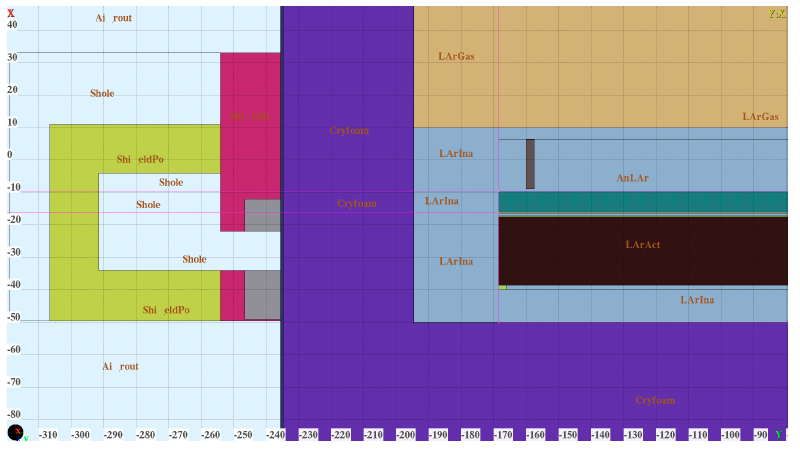}
  \centering  \includegraphics[width=0.46\columnwidth]{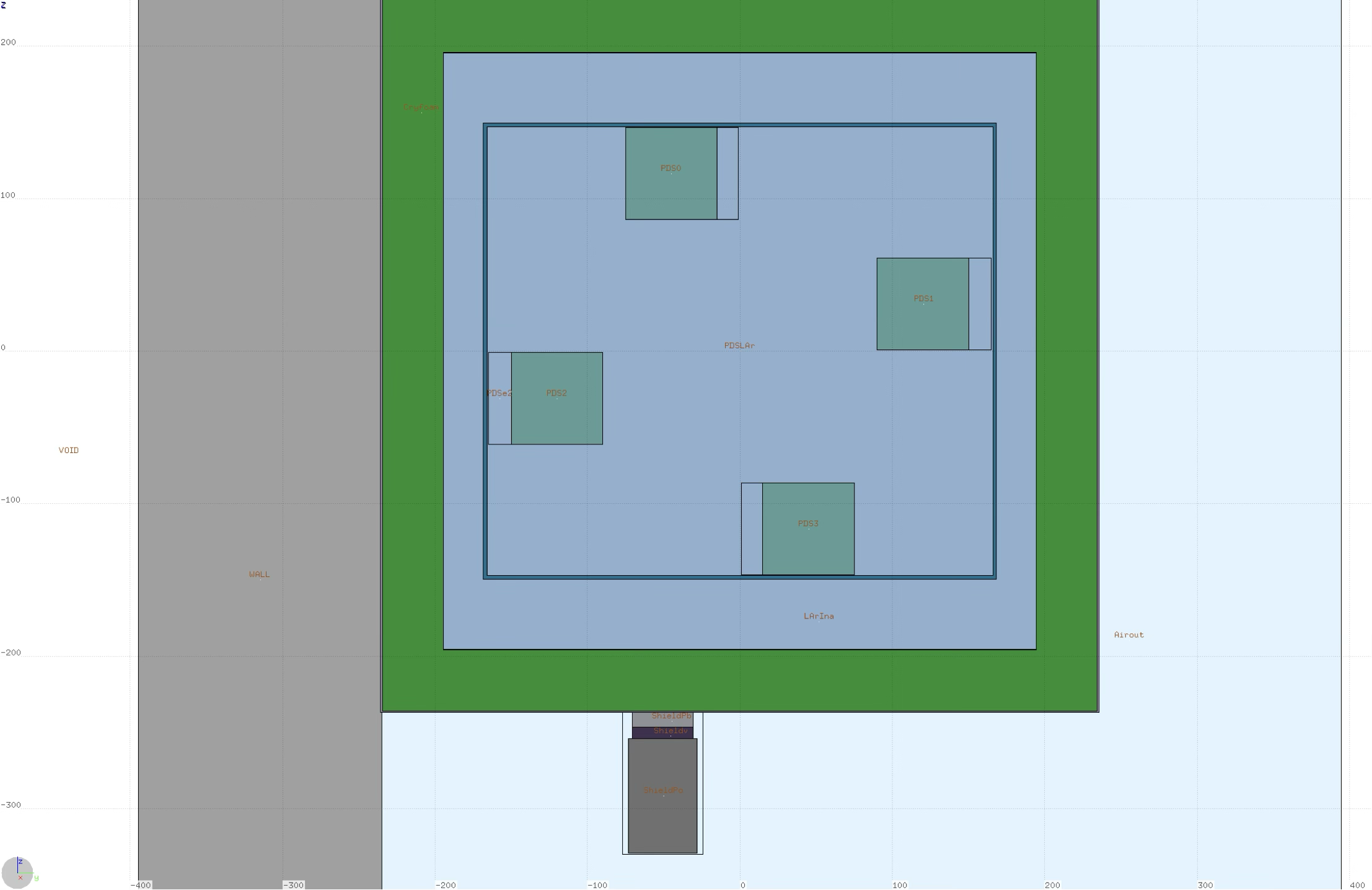}
  \caption{Side cross-sectional view (left) and top view (right) of the ColdBox and PNS geometry implemented in Fluka simulation. }
  \label{fig:Flukageo}
\end{figure}

The simulation of the PNS at the CB is implemented in Fluka2024.1~\cite{Fluka} as shown in Fig.~\ref{fig:Flukageo}, taking advantage of Fluka's detailed modeling of neutron transport and nuclear deexcitation. In the simulation, each event corresponds to a single neutron from the DDG, and all neutrons are emitted at time zero. Optical photons are produced and transported both in the active readout LAr volume and the non-active regions. The adopted scintillation light yield is 2.55$\times$10$^{4}$ photons/MeV, as estimated by LArQL\cite{LArQL} for a Minimum Ionizing particle (MIP) in an electric field of 454 V/cm. The same scintillation photon yield is used for both the active and inactive LAr regions. LArQL has been validated on data at 500 V/cm, where the expected light yield is 2.4$\times$10$^{4}$ photons/MeV. The values of the fast and slow time constants of Ar scintillation light used are 6 ns and 1.5 $\mu$s, respectively. A Rayleigh length of 90 cm is adopted as well as an infinite absorption length~\cite{Fluka_validation}. A photon detection efficiency (PDE) of 3\% is used for all XAs. To decrease CPU time, only 1/10 of the optical photons are tracked. These scale factors are taken into account later in the analysis of the MC sample.

Typical physics processes relevant for neutrons from the PNS include neutron inelastic interactions, neutron elastic interactions, and neutron captures. Each process could leave energy deposits in LAr and generate scintillation light. Events where there is either energy deposition in active LAr, or one neutron capture in active LAr, or scintillation photons arriving at the XAs from neutron interactions on any inactive materials, are recorded. For each recorded event, interactions occurring in all regions, the associated energy depositions, and optical photons are all stored. The recorded events correspond to approximately 3.2\% of generated neutrons. Events with one neutron capture are 0.04\% of the generated neutrons. 

\begin{figure}[h!]

\includegraphics[width=0.49\columnwidth]{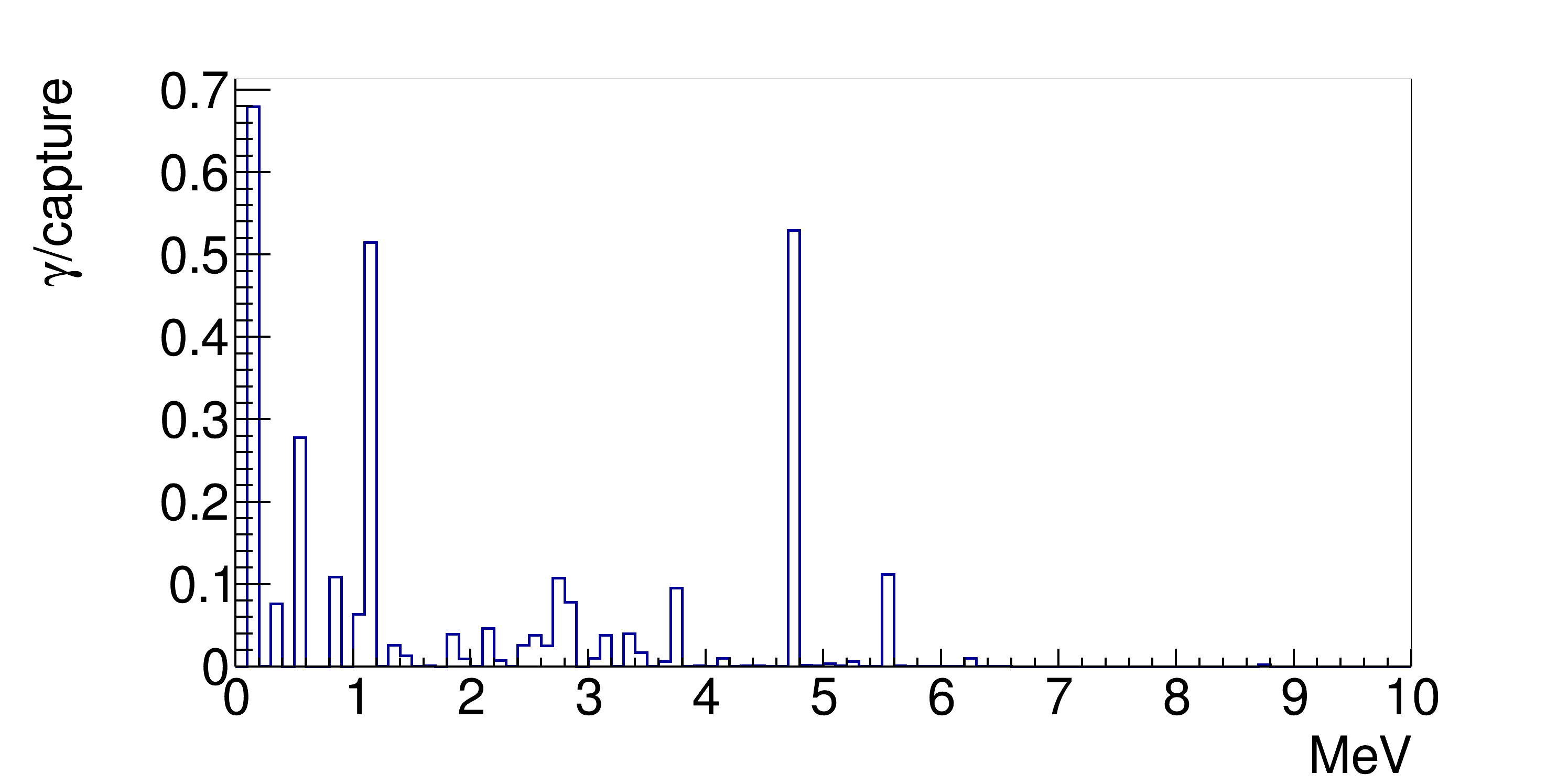}
\includegraphics[width=0.49\columnwidth]{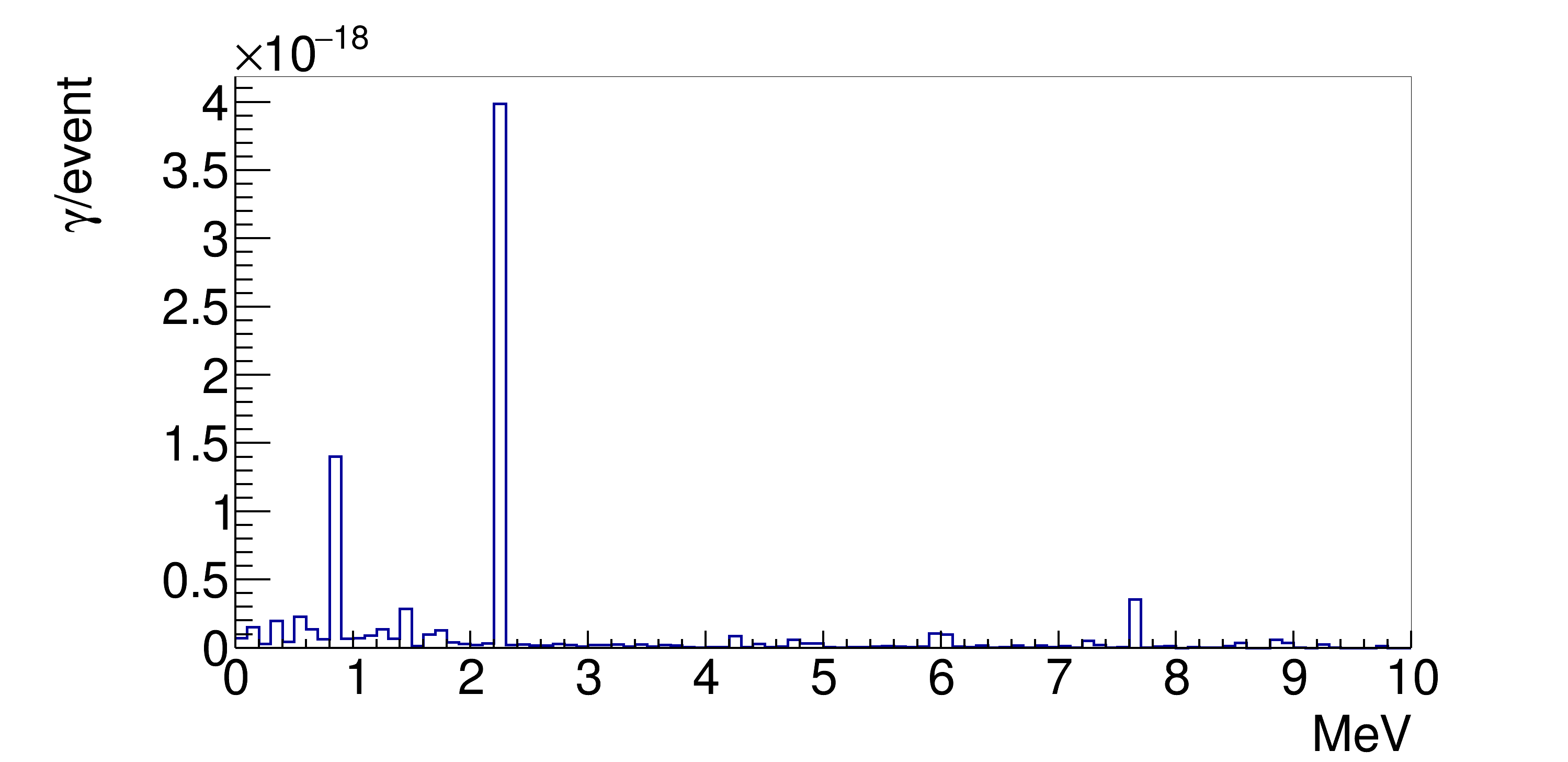}
\centering
\includegraphics[width=0.6\columnwidth]{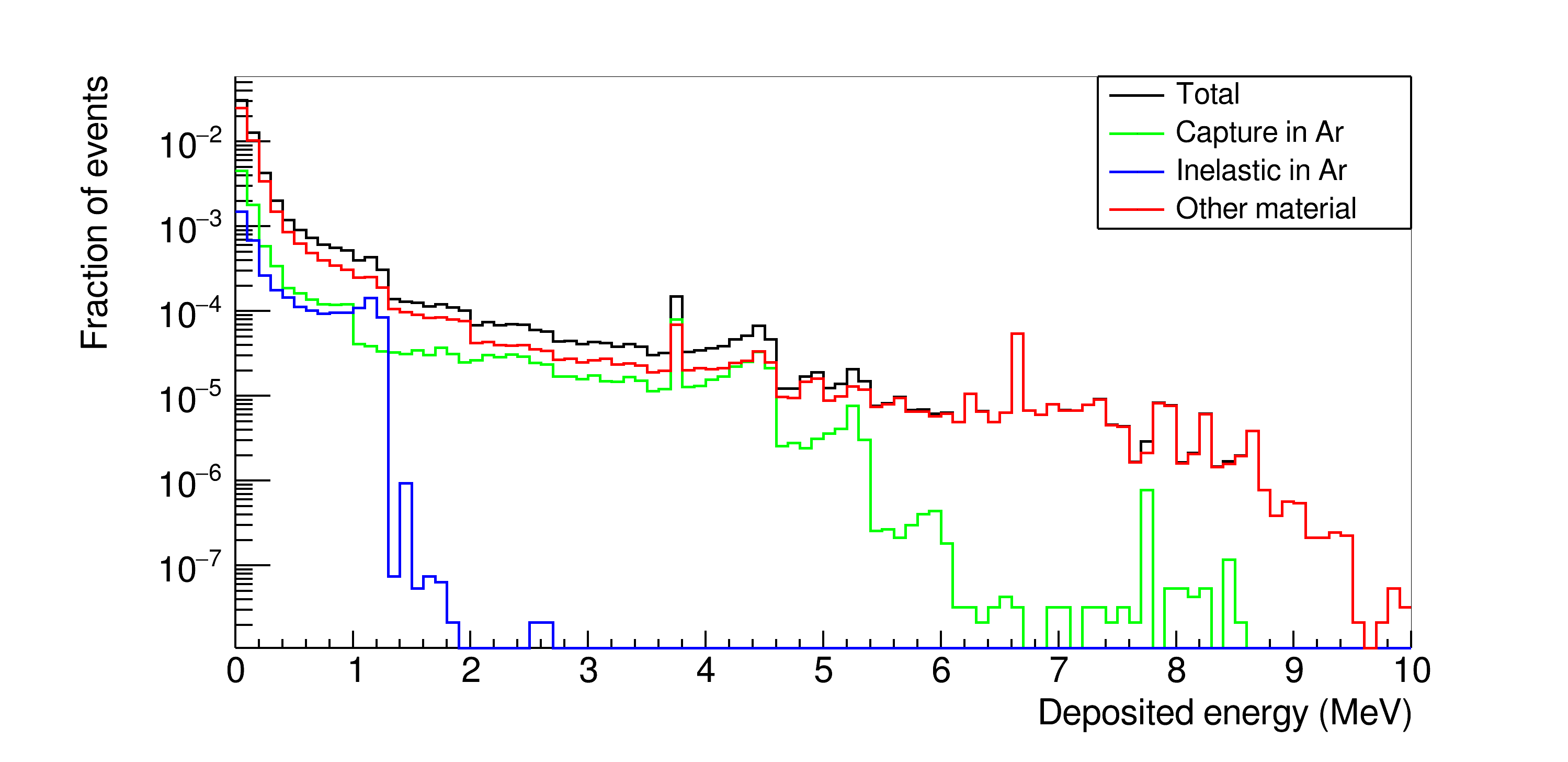}
  \caption{Fluka simulated energy spectrum of $\gamma$ rays (top), from neutron captures in active LAr region of the ColdBox (top left), from neutron interactions in materials different from Ar (top right), and the total energy deposition distribution (bottom) in the ColdBox from processes in the active LAr and all other materials.
  }
  \label{fig:capgamma}
\end{figure}

Fluka performs a detailed simulation of neutron transport and interaction, including the description of gamma ray cascades according to the available nuclear data \cite{Fluka}. Fig.~\ref{fig:capgamma} shows the $\gamma$ spectrum following neutron capture in Ar in the Fluka simulation. The most intense $\gamma$ emissions, at 4.7 MeV, 1.2 MeV, and 167 keV, are clearly visible. Due to the small active volume of the detector, neutron interactions in the surrounding inactive materials have to be carefully modeled. Fig.~\ref{fig:capgamma} also shows the detailed simulated $\gamma$ spectrum from non-Ar materials, showing, for instance, the 2.2~MeV line from capture on hydrogen, the 7.6~MeV line from capture on Fe, and the 876~keV line from inelastic scattering on Fe.

A good representation of what happens outside the readout volume is particularly important in the present setup, where the readout volume is relatively small. Table~\ref{tab:origin} shows the fraction of events associated with the interaction region. Indeed, most of the signals collected by the most exposed XA (C4) originate from neutron interactions occurring outside the active LAr region.

\begin{table}[tbh]
\centering
\begin{tabular}{|c|c|}
\hline\hline
Region    & Fraction \\
    \hline
Active LAr	&0.29\\    
Non-active LAr	&0.26\\
PNS shielding polyethylene	&0.19\\
Cryostat outer wall	&0.17\\
PNS tube	&0.03\\
Cryostat inner wall 	&0.025\\
Other  &0.03\\
\hline\hline
\end{tabular}
\caption{Fraction of events that have light signals on the C4 XA module (nearest to the PNS) originated from neutron interactions in different regions of the ColdBox facility. Only light signals with more than 100 PE are considered. Only the regions counting for more than 1\% of the total light have been detailed. Statistical errors are at or below 1\%.}
\label{tab:origin}
\end{table}

\section{Calibration and Event Selection}
\label{sec:lightana}

In this section, we present studies on data quality, calibration procedures for the amplitude of light signals on each of the four XAs on the cathode, and the minimal event selection for both the simulation and data samples. 

\begin{figure}[h!]
  \centering  \includegraphics[width=0.42\columnwidth]{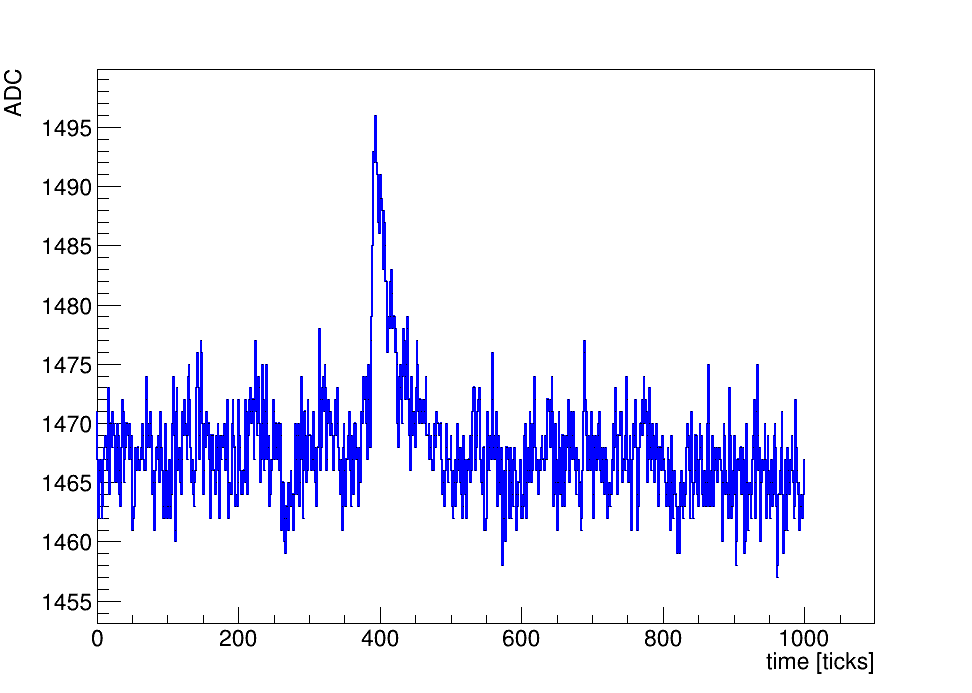}
  \centering  \includegraphics[width=0.48\columnwidth]{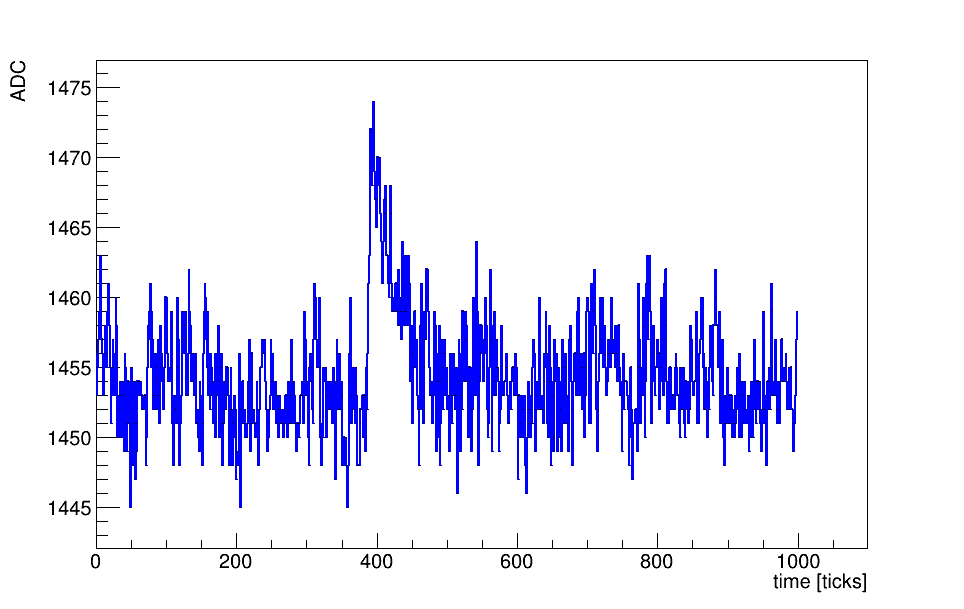}
  \caption{Example raw waveforms from each of the two channels on the C4 XA module. Each time tick represents 16 ns.}
  \label{fig:rawwfm}
\end{figure}

\begin{figure}[h!]
  \centering  \includegraphics[width=0.45\columnwidth]{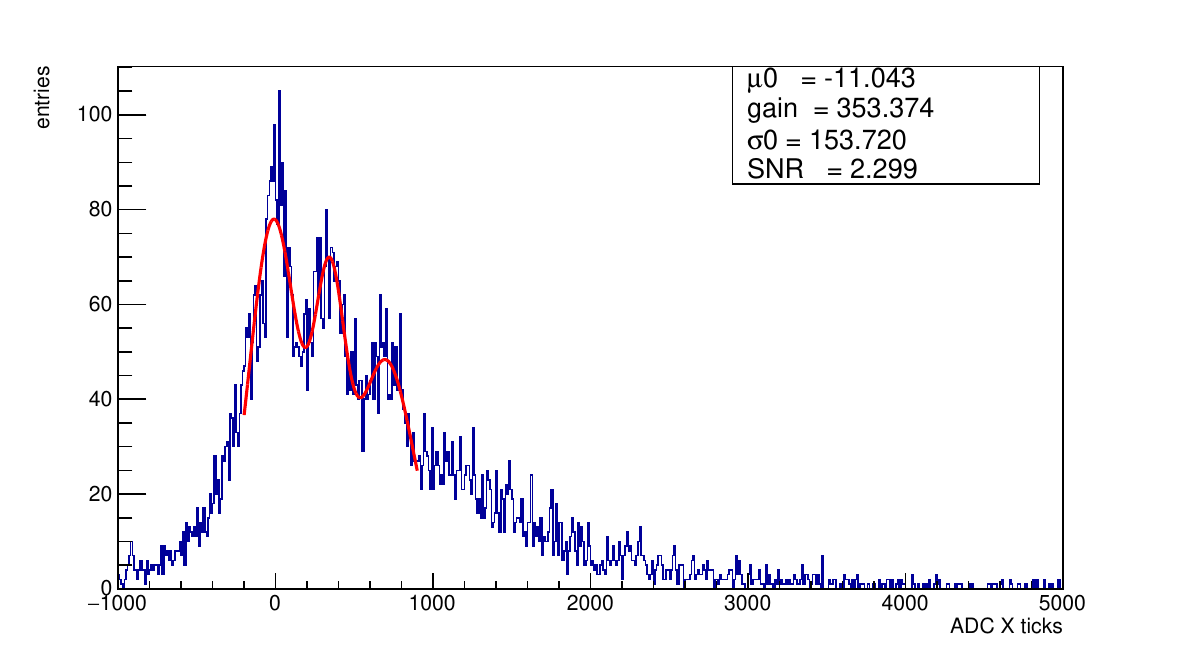}
  \centering  \includegraphics[width=0.45\columnwidth]{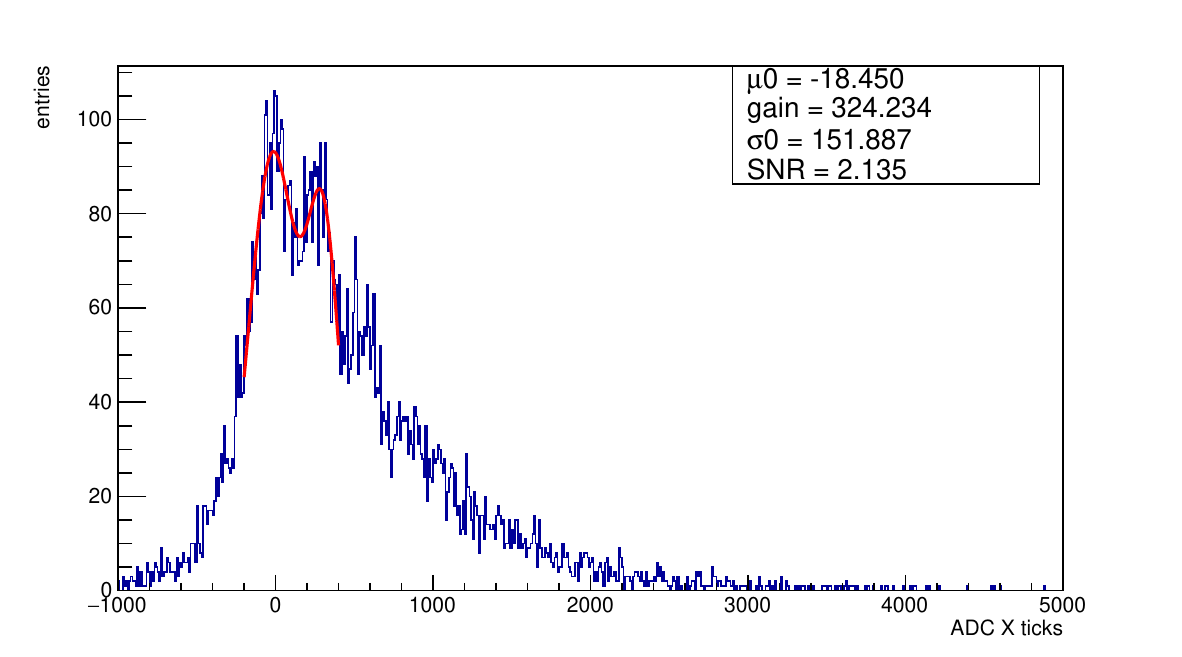}
  \centering  \includegraphics[width=0.45\columnwidth]{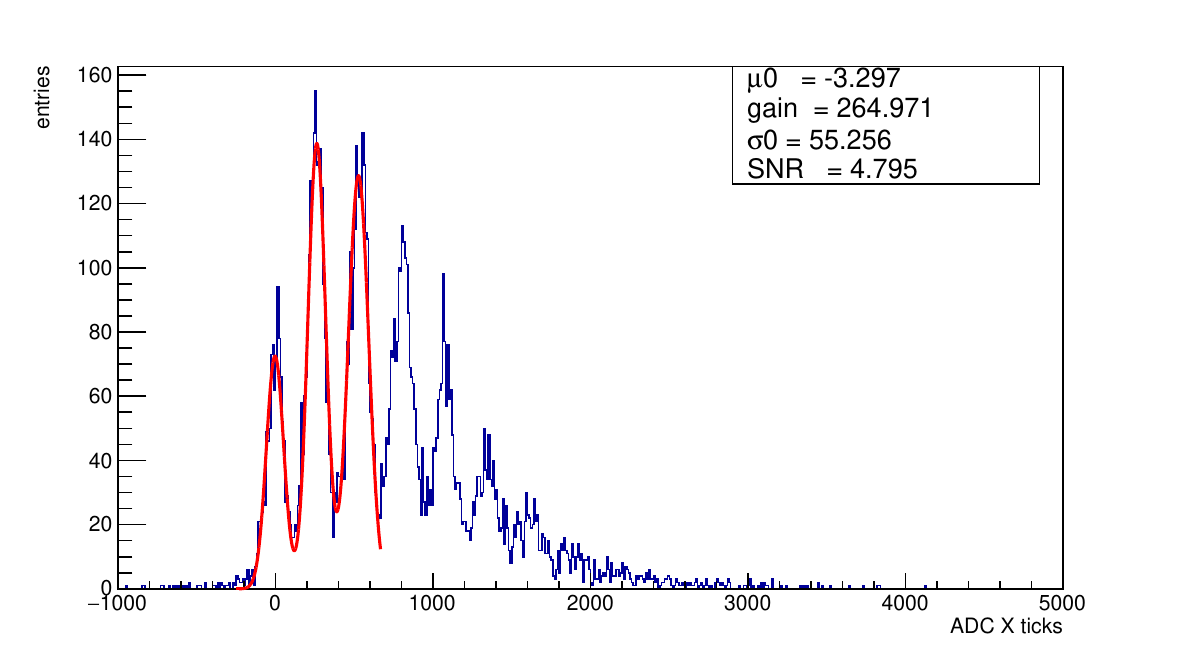}
  \centering  \includegraphics[width=0.45\columnwidth]{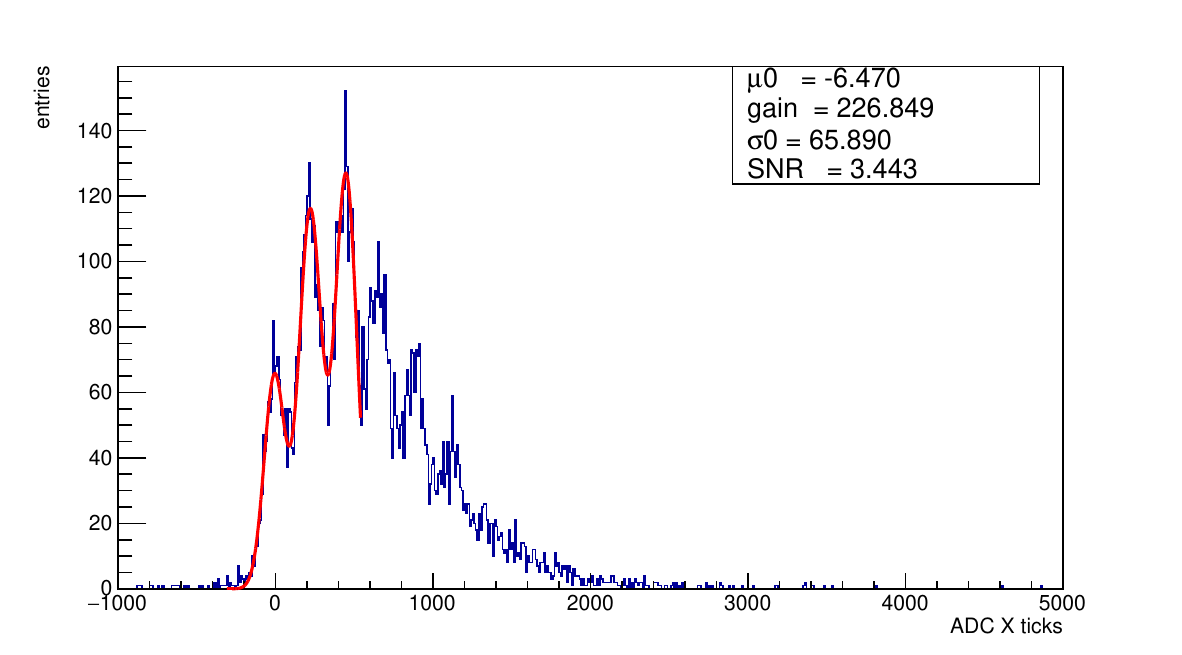}
  \centering  \includegraphics[width=0.45\columnwidth]{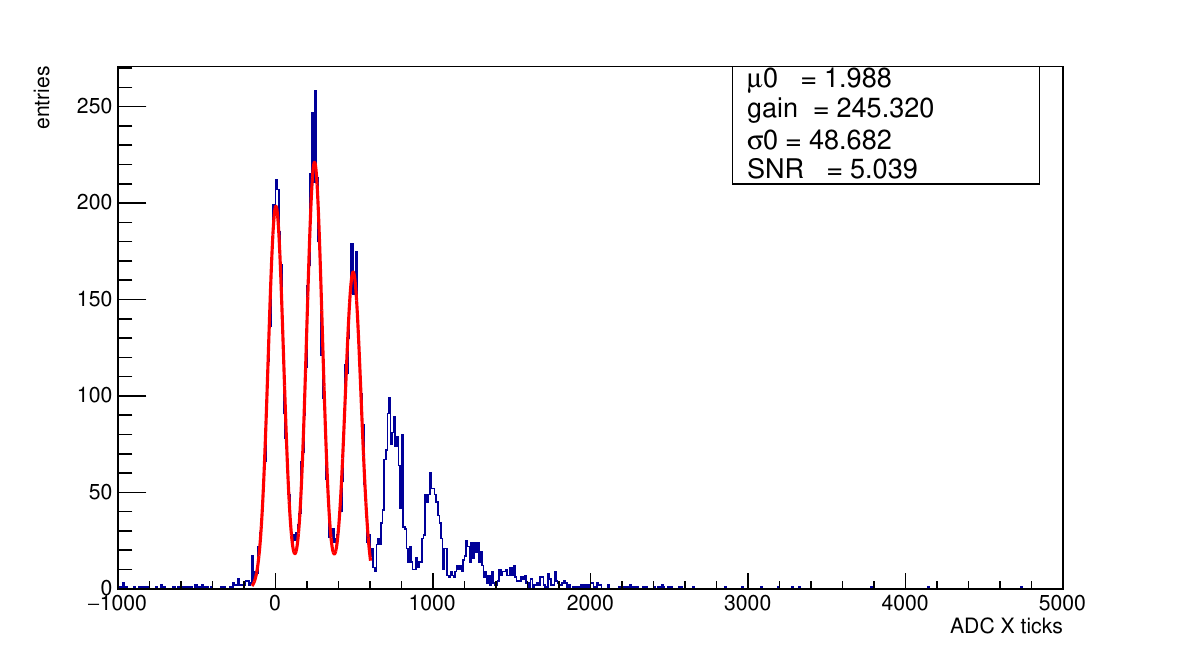}
  \centering  \includegraphics[width=0.45\columnwidth]{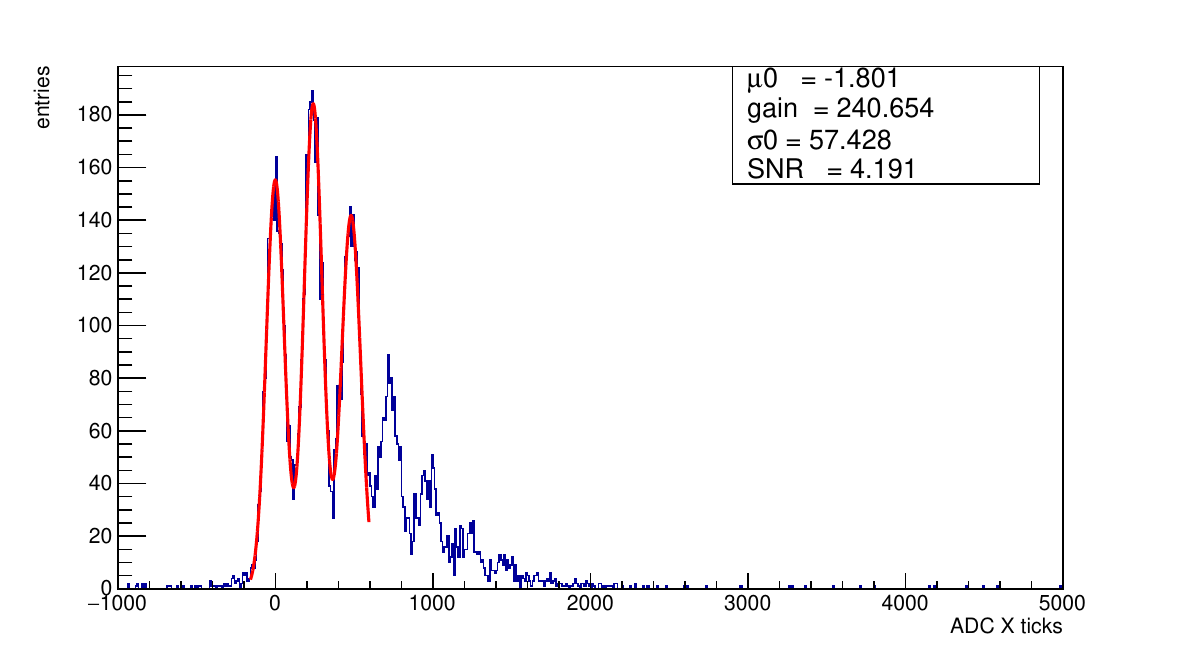}
  \centering  \includegraphics[width=0.45\columnwidth]{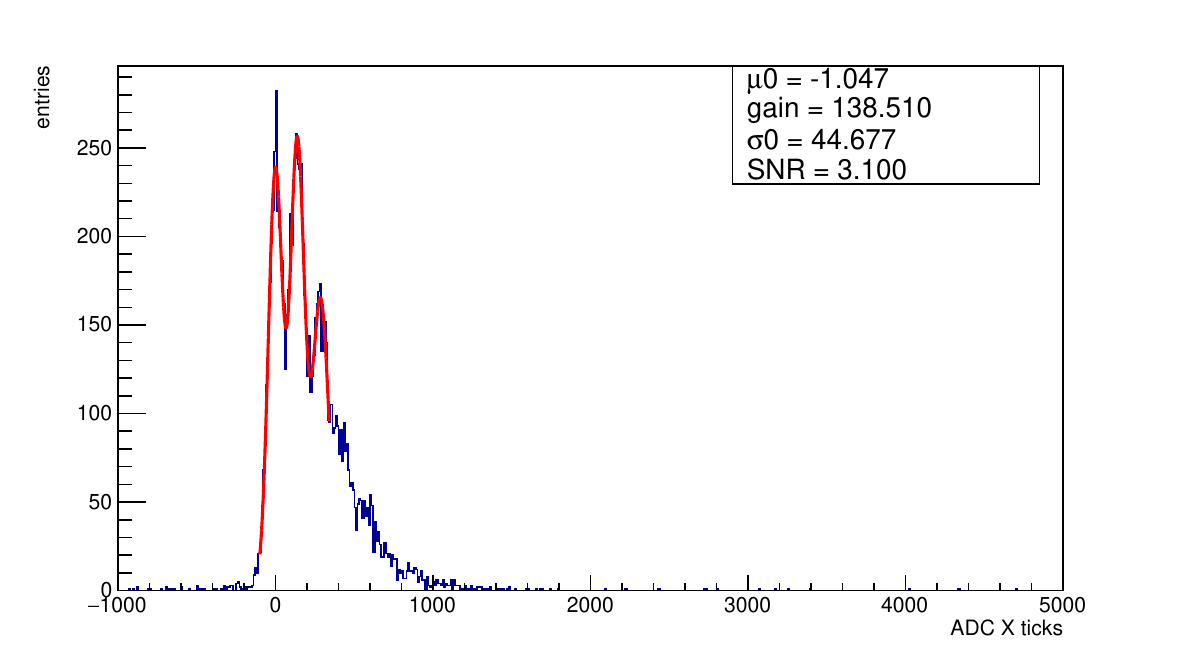}
  \centering  \includegraphics[width=0.45\columnwidth]{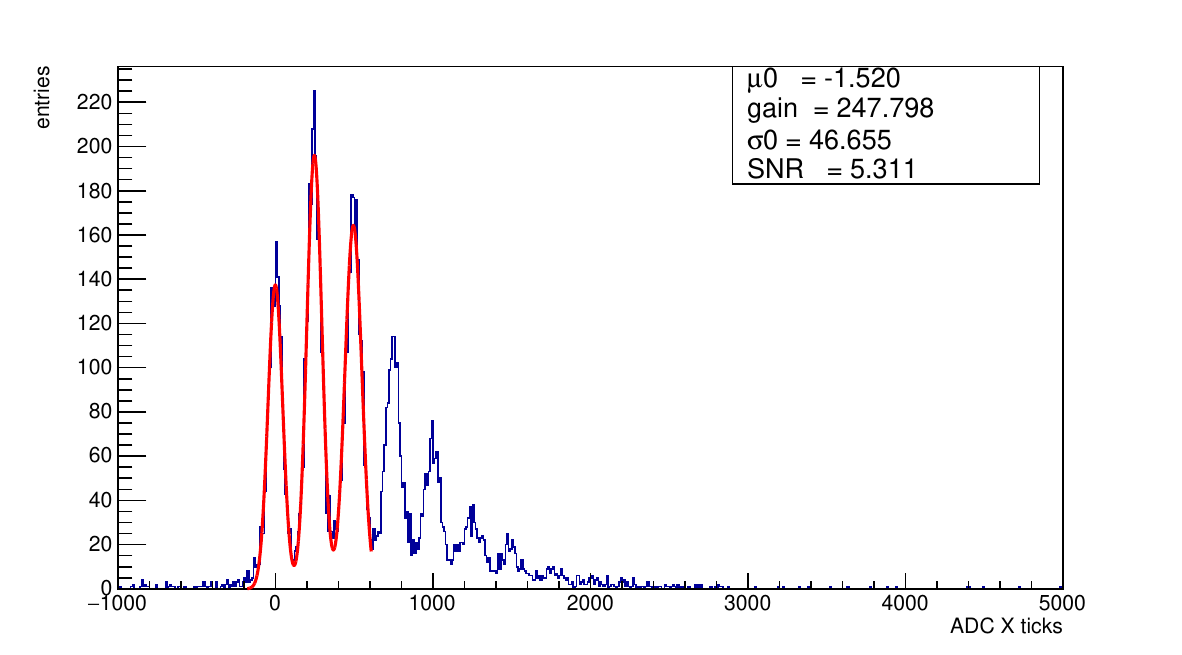}
  \caption{From the top row to the bottom row: small PE responses of both channels on C1, C2, C3, and C4 XA modules.}
  \label{fig:fingerplot}
\end{figure}

Example raw waveforms from the two channels of the C4 XA module are shown in Fig.~\ref{fig:rawwfm}. Except for the C1 module, the other modules show good signal-to-noise performance, as demonstrated by the small PE responses shown in Fig.~\ref{fig:fingerplot}. We note the runs at this ColdBox were taken at an early stage of prototyping, and the detector and readout electronics are far from the final version described in the DUNE VD TDR~\cite{DUNE_VD_TDR}. The single photoelectron templates for all channels are displayed in Fig.~\ref{fig:SPEtemplate}. 

\subsection{ADC to Photoelectron Calibration in Data}
\label{sec:adc2pe}
The digitized counts from analog-to-digital converters (ADC) on XA readout are converted to photoelectrons (PE) via a calibration constant. The calibration constant is dependent on run and channel for a specific XA. They are obtained by finding the average ADC counts for many single PE signals on each XA channel. 


\subsection{Relative PDE Calibration in Data}
\label{sec:pdecali}

Different XA configurations, SiPMs, and dichroic filters from different manufacturers could cause variations in the PDE among XA modules. The XA modules in these ColdBox runs are prototypes and don't represent the final detector design for DUNE. Variations in the PDE of all XAs are calibrated by studying cosmic tracks passing through the center of the CB at close to zero zenith angle (within 5 cm between top and bottom crossing points). For these cosmic tracks, we expect a similar number of photons to arrive at the surface of each of the four XAs on the cathode. Therefore, the detected PEs on each XA provide a calibration of the relative PDE among the four XAs. Tab.~\ref{tab:pdecali} shows the average detected PEs on the four XAs from these cosmic tracks. The relative PDE translates to a scale factor in PE, later applied to data on each XA when comparing to MC. The C3 XA module is used as the reference detector because the detected PE on the two channels is the closest, which is expected for a good detector. Therefore, we set the PE scale factor to one for the C3 XA module. The other XAs have a scale factor derived from the ratio of their total detected PE to that on C3. 

\begin{table}[h!]
\centering
\begin{tabular}{r|c|c|c|c}
\hline\hline
    XA Module     & ch0 detected PE & ch1 detected PE & Tot. detected PE & PE scale factor \\
    \hline
    C1         & 125.5  & 130.8  & 256.3  & 1.85   \\
    C2         & 68.2   & 62.97  & 131.2  & 0.947  \\
    C3         & 69.14  & 69.4   & 138.5  & 1.0 (reference XA)  \\
    C4         & 57.42  & 48.4   & 105.8  & 0.764 \\
\hline\hline
\end{tabular}
\caption{Relative PDE calibration using cosmics passing through the center of CB at close to zero zenith angle. Here, C3 is used as the reference so that all detected PEs on other XAs are scaled using the listed scale factors.}
\label{tab:pdecali}
\end{table}

\subsection{Photon to Peak ADC Calibration in MC}
\label{sec:ph2adccali}
\begin{figure}[h!]
  \centering  \includegraphics[width=0.45\columnwidth]{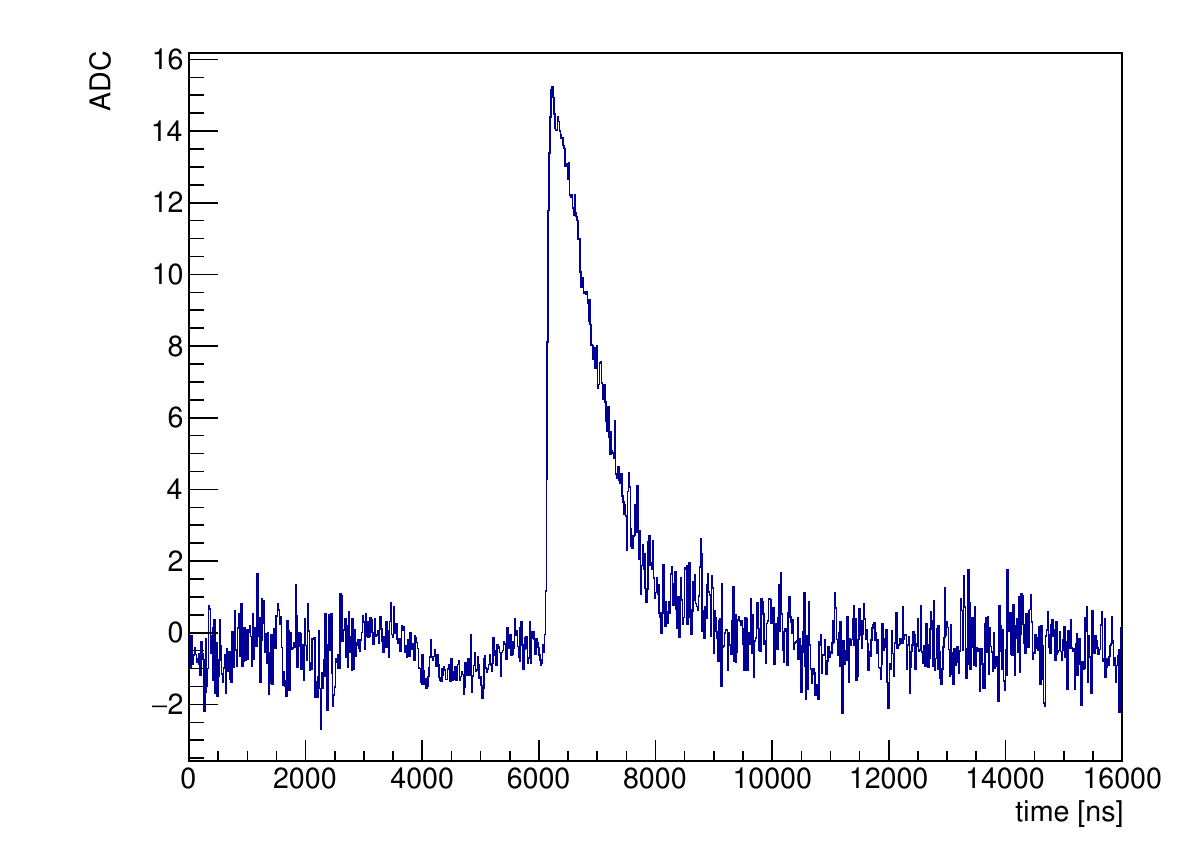}
  \centering  \includegraphics[width=0.45\columnwidth]{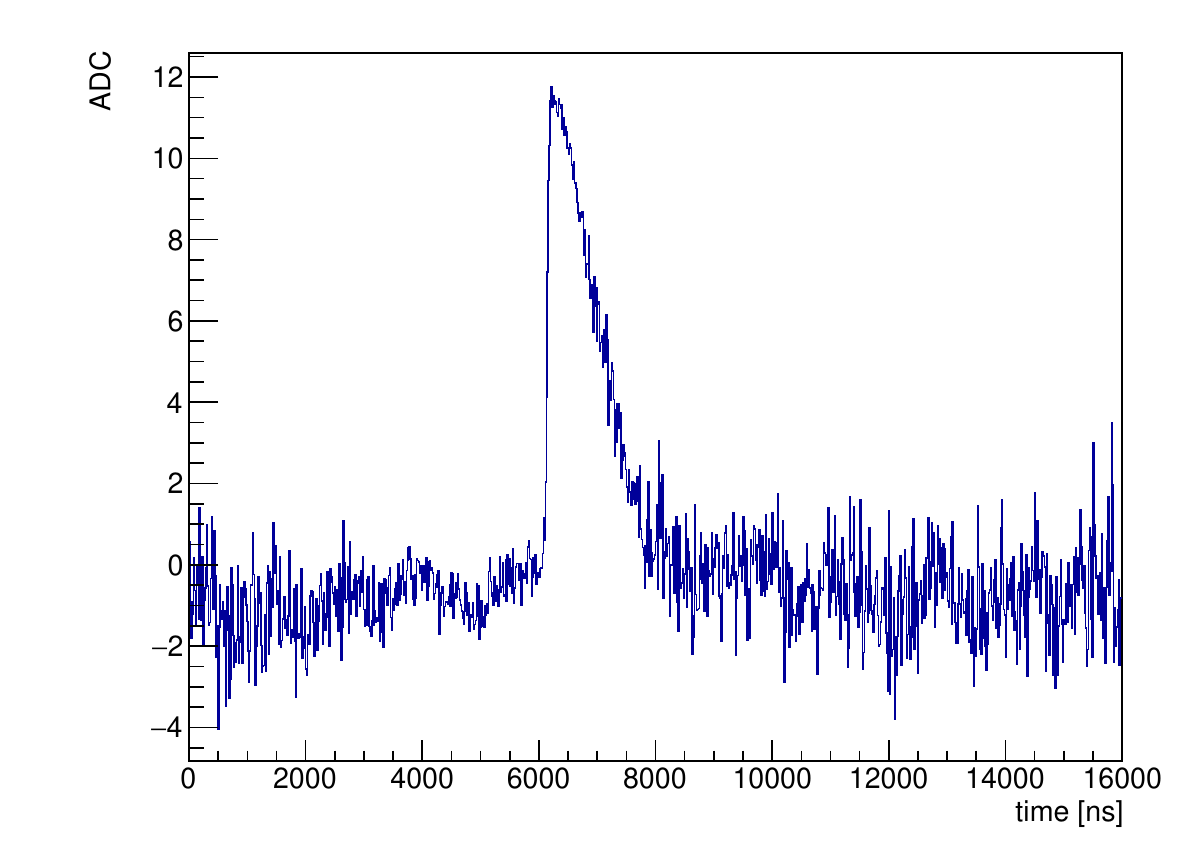}
  \centering  \includegraphics[width=0.45\columnwidth]{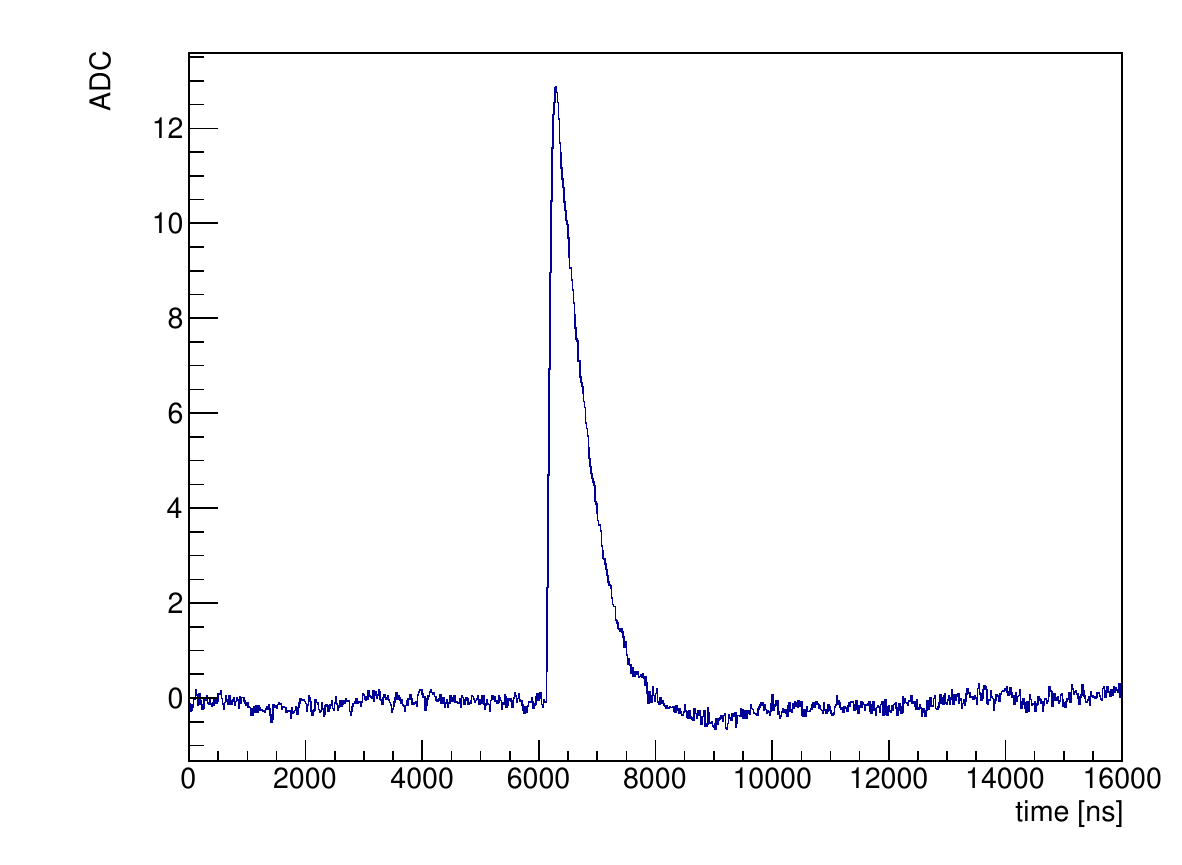}
  \centering  \includegraphics[width=0.45\columnwidth]{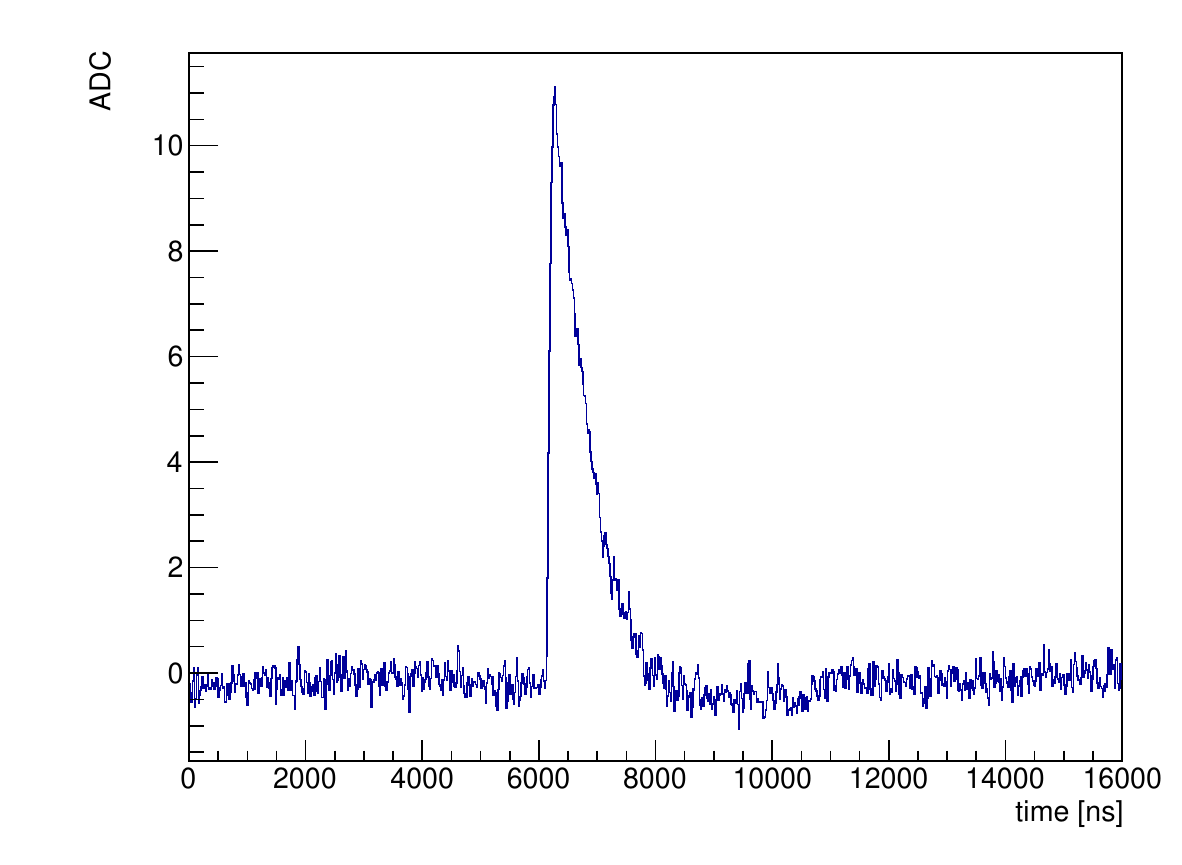}
  \centering  \includegraphics[width=0.45\columnwidth]{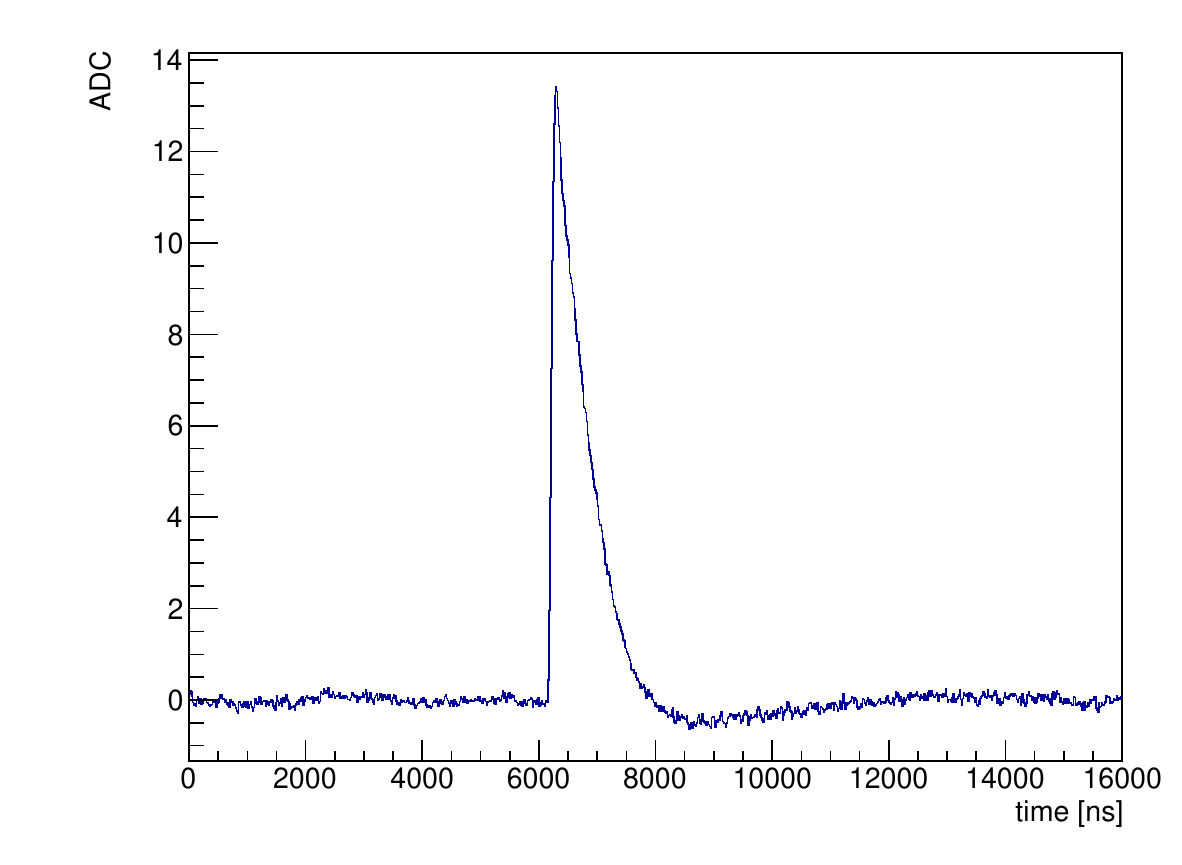}
  \centering  \includegraphics[width=0.45\columnwidth]{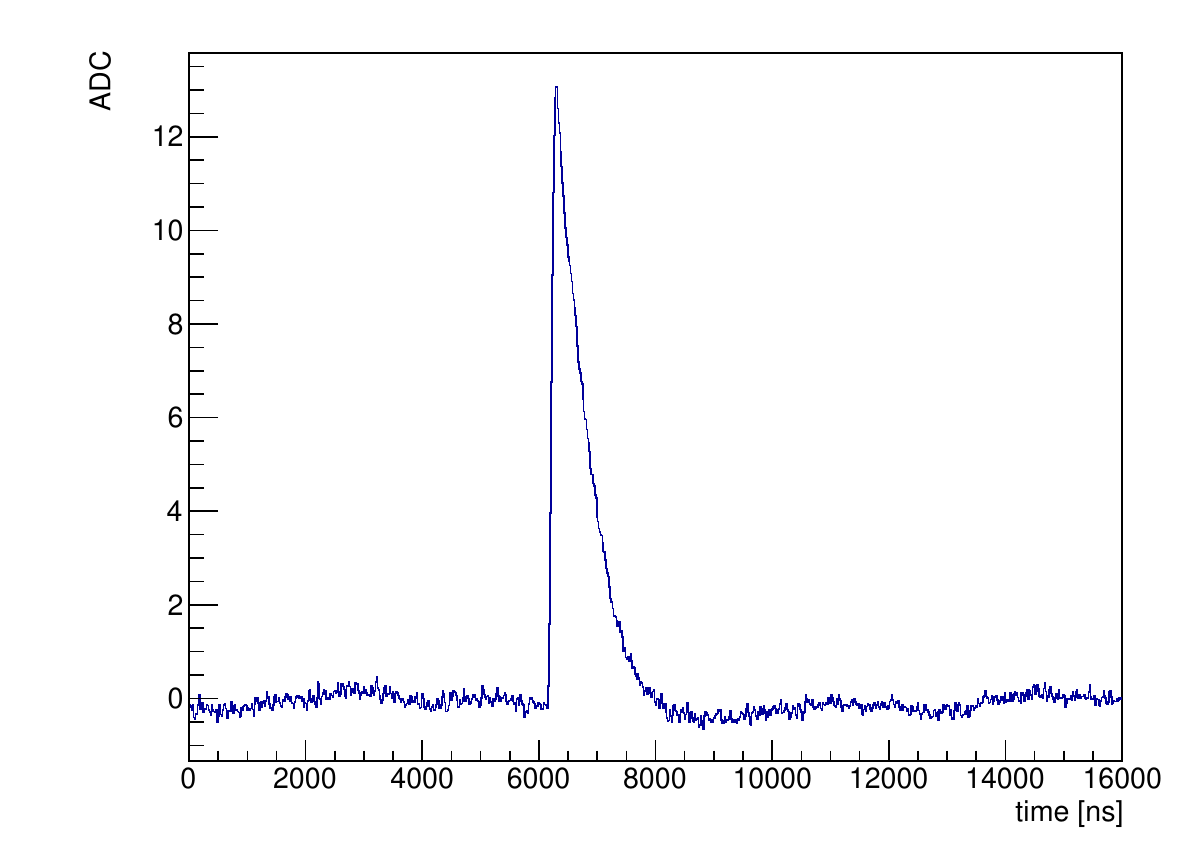}
  \centering  \includegraphics[width=0.45\columnwidth]{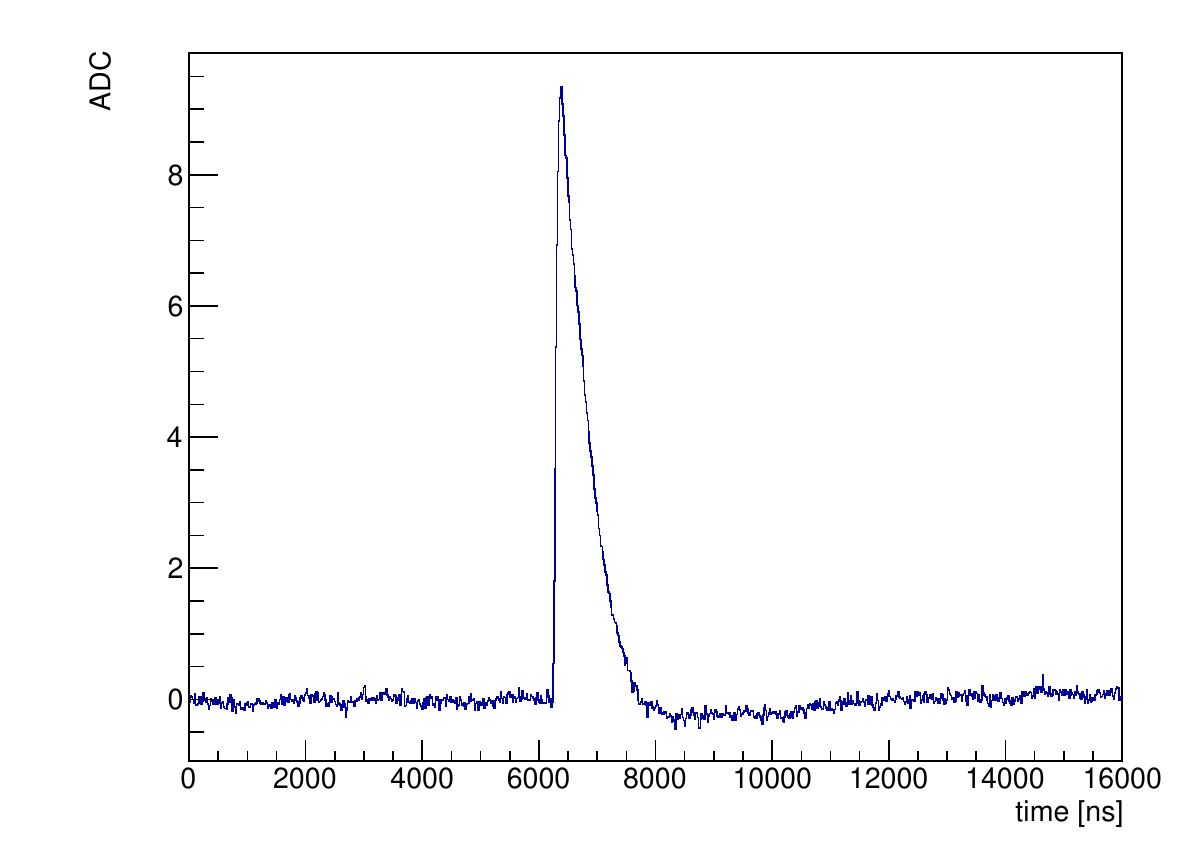}
  \centering  \includegraphics[width=0.45\columnwidth]{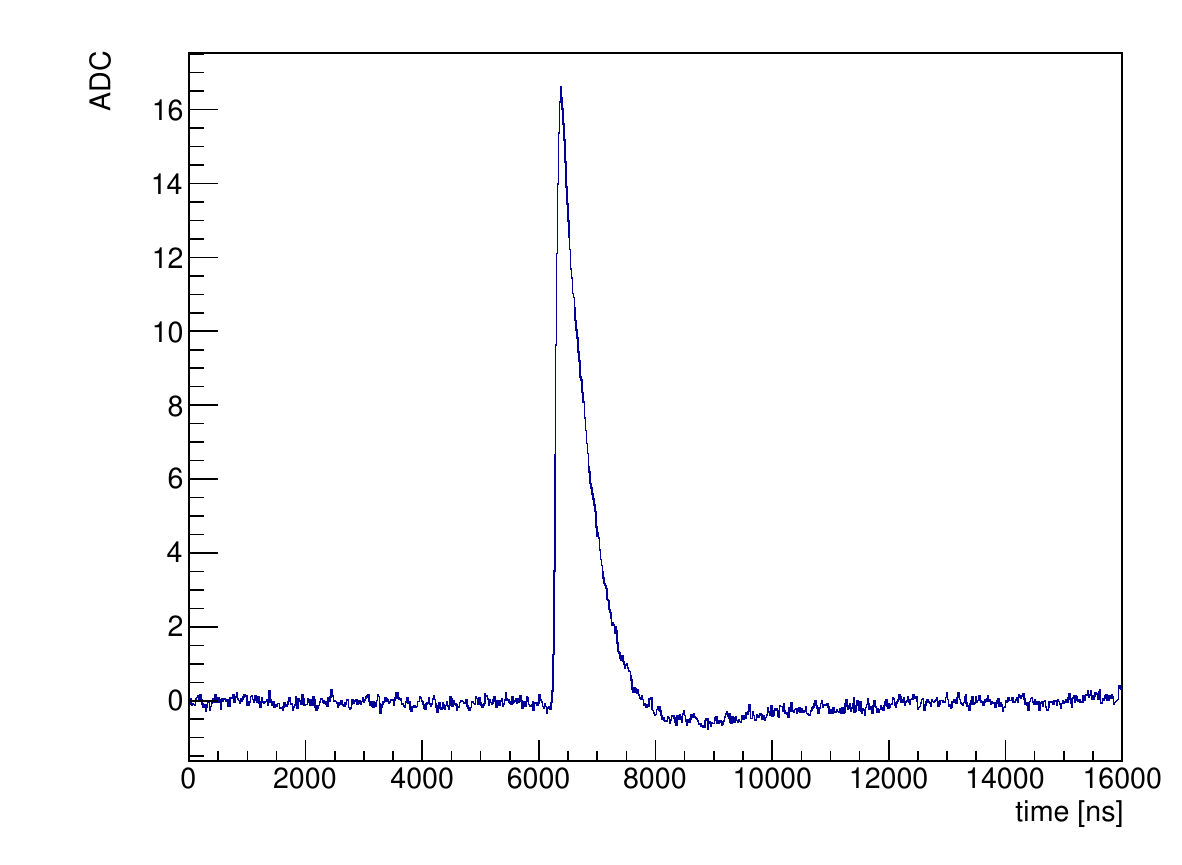}
  \caption{From the top row to the bottom row: single PE templates for two channels on C1, C2, C3, and C4 XA modules.}
  \label{fig:SPEtemplate}
\end{figure}

As described in Sec.~\ref{sec:adc2pe}, the number of photoelectrons in data is derived from the peak ADC after applying an ADC to PE scale factor. The Fluka simulation provides the number of photons generated in a typical neutron event. These photons are summed and then convolved with the single photoelectron templates (Fig.~\ref{fig:SPEtemplate}) obtained from LED calibration runs. The peak ADC is then used to calculate the number of photoelectrons, as consistently performed in data. The fraction of the measured PE from the amplitude in the simulation relative to the total detected simulated photons is 0.3, consistent with the fast component of LAr scintillation light. 

\subsection{Event Selection}
\label{sec:lightsel}

We use the Lardon framework~\cite{lardon} for this CB data analysis. A peak finding algorithm is used to find photon signals larger than $\sim$15 PE. For two channels on each XA, the signals whose peak times are within 80 ns (5 PD time ticks, each time tick represents 16 ns) are used to calculate the total detected signal for a specific activity. This time coincidence requirement is derived from the peak time difference between all possible pairs of PD signals from two channels on the XA from a cosmic run. For each XA channel, the ADC-PE calibration constants are used to calculate the PE. The total detected PE on an XA is then obtained by summing up the PE from two channels. 

Because of geometric acceptance, activities closer to the XA detector produce a larger number of PE, while those at the center of the CB generate weaker signals. At a small number of PE, the PE count is subject to fluctuations in the waveform baseline from prototyping electronics, SiPM dark counts, and environmental background light noise. As a result, we restrict the region of interest to 100 PE and above for each XA module to avoid analyzing ambiguous low PE pulses. A maximum cut of 2100 PE is also applied to avoid looking at saturated signals.

The same cuts used in data selection are applied to the Fluka simulation sample. To summarize, these include the requirement that 1) the simulated light signal has to be produced between the start of the DAQ window and 1050 $\mu$s, the maximum DAQ window, and 2) the simulated light signal amplitude is between 100 PE and 2100 PE for each XA module. The time coincidence between two channels on each XA module isn't needed because there is no split of two channels per XA module in the Fluka simulation. We note that the simulation didn’t run through the same detector simulation and reconstruction due to difficulty in integrating Fluka with the data analysis framework used for this paper. However, this is taken into account by the photon to peak ADC calibration detailed in Sec~\ref{sec:ph2adccali}. We consider the above cuts on light signal timing and amplitude as a minimal selection for a minimally biased data-MC comparison. The value of this work is to offer a starting point for the modeling of neutrons with LAr and typical materials in the DUNE cryostat, and the validation of the simulation. This simulation is now extended into a bigger cryostat, and work is underway to interface with the official DUNE software simulation. These will be reported in a future publication.

\section{Result}
\label{sec:mainresult}

In this section, we present the main result of the analysis. The light signal amplitude (Sec.~\ref{sec:lightdatamc}) and timing (Sec.~\ref{sec:timing}) are compared between data and simulation.

\subsection{Light Signal Amplitude}
\label{sec:lightdatamc}

\begin{figure}[h!]
  \centering  \includegraphics[width=0.6\columnwidth]{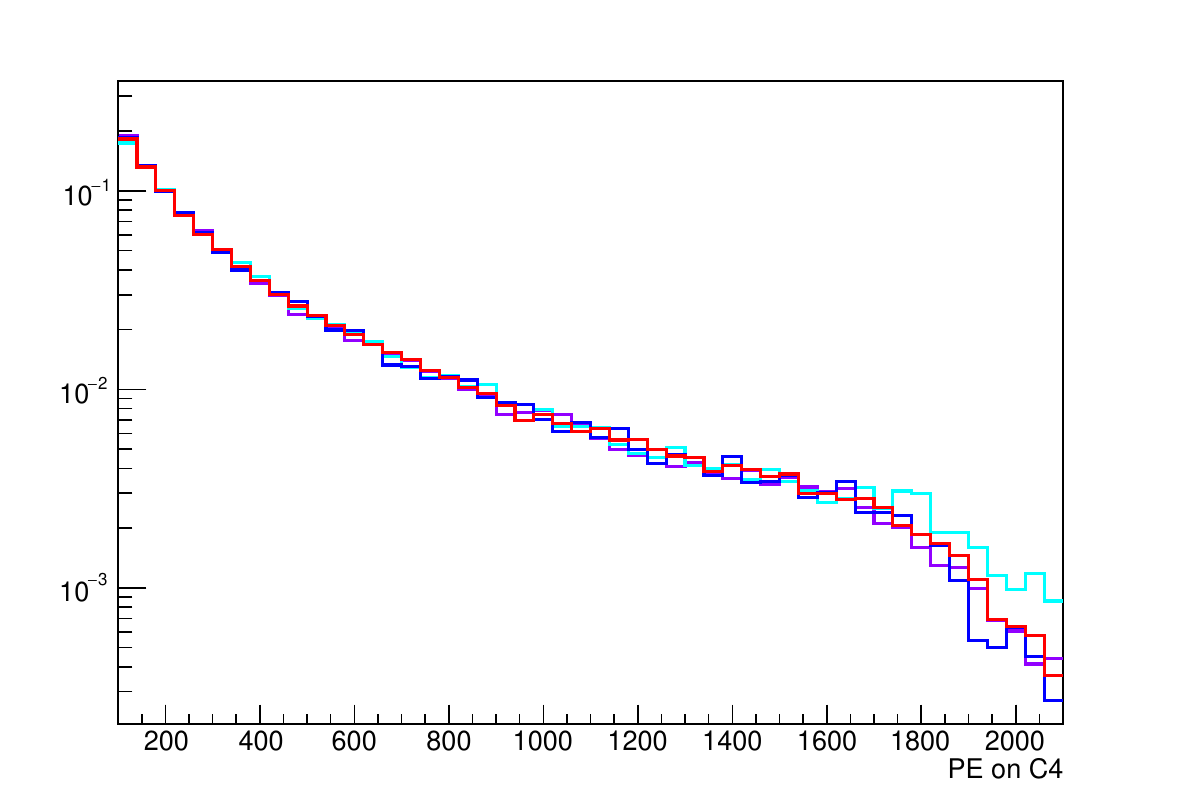}
  \caption{Normalized PE spectrum from four cosmic runs taken during the PNS data taking period.}
  \label{fig:cosmicpestability}
\end{figure}

The cosmic background is modeled with a data-driven method. Only the first millisecond of data is used for all cosmic runs to be consistent with the PNS run DAQ window. The total number of collected triggers from all cosmic runs is normalized to the total PNS triggers in Tab.~\ref{tab:runinfo}. During all data taking time, the cosmic PE spectrum is stable, as shown in Fig.~\ref{fig:cosmicpestability} for four different cosmic runs during the PNS data taking period. 

\begin{figure}[b!]
  \centering  \includegraphics[width=0.49\columnwidth]{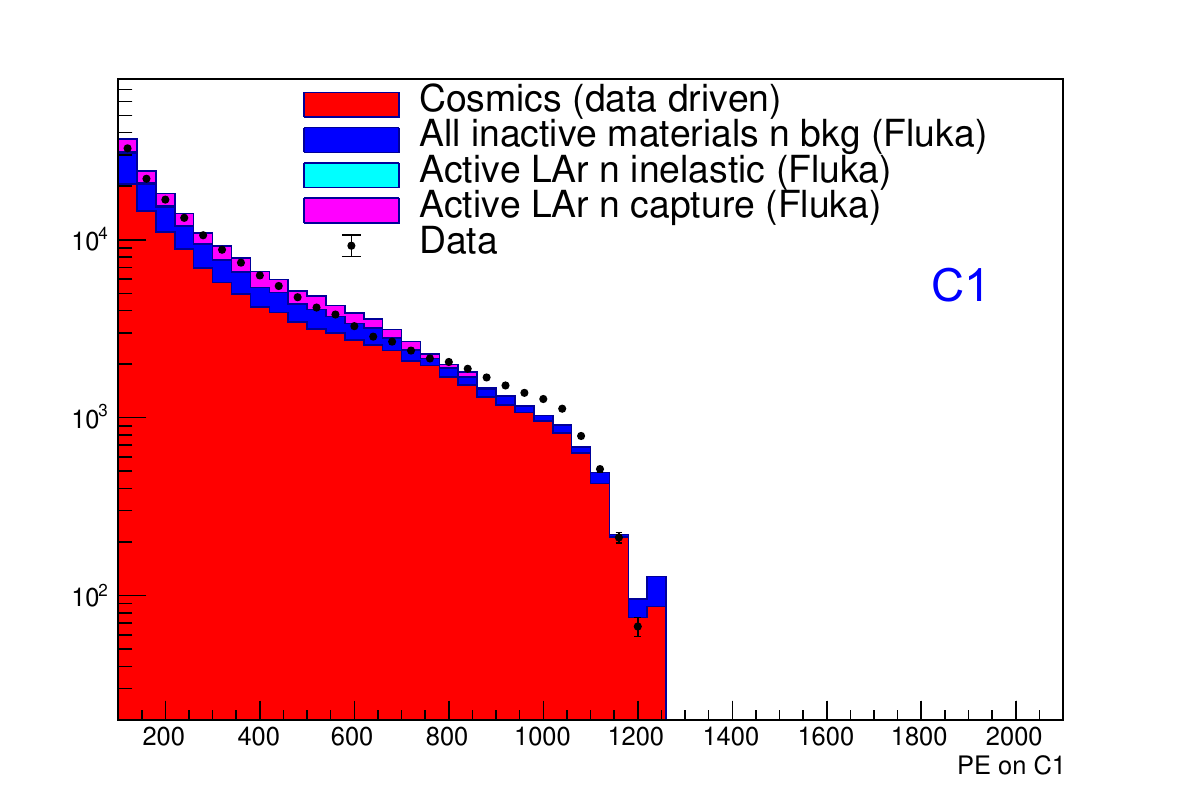}
  \centering  \includegraphics[width=0.49\columnwidth]{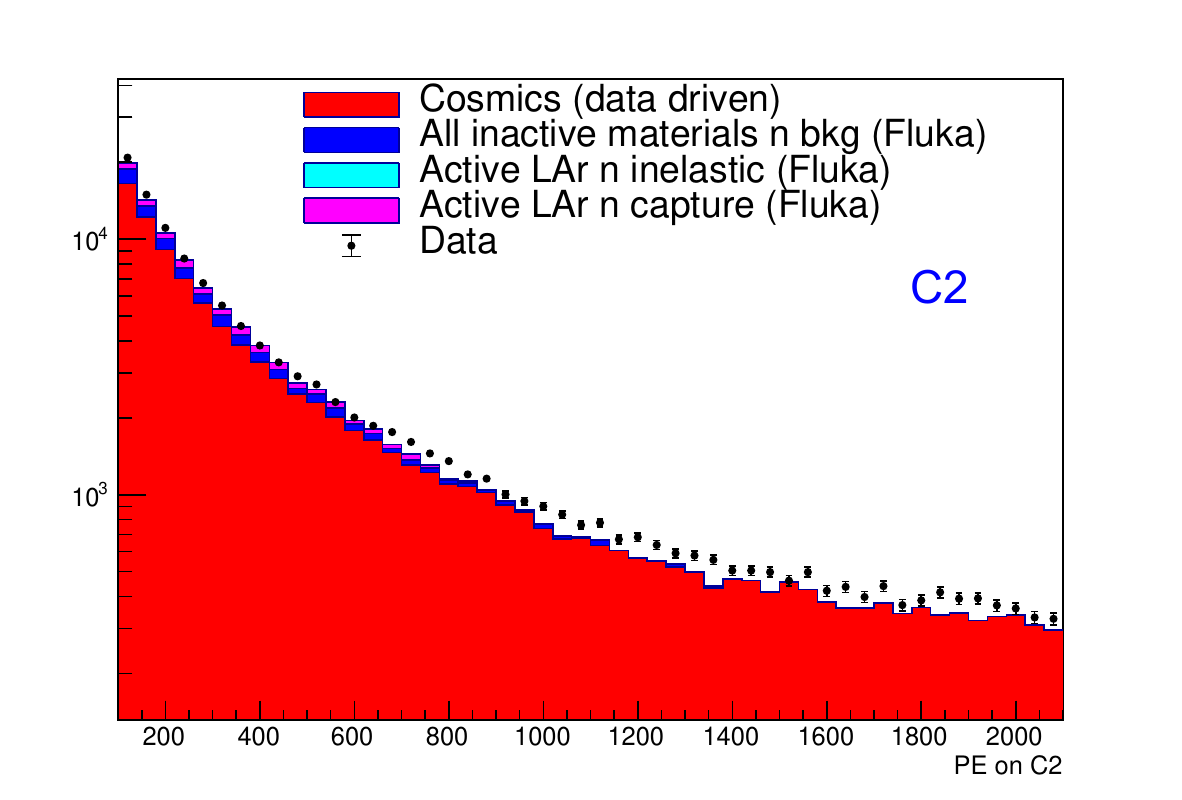}
  \centering  \includegraphics[width=0.49\columnwidth]{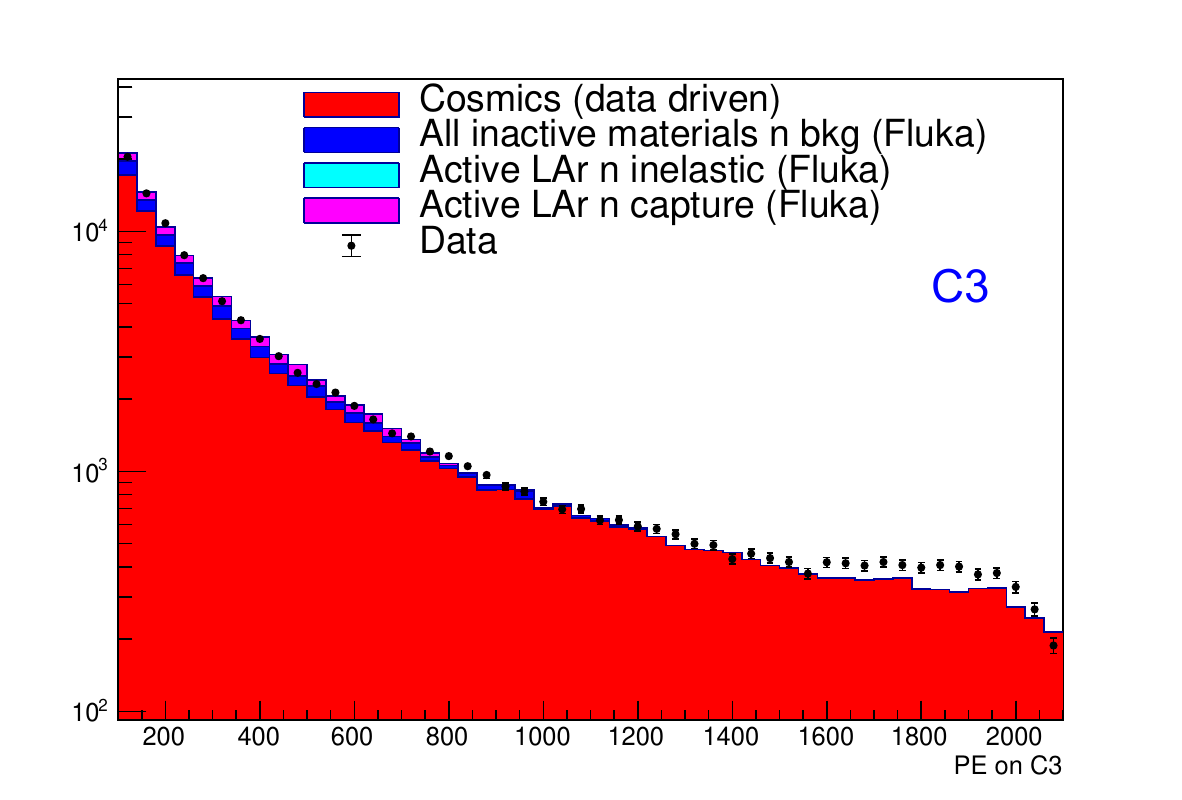}
  \centering  \includegraphics[width=0.49\columnwidth]{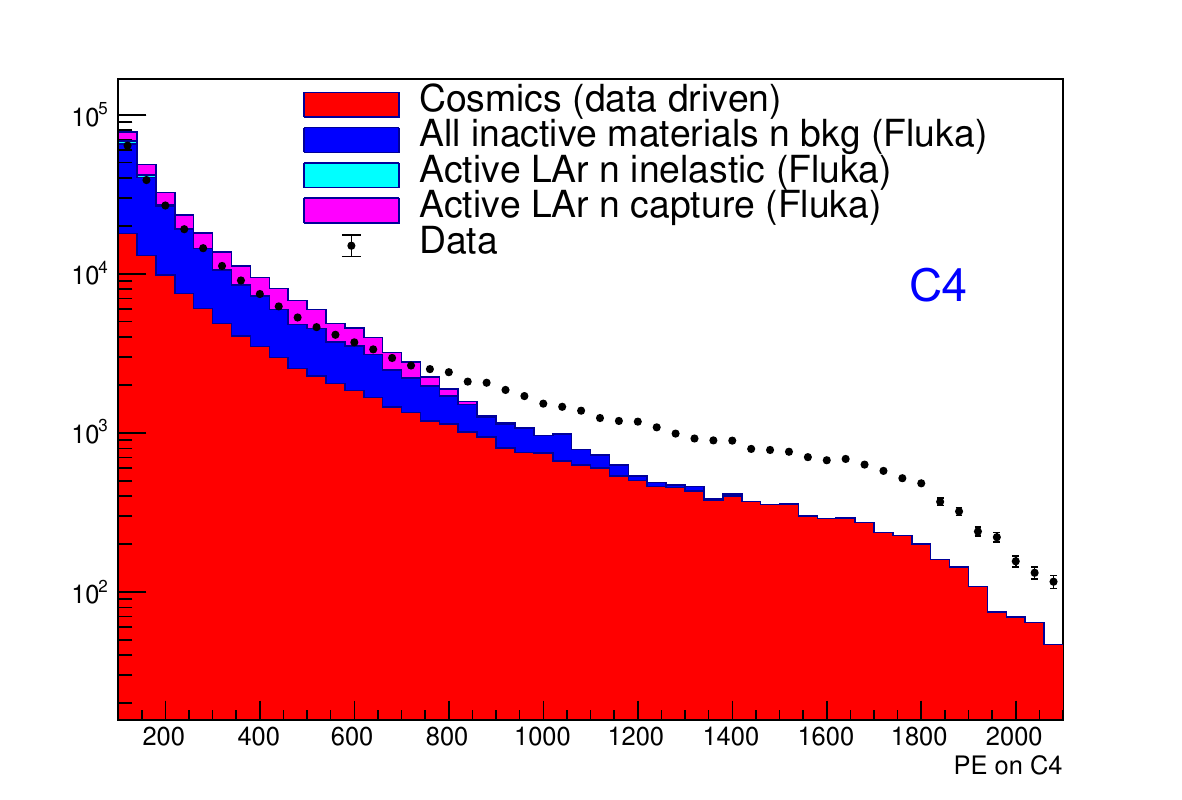}
  \caption{Comparison of light signals in Fluka simulation and PNS run data for all four XA modules on the cathode. Top left: C1. Top right: C2. Bottom left: C3. Bottom right: C4. }
  \label{fig:fulldatamccompare}
\end{figure}

A full comparison of data from all PNS runs in Tab.~\ref{tab:runinfo} to the Fluka simulation is shown in Fig.~\ref{fig:fulldatamccompare}. A scale factor of 0.3 is applied to the total photons in each simulated neutron event to be consistent with the photoelectron calculation method in data as described in Sec.~\ref{sec:ph2adccali}. Because of the unknown neutron beam intensity, we need to normalize the Fluka simulation to the rest of the data after the cosmic backgrounds are subtracted. We derive this normalization factor based on the C3 module because its two channels show the best agreement in PE response to cosmic rays, as demonstrated in Tab.~\ref{tab:pdecali} and detailed in Sec.~\ref{sec:pdecali}. The normalization factors are also calculated separately for runs with different numbers of neutron beam bunches observed and then applied to simulation events for all XAs modules consistently.

\begin{figure}[t!]
  \centering  \includegraphics[width=0.49\columnwidth]{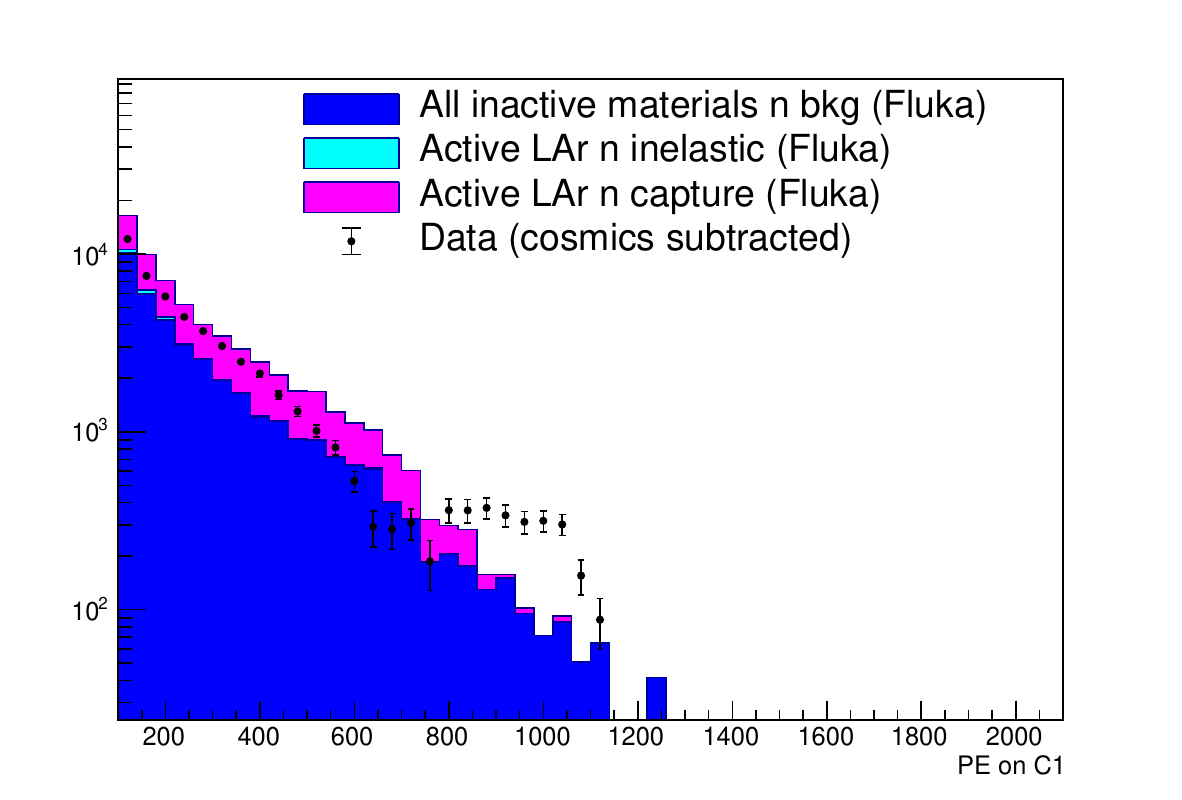}
  \centering  \includegraphics[width=0.49\columnwidth]{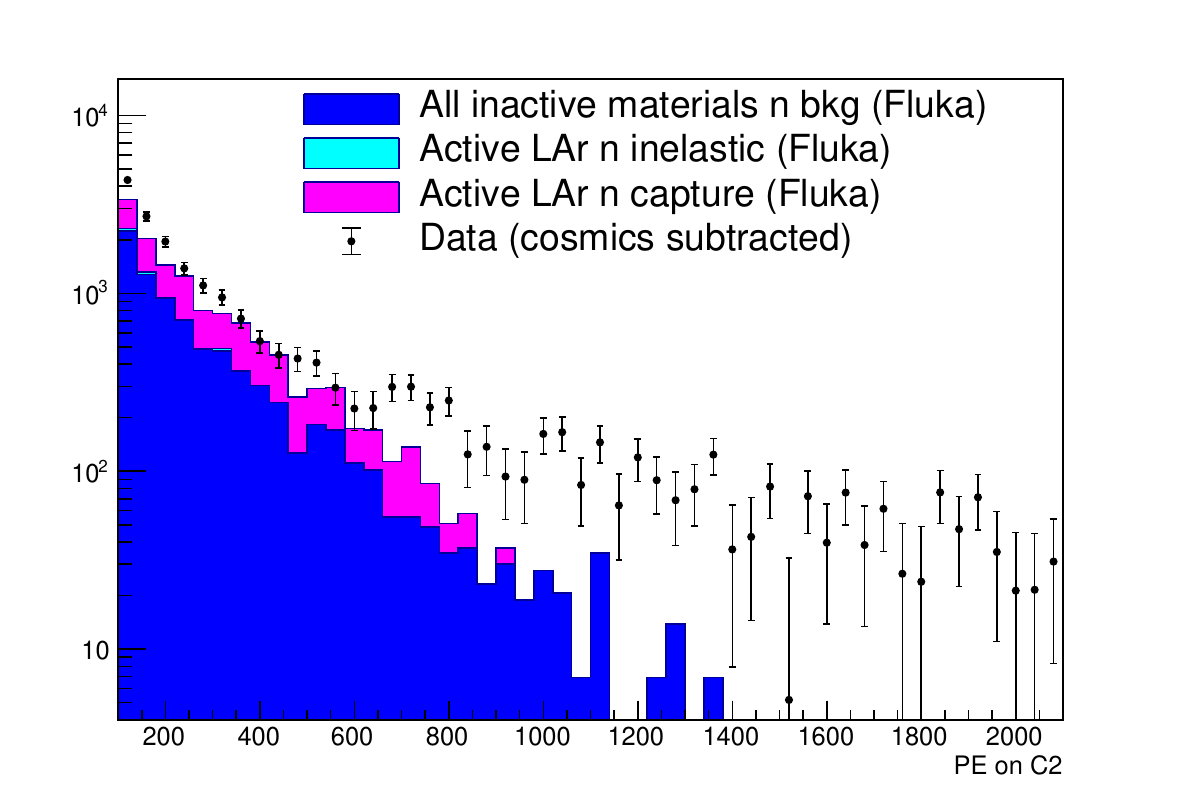}
  \centering  \includegraphics[width=0.49\columnwidth]{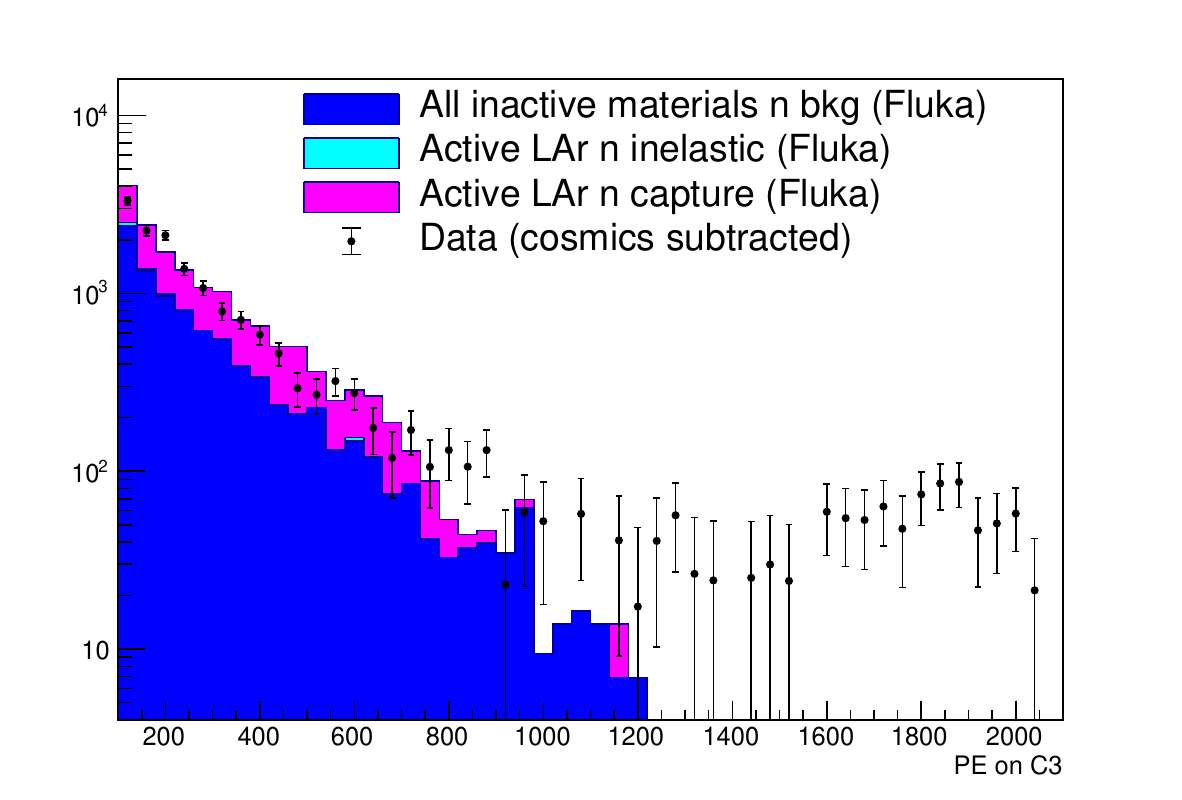}
  \centering  \includegraphics[width=0.49\columnwidth]{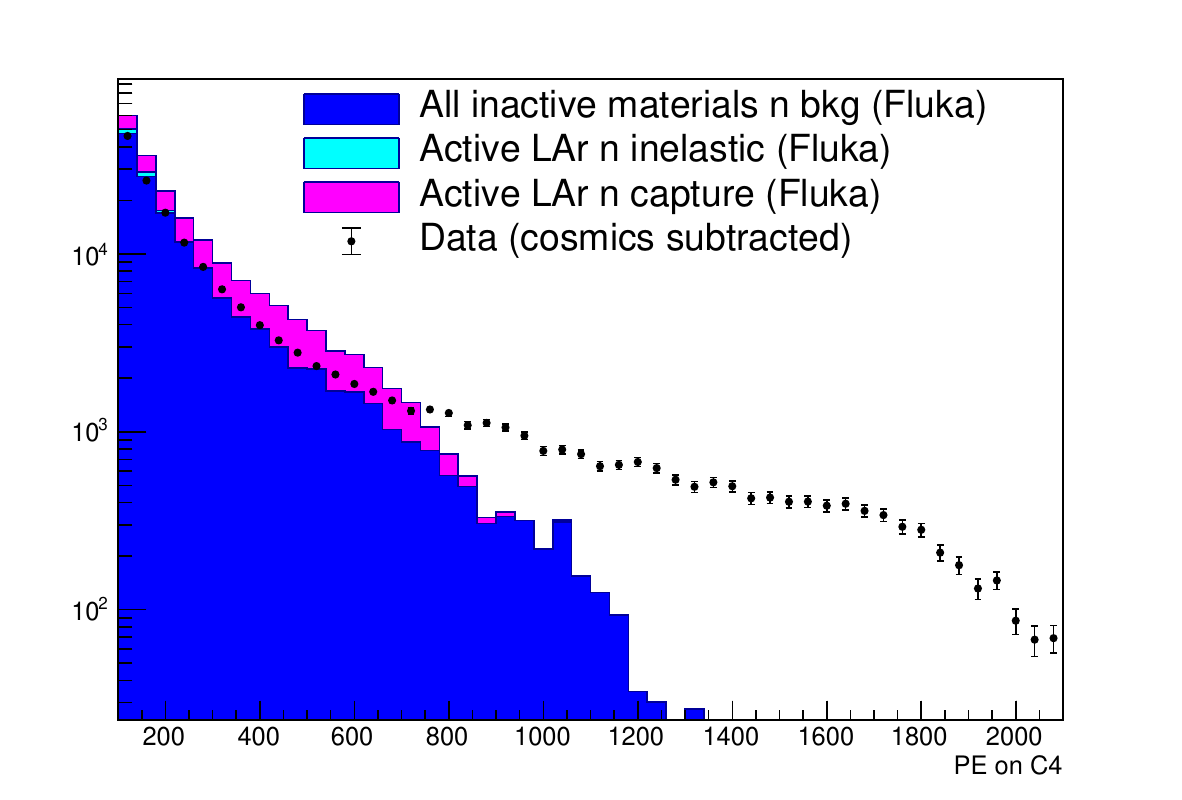}
  \caption{Comparison of neutron-related light signals amplitude (PE) in Fluka simulation and PNS run data for each XA module on the cathode. Predicted cosmic background is subtracted from the data distribution. Top left: C1. Top right: C2. Bottom left: C3. Bottom right: C4.}
  \label{fig:datamccomparenocosmics}
\end{figure}

The data-to-MC comparison after subtracting cosmic backgrounds from the PNS data is shown in Fig.~\ref{fig:datamccomparenocosmics}. An overall good agreement between data and MC is observed for all XA modules on the cathode up to about 650 PE. Above 650 PE, an excess of data is observed on all XA modules that is not accounted for by the simulation. This could originate from several possible sources, and further study is presented in Sec.~\ref{sec:highPEexcess}.

The small disagreement below 650 PE could come from minor displacement of detector positions or the position of the neutron source in the simulation, as part of the systematic effects discussed in Sec.~\ref{sec:syst}. For modules that are closer to the PNS (C4 being the closest), a larger neutron-induced background from interactions on materials outside active LAr is observed. The neutron inelastic scattering background in the active LAr region is small on all modules, as we expect most inelastic interactions to happen earlier in the CB wall and the passive LAr region. For the C1 module, the data stops around 1100 PE due to the applied relative PDE calibration factor of 1.85 as described in Sec.~\ref{sec:pdecali}. The total number of entries for each component in Fig.~\ref{fig:fulldatamccompare} is listed in Tab.~\ref{tab:evtrate}.

\begin{table}[h!]
\centering
\begin{tabular}{p{0.025\textwidth}|p{0.085\textwidth}|p{0.085\textwidth}|p{0.09\textwidth}|p{0.09\textwidth}|p{0.1\textwidth}|p{0.085\textwidth}|p{0.1\textwidth}|p{0.08\textwidth}}
\hline\hline
    XA & Cosmics & Inactive LAr (MC) & Active LAr Inelastic (MC) & Active LAr Capture (MC) & Total Predicted (MC) & Data & Data - Cosmics & <PE> \\
    \hline
    C1         & 111983  &  37699   &  1049  &  24627  & 63375   & 161979 & 49996 & 286\\
    C2         & 96803   &  8158    &  135   &  4884   & 13177   & 115962 & 19159  & 401\\
    C3         & 92630   &  8957    &  176   &  6747   & 15881   & 108511 & 15881  & 366\\
    C4         & 98222   &  143724  &  4759  & 48205   & 196688  & 257611 & 159389 & 343\\
\hline\hline
\end{tabular}
\caption{Predicted number of light signals for each component in PNS data. The Fluka MC simulation is normalized to the data on the C3 module after the predicted cosmic background is subtracted. The same MC scale factor is applied to all XA modules. The average photoelectrons from the cosmic subtracted data in Fig.~\ref{fig:datamccomparenocosmics}, $<\text{PE}>$, is shown in the last column.}
\label{tab:evtrate}
\end{table}

\subsection{Light Signal Time}
\label{sec:timing}

\begin{figure}[h!]
  \centering  \includegraphics[width=0.6\columnwidth]{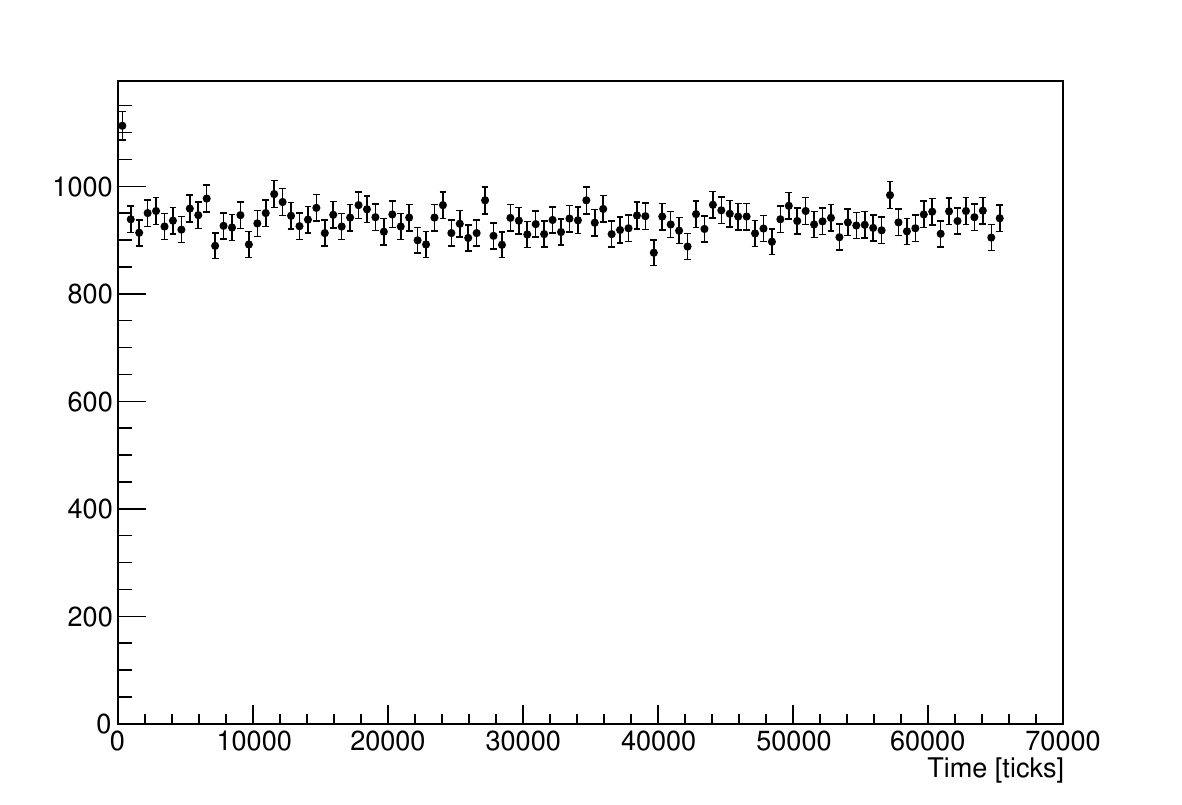}
  \caption{Observed light signal peak time on XA C4 module in cosmic runs.  Each time tick represents 16 ns.}
  \label{fig:datatimingcosmics}
\end{figure}

The light signal timing on the C4 module from all cosmic runs is shown in Fig.~\ref{fig:datatimingcosmics}. A flat timing profile is observed as expected. Similar timing profiles are observed on the other XA modules. 

\begin{figure}[h!]
  \centering  \includegraphics[width=0.42\columnwidth]{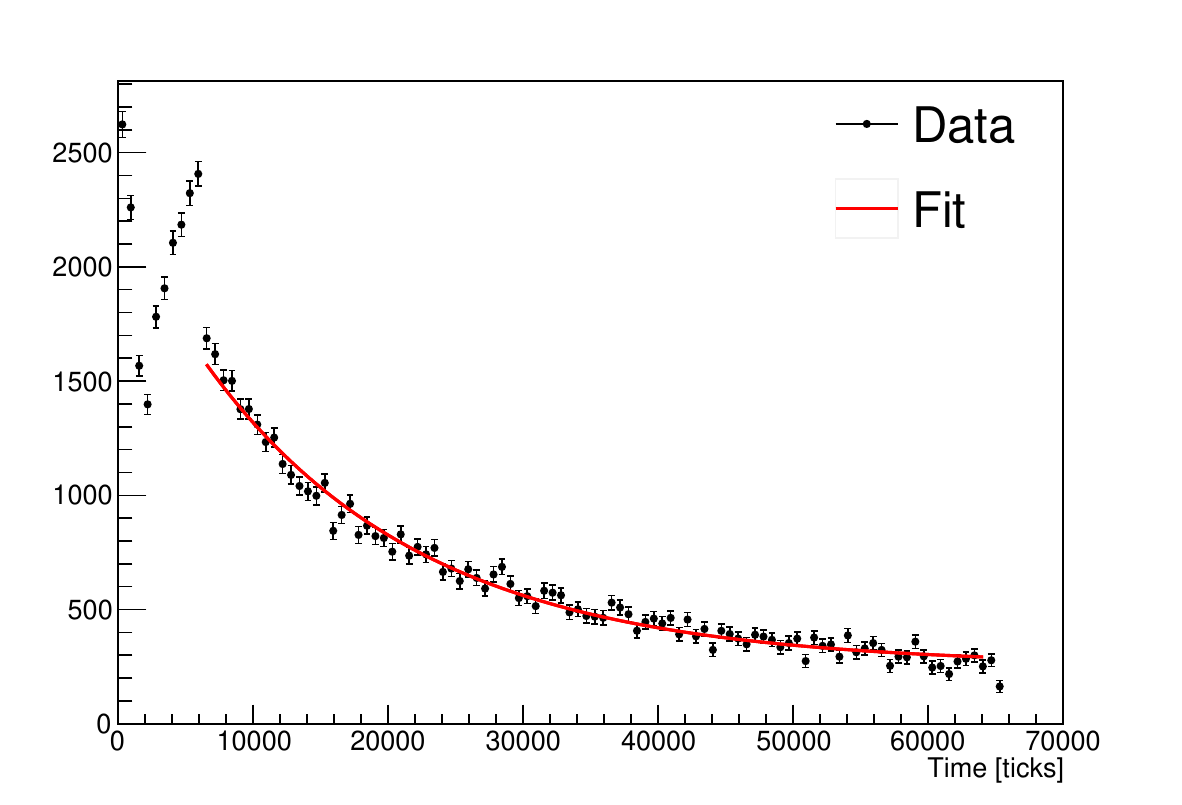}
  \includegraphics[width=0.48\columnwidth]{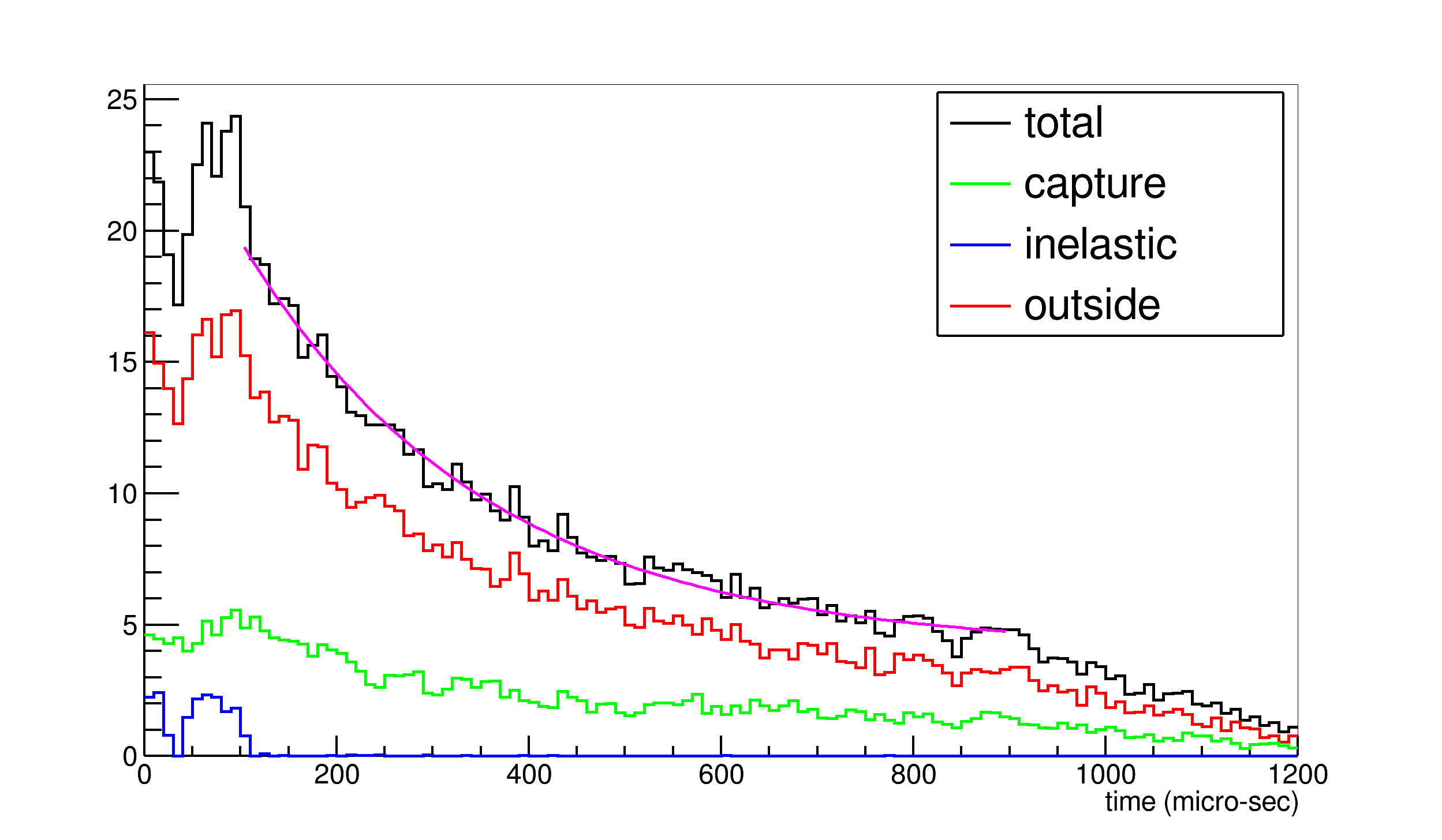}
  \caption{Left: Observed light signal peak time on XA C4 module during a PNS run with the highest statistics. More than one of the five beam bursts is observed in the 1-ms trigger window. The red line shows the fit to obtain the decay time constant in the data. Each time tick represents 16 ns. Right: The time distribution of optical photons (in microseconds) in the Fluka simulation from events with a neutron capture in active LAr (green), from events with inelastic neutron interactions in active LAr (yellow), and from events with no neutron interaction in active LAr (blue). The total distribution is shown in black. The magenta line shows the fit to obtain the decay time constant in the Fluka simulation. }
  \label{fig:datatiming}
\end{figure}

Fig.~\ref{fig:datatiming} shows an observed neutron beam bunch structure in a PNS run on the C4 module closest to the neutron source. A similar beam time profile is observed in all PNS runs, with a slight variation in the observed number of bunches that could be related to the shift of the TTL signal from the PNS hardware or a changed configuration of the time that is read out before the TTL trigger. After the neutron beam stops, we observe a drop in the light signals due to the expected decrease in neutron activity over time. An exponential function is fitted to the decay part of the spectrum between 6300 and 64000 time ticks (i.e., 100.8 - 1024 $\mu$s), and a decay time constant of 257 $\pm$ 8 $\mu$s is obtained.

To further understand the processes that contribute to the exponential decay tail in the timing spectrum, we first reproduce the timing structure observed in the data. In the Fluka simulation, all neutrons are produced at time zero. In the offline analysis, we assign a random time to the time-zero neutron, assuming a time structure with 5 beam bunches, each 60 $\mu$s apart, separated by 20 $\mu$s. An extra time offset is applied depending on how many beam bunches are observed in the DAQ window. Fig.~\ref{fig:datatiming} right plot shows the time distribution of optical photons from all simulated neutron events. The characteristic decay time constant is fit to be 255 $\pm$ 68 $\mu$s in the Fluka simulation from the decay spectrum between 100 and 900 $\mu$s, in good agreement with that obtained in the fit to data within one standard deviation. Fluka simulation also suggests the decay time spectrum receives primary contribution from the neutron interactions outside the active LAr volume.

\begin{figure}[t!]
  \centering  \includegraphics[width=0.49\columnwidth]{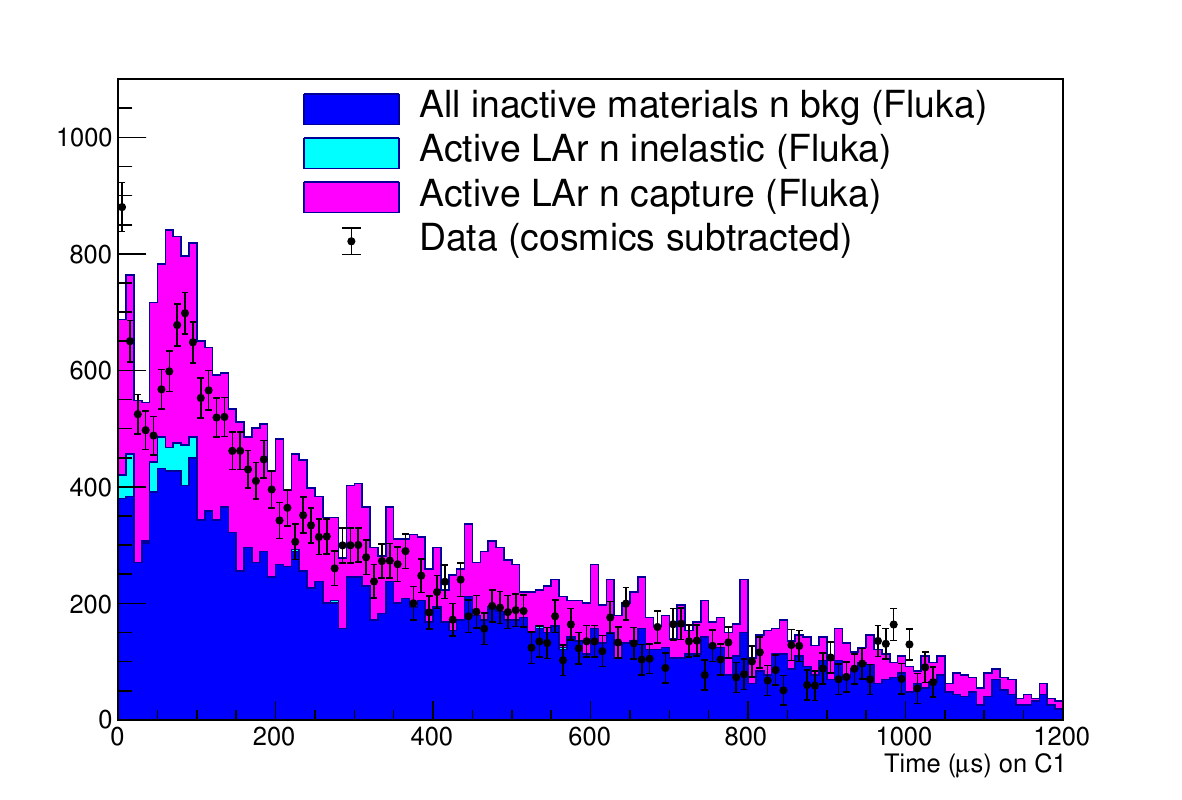}
  \centering  \includegraphics[width=0.49\columnwidth]{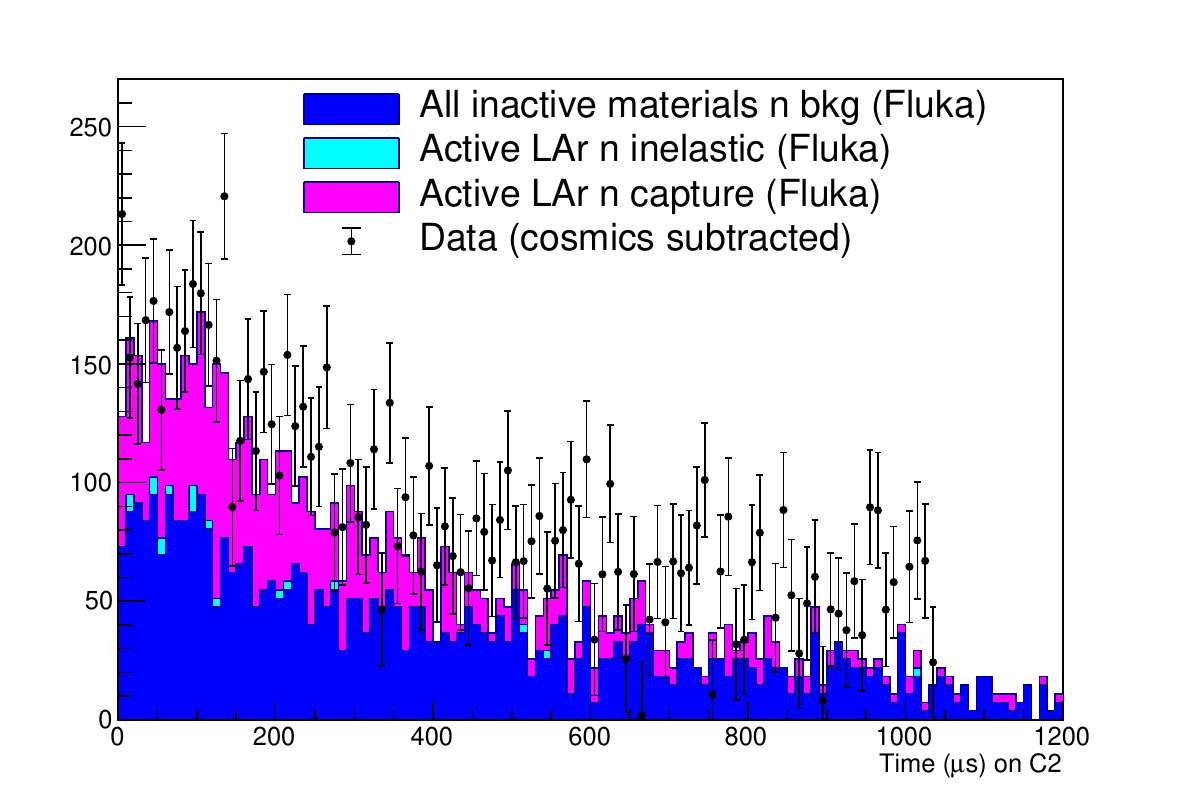}
  \centering  \includegraphics[width=0.49\columnwidth]{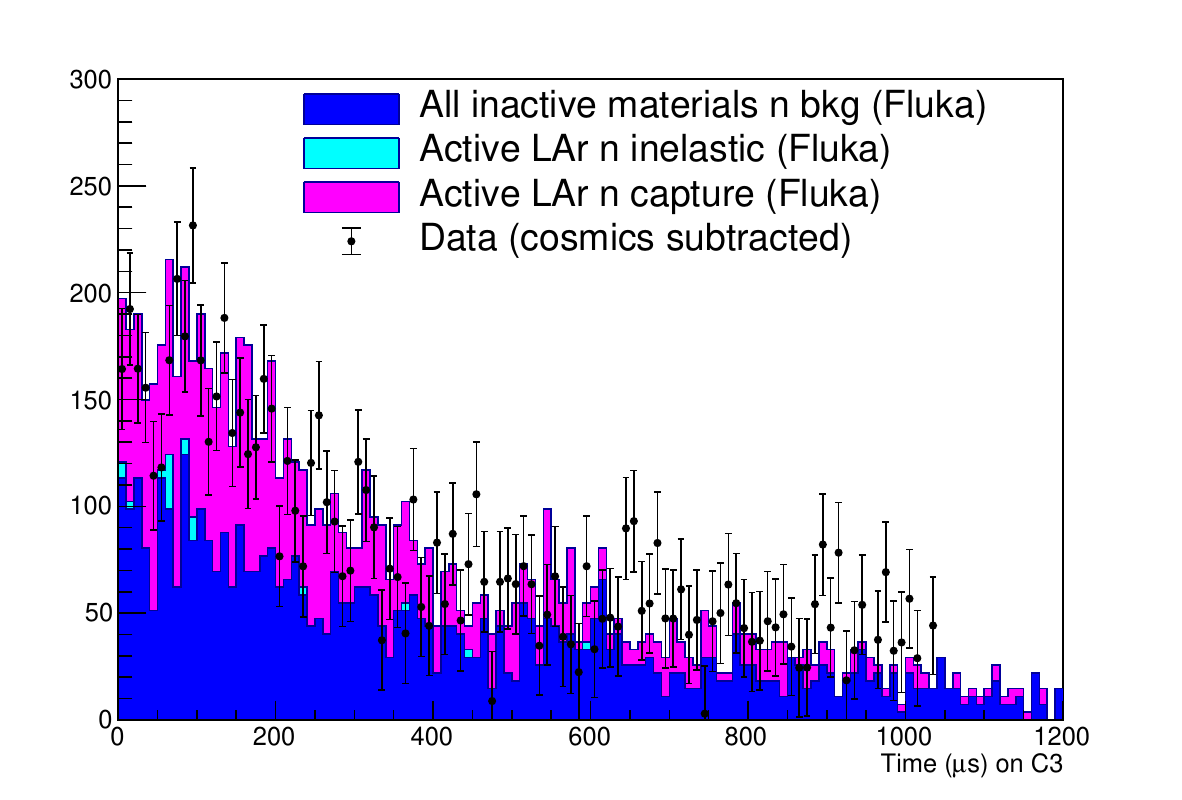}
  \centering  \includegraphics[width=0.49\columnwidth]{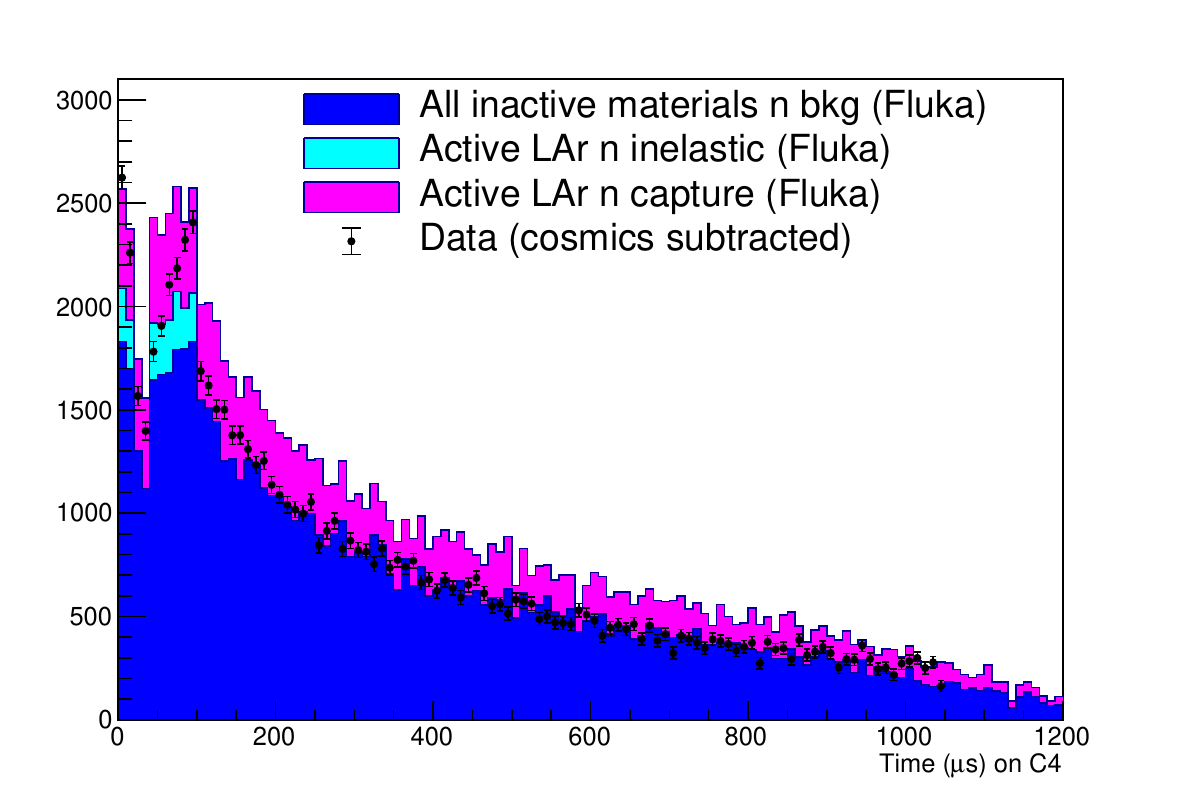}
  \caption{Comparison of neutron-related light signal timing in Fluka simulation with a PNS run where \textit{one and a half} neutron beam bunches are observed in the DAQ window. Predicted cosmic background is subtracted from the data distribution. Top left: C1. Top right: C2. Bottom left: C3. Bottom right: C4.}
  \label{fig:datamccomparenocosmicstiming1nbunch}
\end{figure}

\begin{figure}[t!]
  \centering  \includegraphics[width=0.49\columnwidth]{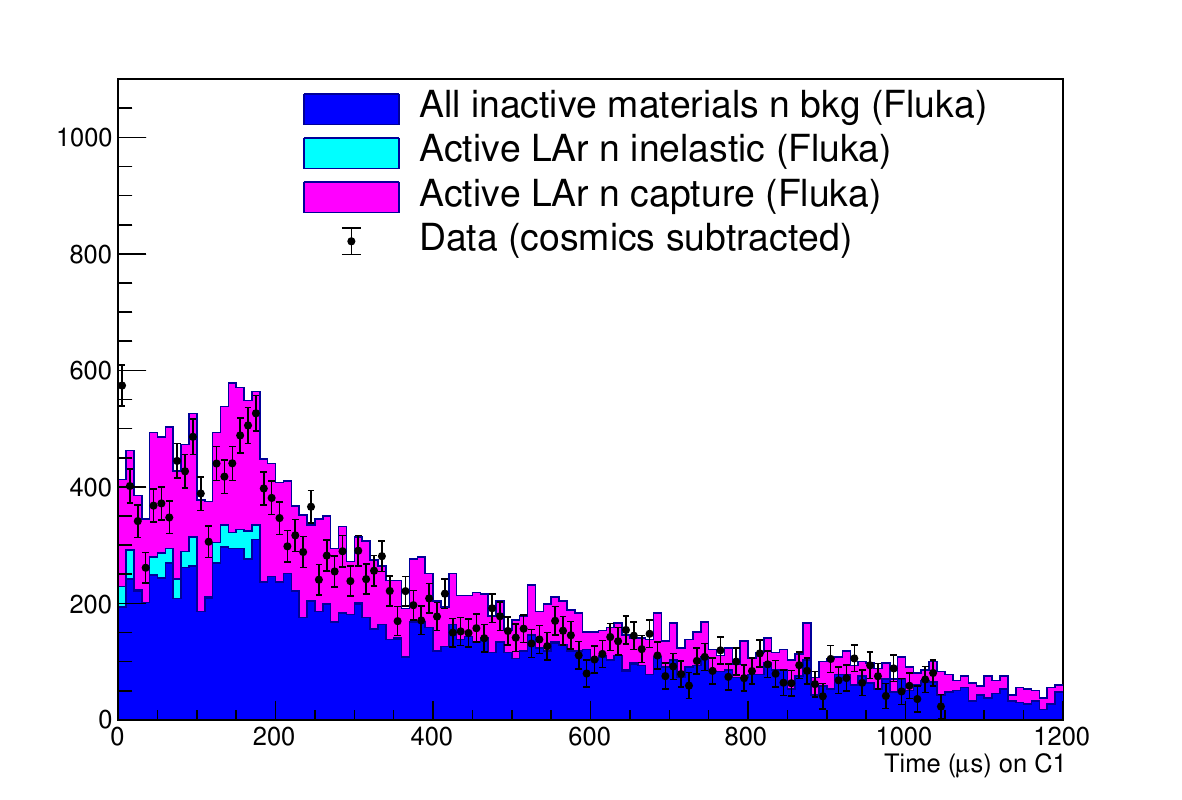}
  \centering  \includegraphics[width=0.49\columnwidth]{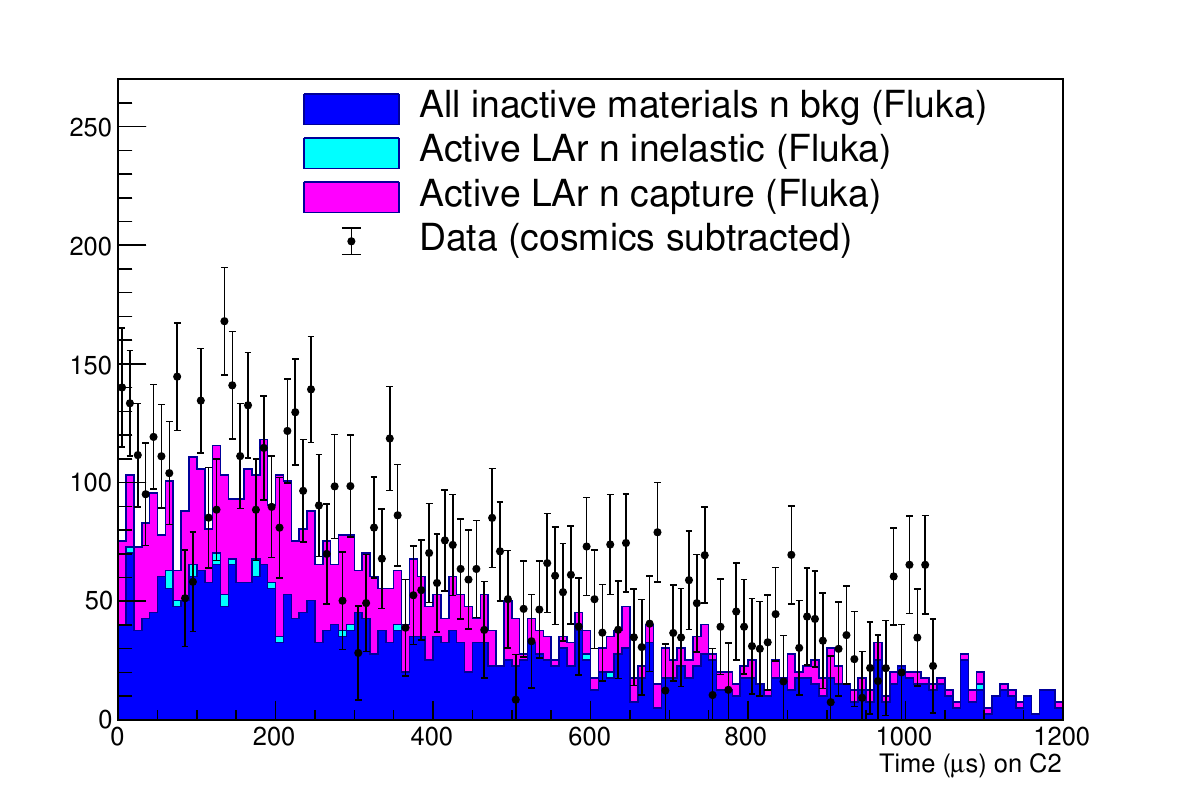}
  \centering  \includegraphics[width=0.49\columnwidth]{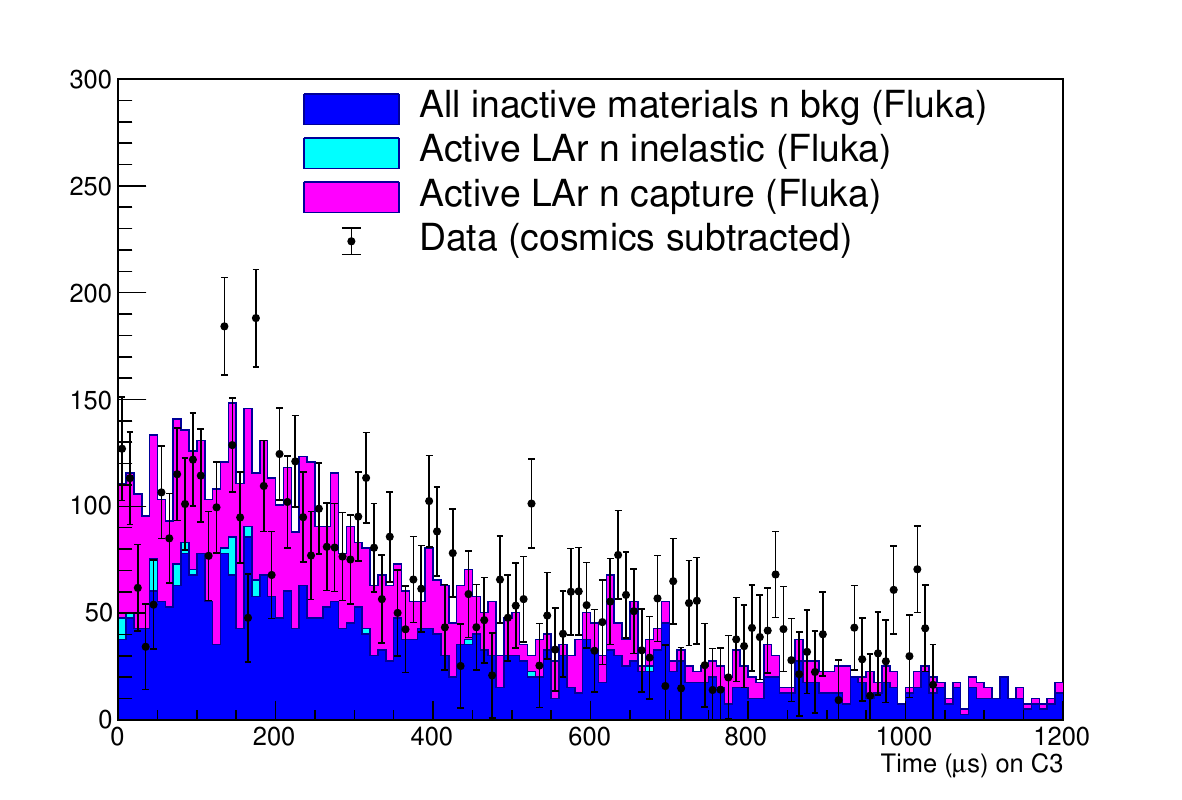}
  \centering  \includegraphics[width=0.49\columnwidth]{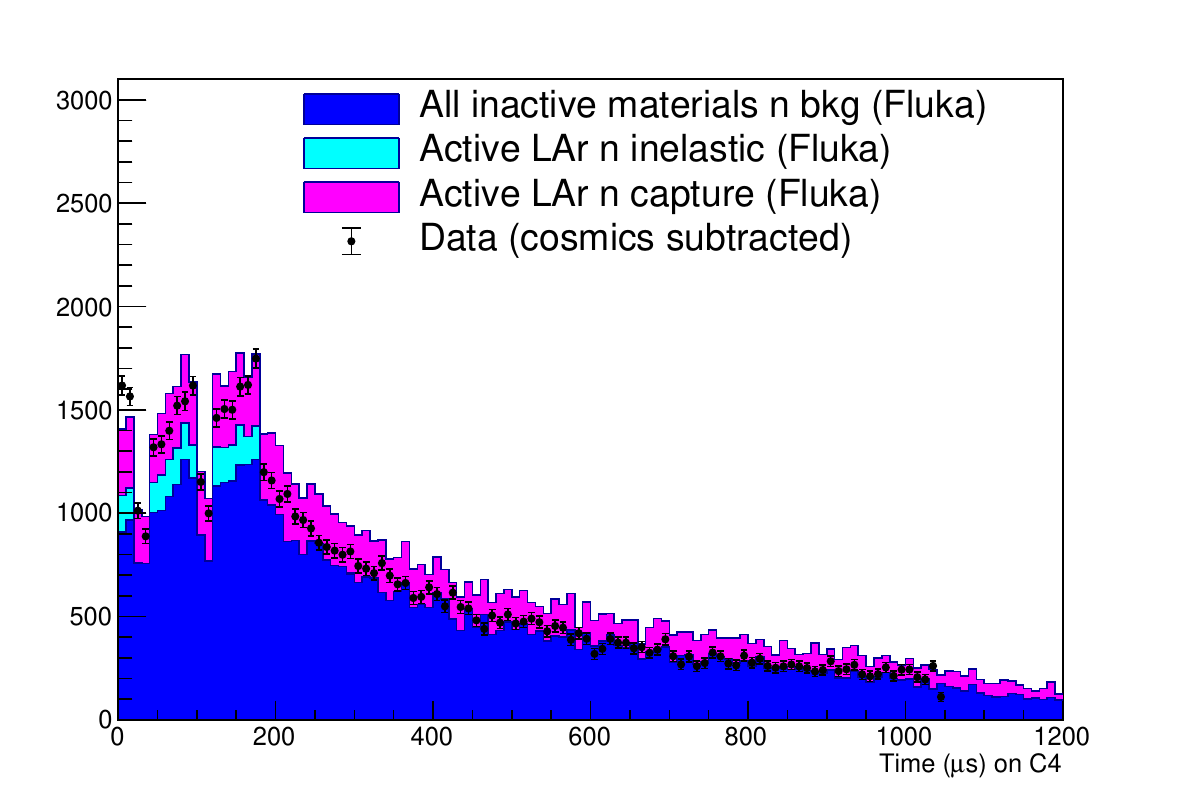}
  \caption{Comparison of neutron-related light signal timing in Fluka simulation with a PNS run where \textit{two and a half} neutron beam bunches are observed in the DAQ window. Predicted cosmic background is subtracted from the data distribution. Top left: C1. Top right: C2. Bottom left: C3. Bottom right: C4.}
  \label{fig:datamccomparenocosmicstiming2nbunch}
\end{figure}

The comparison of the timing distribution with simulation for all four XA modules is shown in Fig.~\ref{fig:datamccomparenocosmicstiming1nbunch} and Fig.~\ref{fig:datamccomparenocosmicstiming2nbunch}. The same simulation-to-data normalization factors used in the PE comparison are applied here. For the PNS run in Fig.~\ref{fig:datamccomparenocosmicstiming1nbunch}, one and a half neutron beam bunches are in the DAQ time window. In Fig.~\ref{fig:datamccomparenocosmicstiming2nbunch}, two and a half neutron beam bunches are in the DAQ time window. In both runs, the simulation successfully reproduces the shape of the timing distribution in the data. The simulation accurately captures the exponential decay of neutron interactions. Some shape discrepancy is observed in the beam-on time period as seen on C4 and C1 modules, which are close to the neutron source. Overall, the simulation predicts more signals than data on these two modules than on the other modules. This suggests further understanding is needed for the neutron source stability and intensity, and the interactions during the neutron beam time.

\subsection{Discussion on Excess at High PE in Data}
\label{sec:highPEexcess}

\begin{figure}[t!]
  \centering  \includegraphics[width=0.49\columnwidth]{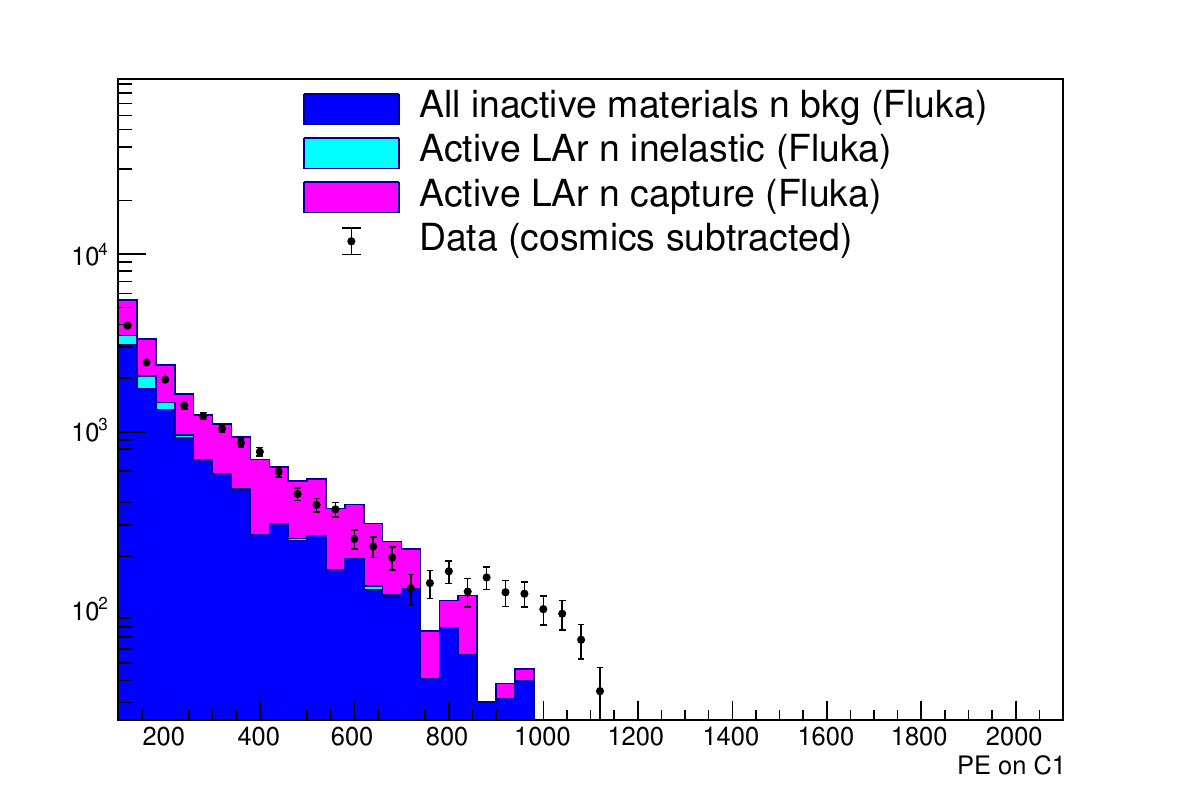}
  \centering  \includegraphics[width=0.49\columnwidth]{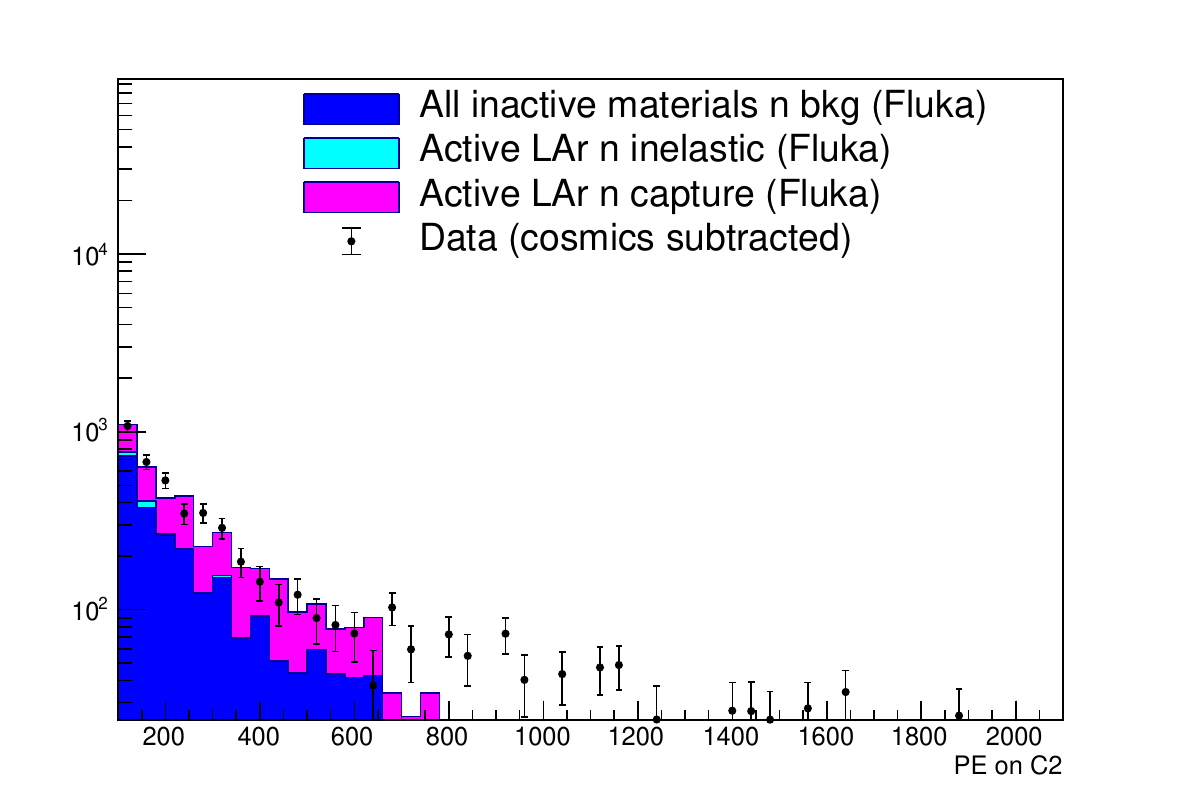}
  \centering  \includegraphics[width=0.49\columnwidth]{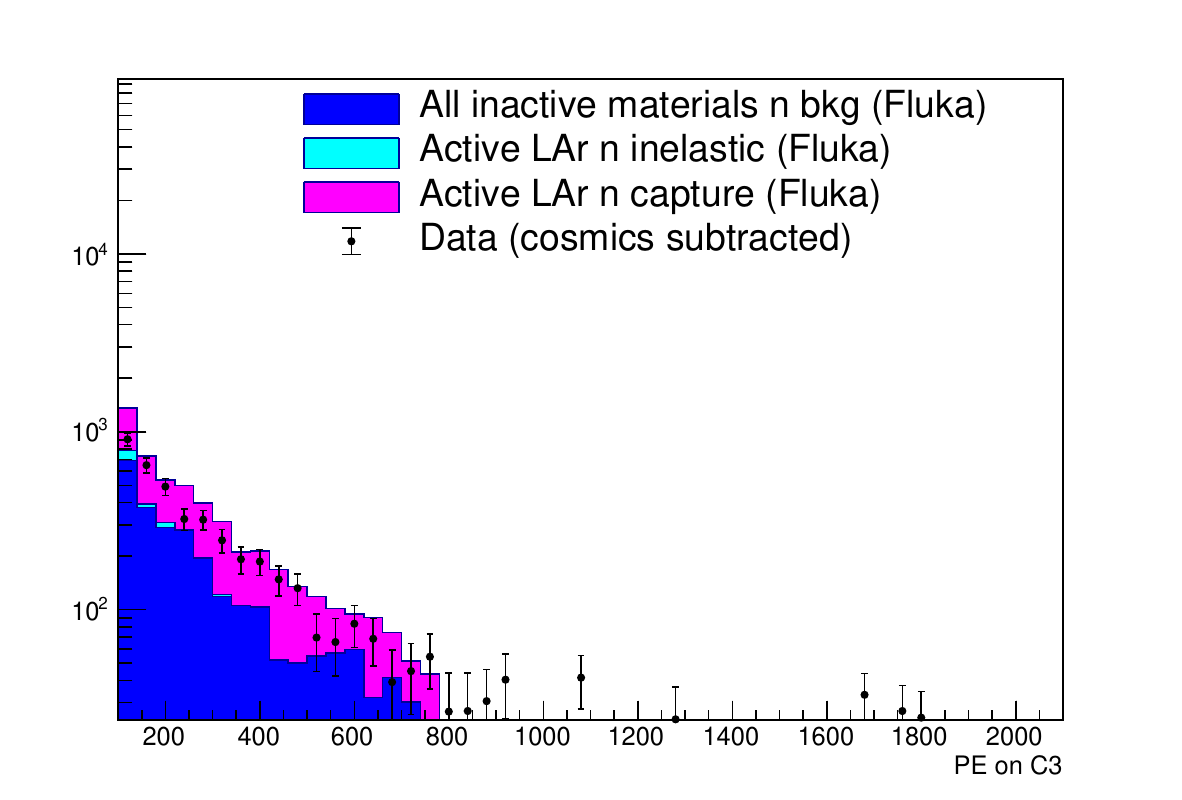}
  \centering  \includegraphics[width=0.49\columnwidth]{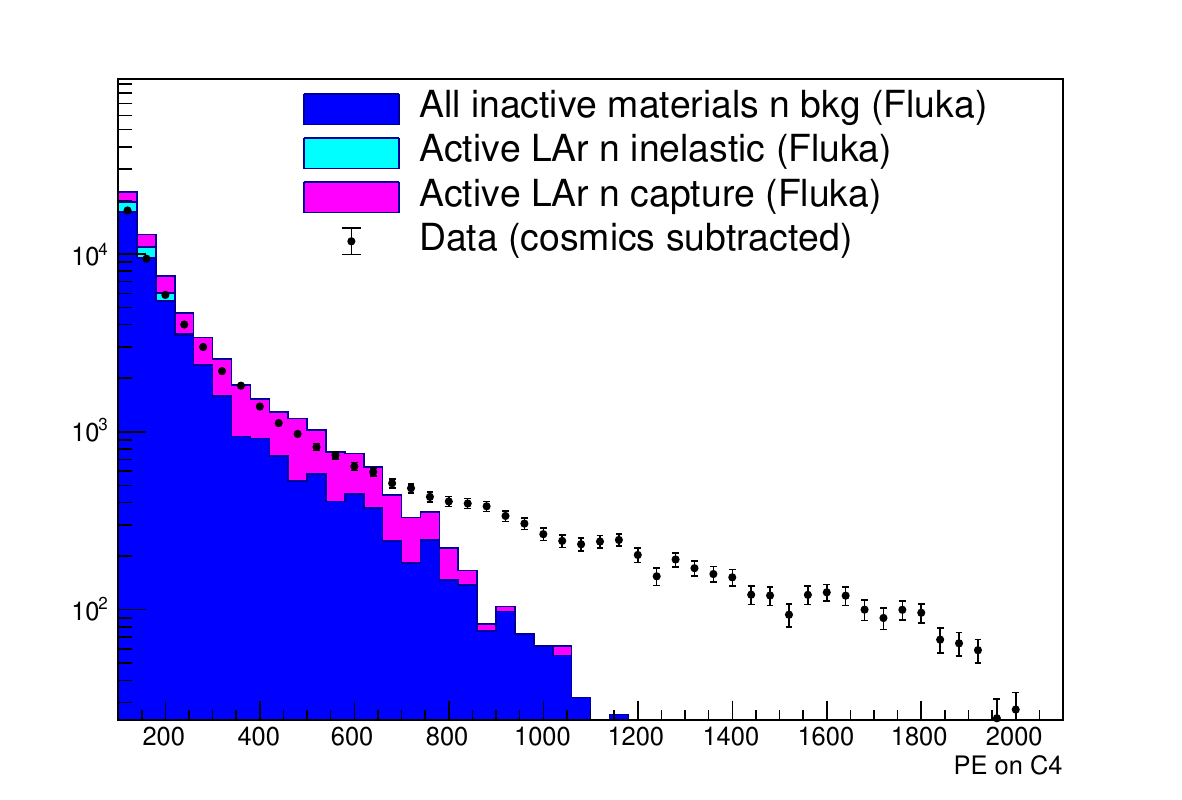}
  \caption{Comparison of neutron-related light signals in Fluka simulation and PNS run data for each XA module on the cathode for the time period 0-160 $\mu$s in the DAQ window. Predicted cosmic background is subtracted from the data distribution. Top left: C1. Top right: C2. Bottom left: C3. Bottom right: C4.}
  \label{fig:datamccomparenocosmicstimeslice0}
\end{figure}

\begin{figure}[t!]
  \centering  \includegraphics[width=0.49\columnwidth]{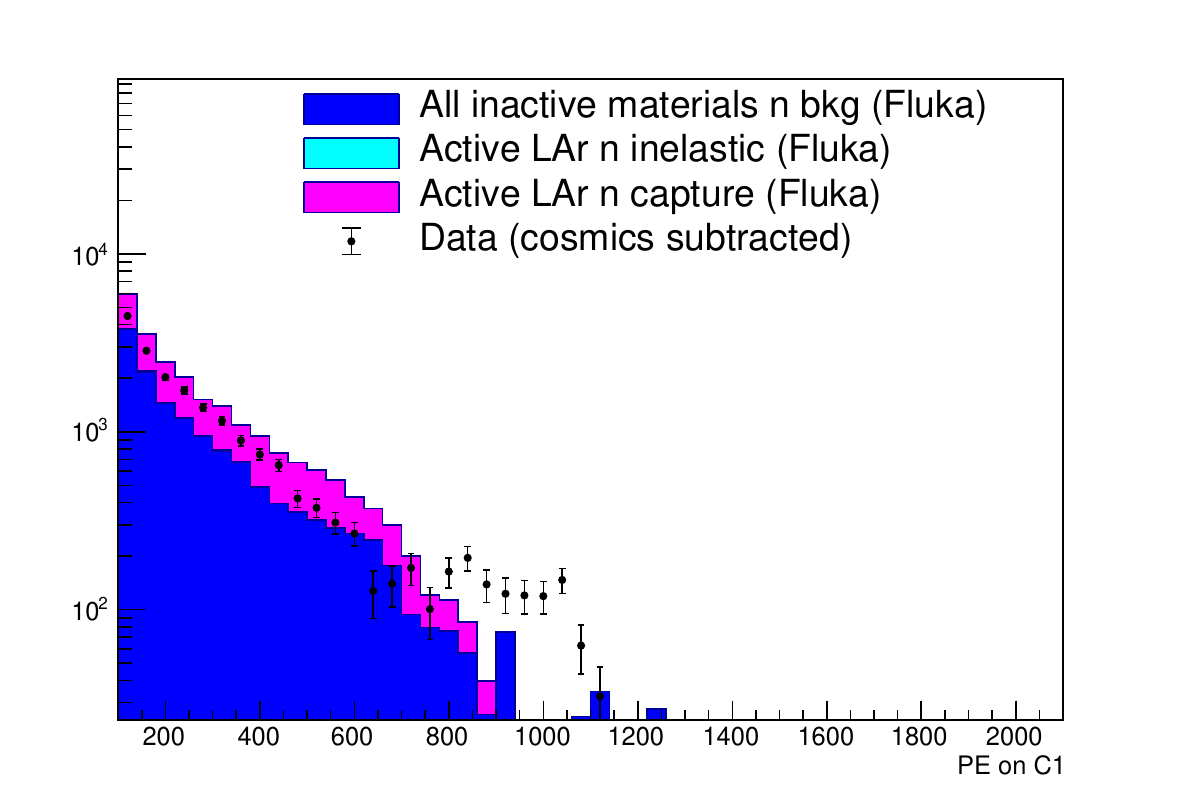}
  \centering  \includegraphics[width=0.49\columnwidth]{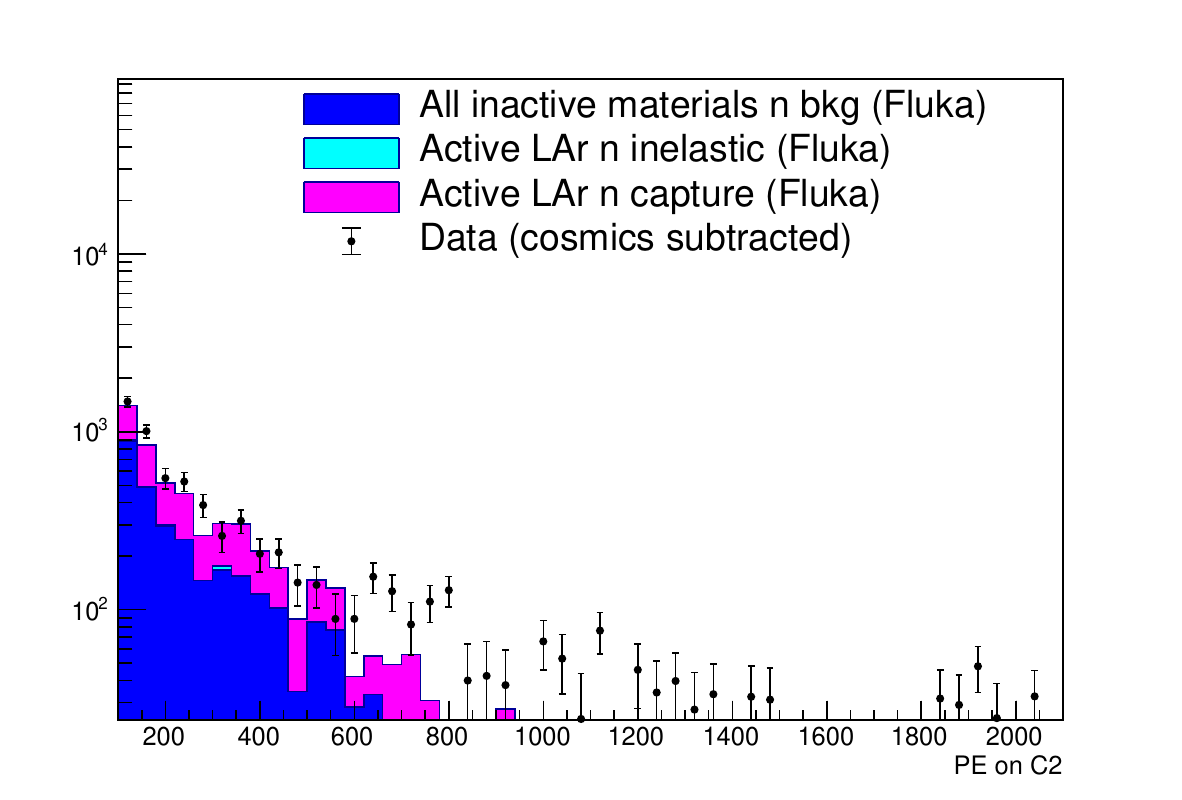}
  \centering  \includegraphics[width=0.49\columnwidth]{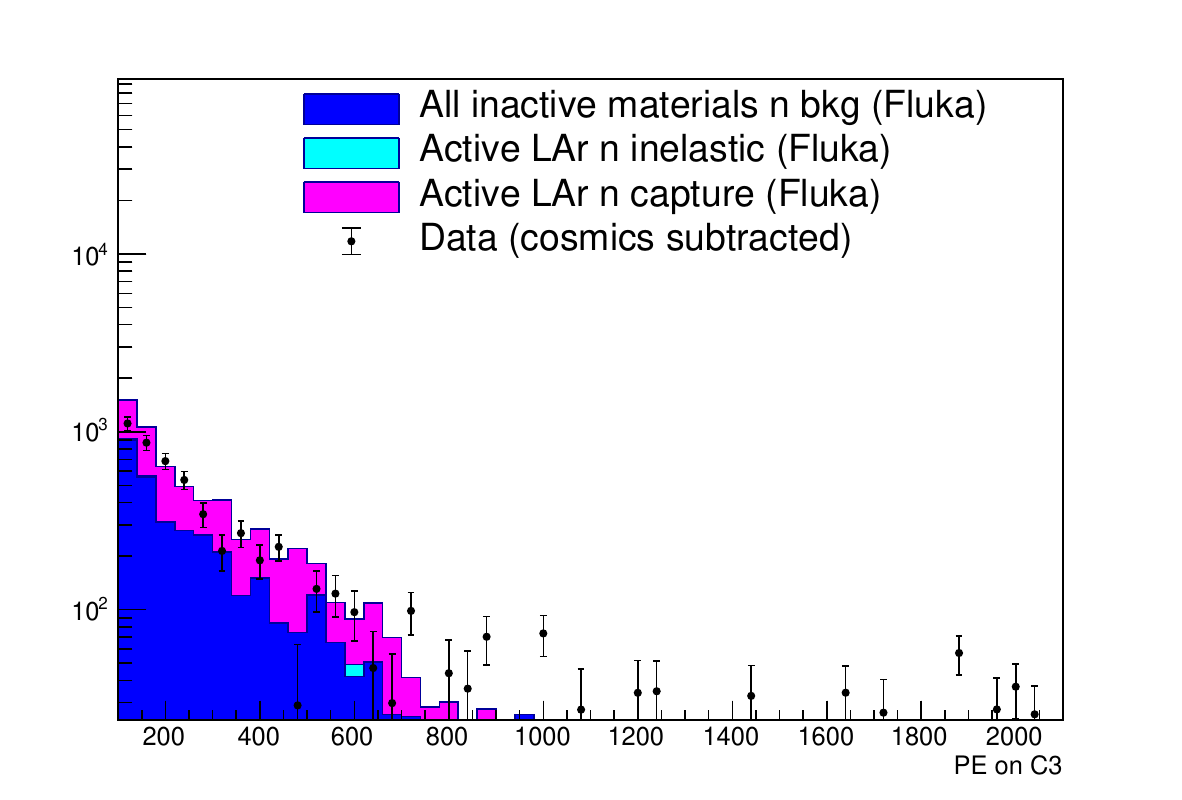}
  \centering  \includegraphics[width=0.49\columnwidth]{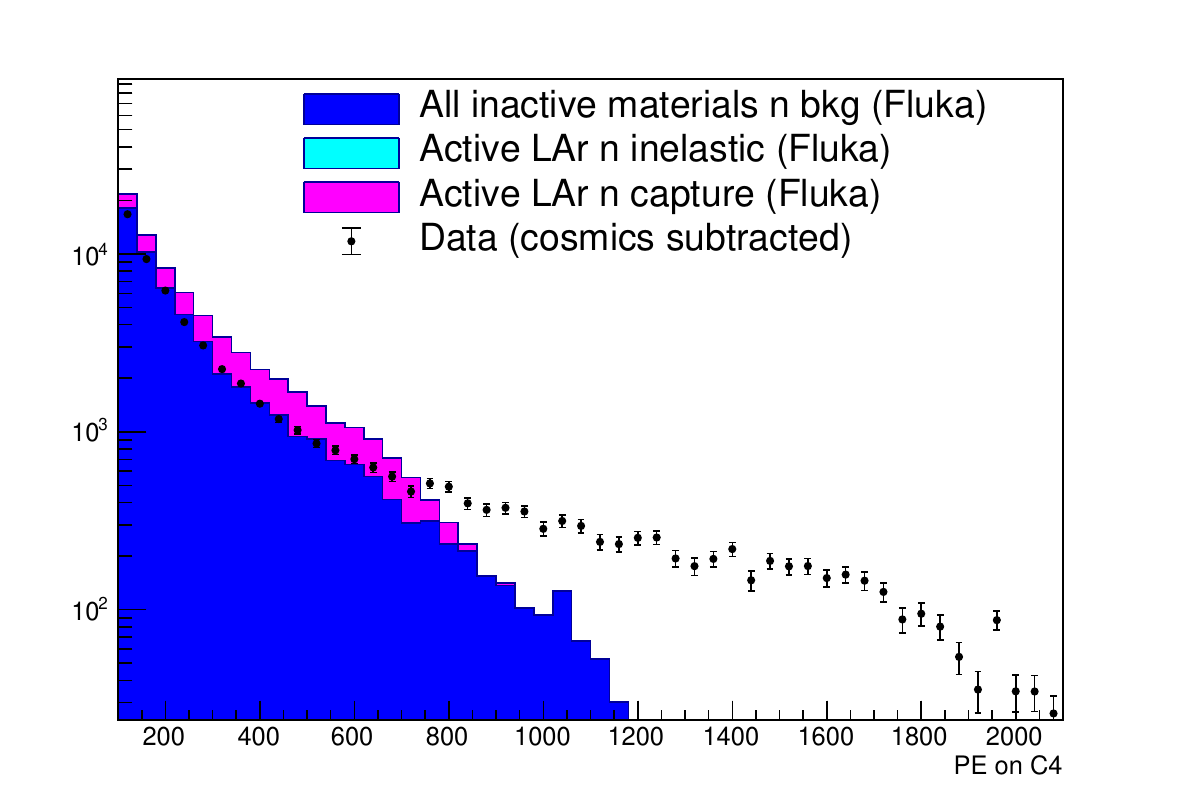}
  \caption{Comparison of neutron-related light signals in Fluka simulation and PNS run data for each XA module on the cathode for the time period 160-480 $\mu$s in the DAQ window. Predicted cosmic background is subtracted from the data distribution. Top left: C1. Top right: C2. Bottom left: C3. Bottom right: C4.}
  \label{fig:datamccomparenocosmicstimeslice1}
\end{figure}

\begin{figure}[t!]
  \centering  \includegraphics[width=0.49\columnwidth]{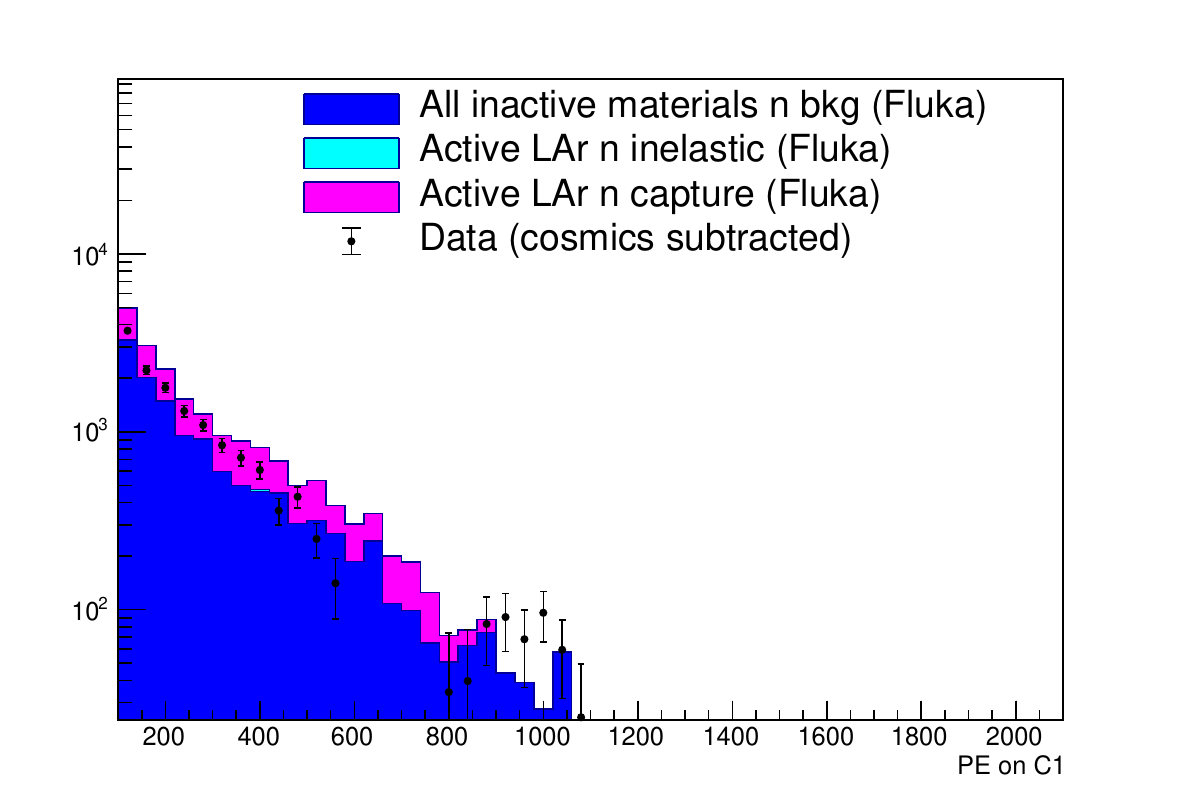}
  \centering  \includegraphics[width=0.49\columnwidth]{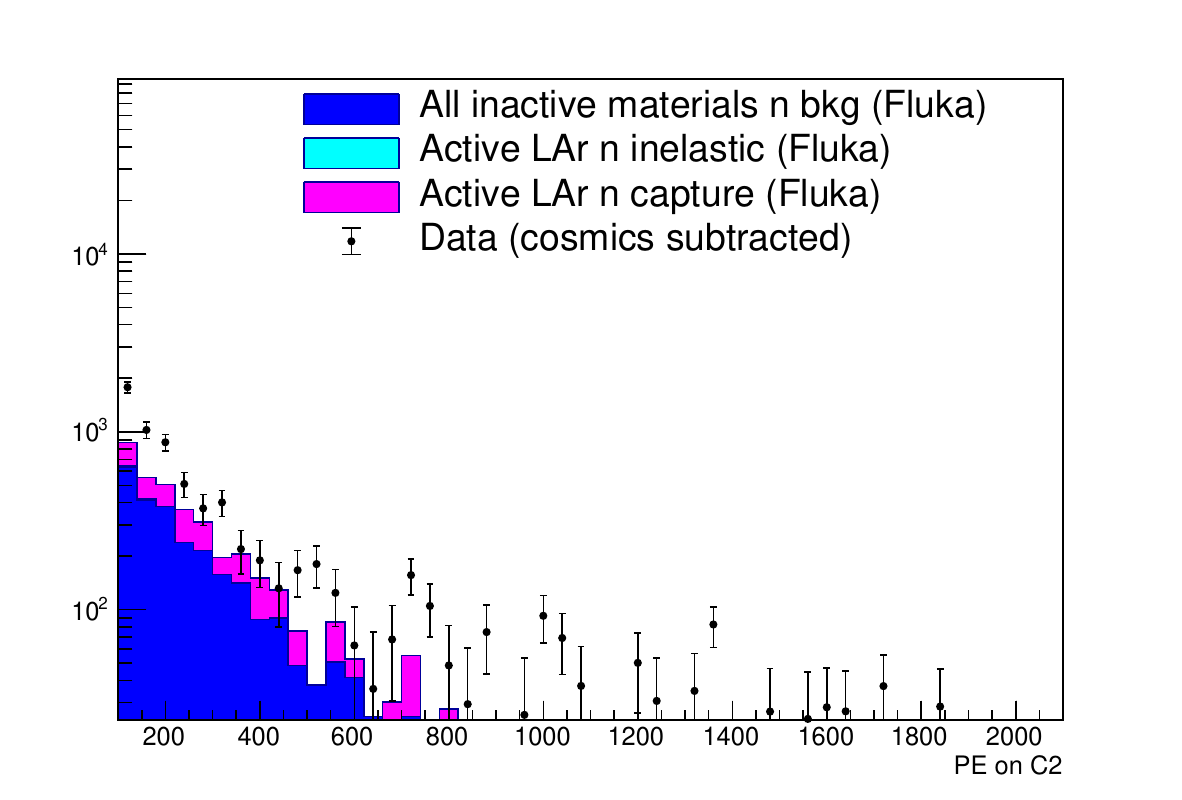}
  \centering  \includegraphics[width=0.49\columnwidth]{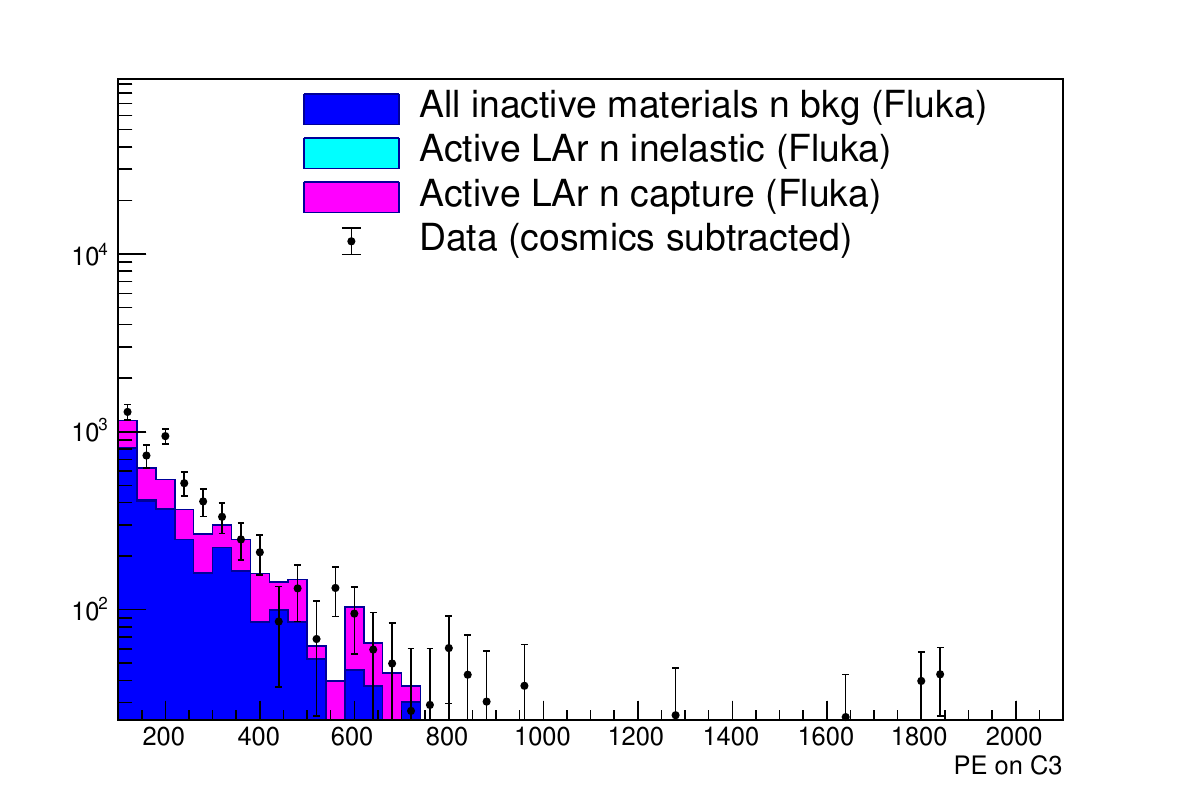}
  \centering  \includegraphics[width=0.49\columnwidth]{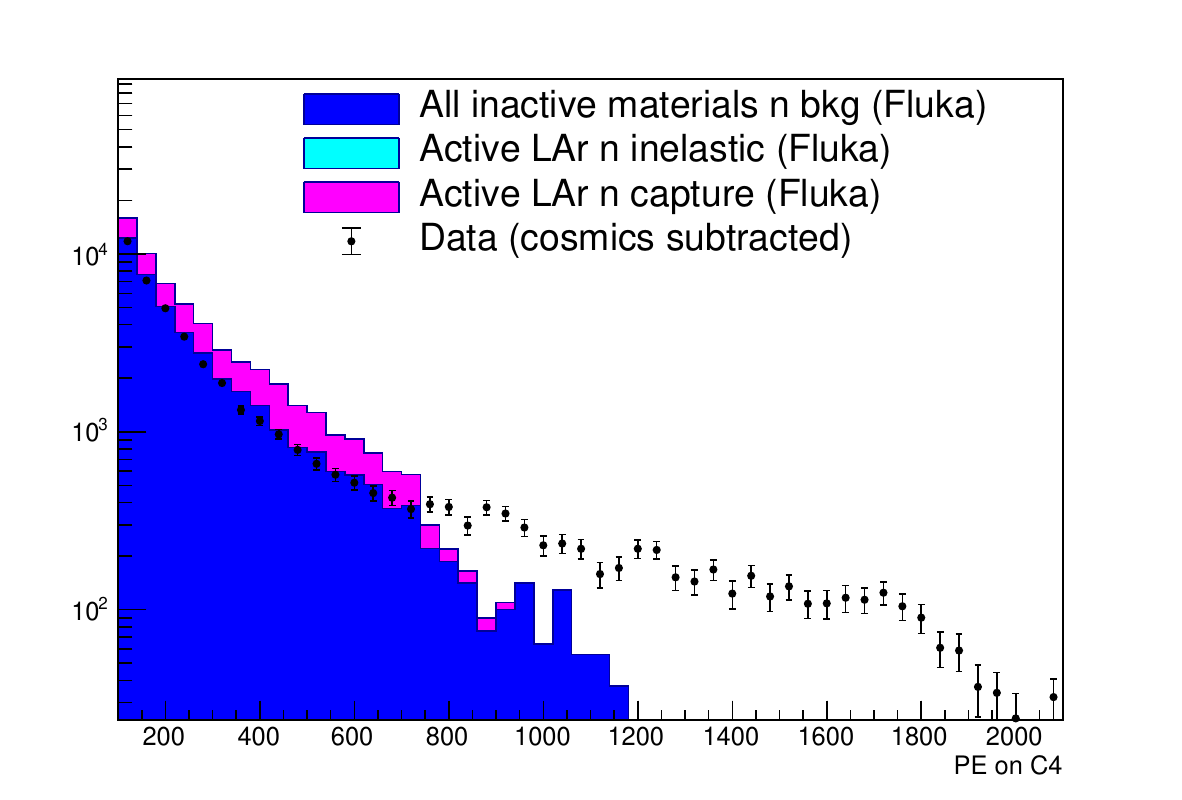}
  \caption{Comparison of neutron-related light signals in Fluka simulation and PNS run data for each XA module on the cathode for the time period 480-1050 $\mu$s in the DAQ window. Predicted cosmic background is subtracted from the data distribution. Top left: C1. Top right: C2. Bottom left: C3. Bottom right: C4.}
  \label{fig:datamccomparenocosmicstimeslice2}
\end{figure}

We further evaluate the PE spectra in three time periods, 0-160 $\mu$s, 160-480 $\mu$s, and 480-1050 $\mu$s, in Fig.~\ref{fig:datamccomparenocosmicstimeslice0}, Fig.~\ref{fig:datamccomparenocosmicstimeslice1}, and Fig.~\ref{fig:datamccomparenocosmicstimeslice2}, respectively.  The first period roughly corresponds to the initial neutron beam-on period, where many neutron interactions, especially the inelastic interactions in active LAr, occurred. The second period (160-480 $\mu$s) corresponds to time just after the neutron beam is off, where there are still many neutron activities. And the last period (480-1050 $\mu$s) corresponds to the rest of the data acquisition window. In all cases, the high-PE excess is observed consistently during the whole DAQ window. Below, we discuss several possible sources that could produce the high-PE excess observed in the data in Fig.~\ref{fig:datamccomparenocosmics}.

\begin{figure}[h!]
  \centering  
  \includegraphics[width=0.42\columnwidth]{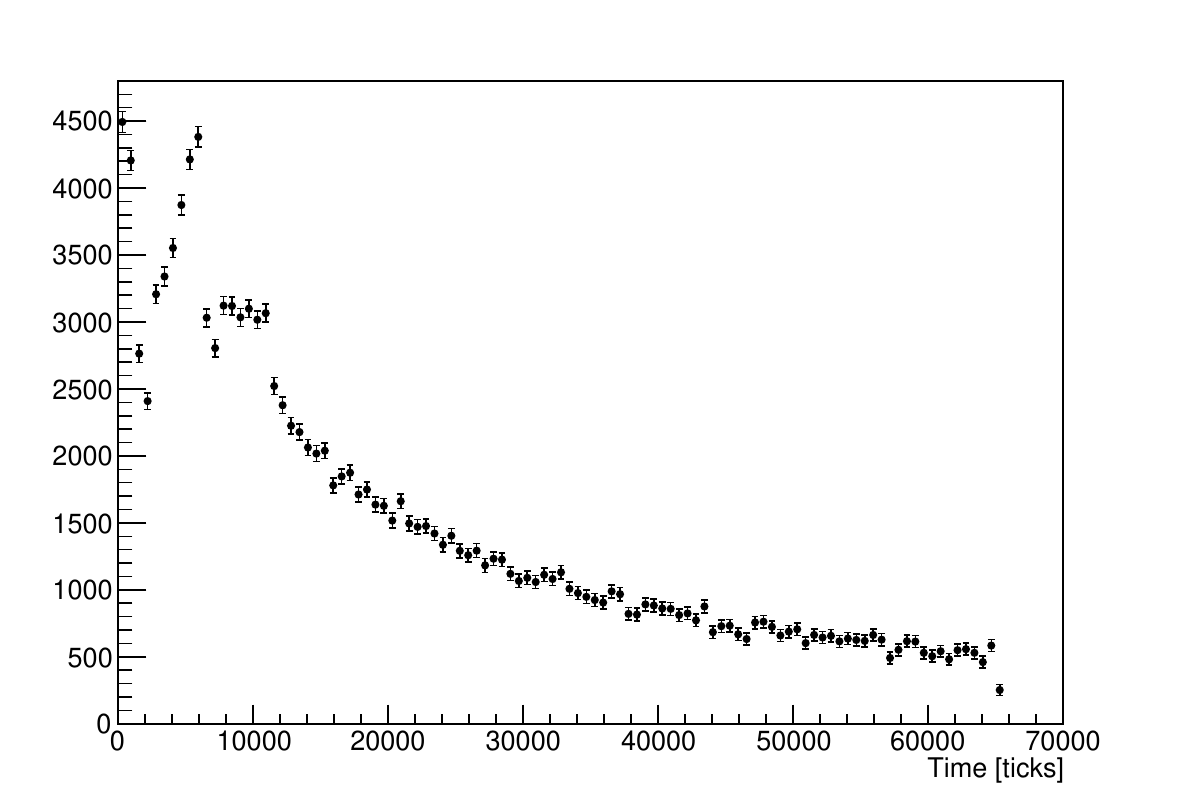}
  \includegraphics[width=0.42\columnwidth]{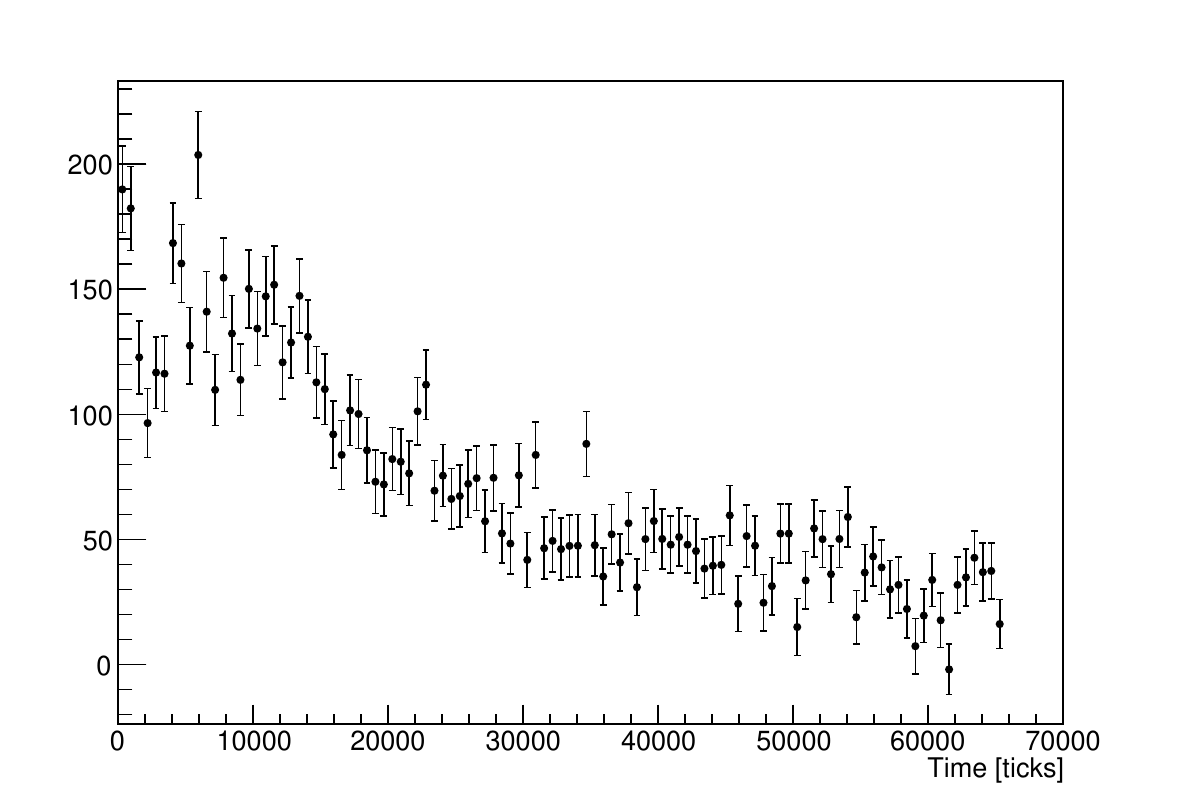}
  \caption{Timing distribution for C4 XA detected events with number of PE below (left) and above (right) 1200 PE.}
  \label{fig:1200PEsplittiming}
\end{figure}

\textit{Cosmic background}.
The excess at high PE region could originate from residual cosmic flux not fully subtracted in PNS run data. To test this hypothesis, we plot the timing distribution separately for events with the number of PE below and above 1200 in Fig.~\ref{fig:1200PEsplittiming}. After subtracting the cosmic background in data, the not well modeled tail above 1200 PE accounts for less than 5\% of the total neutron-induced light signals. For cosmic events, the timing distribution is expected to be flat as shown in Fig.~\ref{fig:datatimingcosmics}. Instead, in the right plot of Fig.~\ref{fig:1200PEsplittiming}, for light signals with PE $>$1200, a decay time profile similar to the events with PE $<$ 1200 is observed (Fig.~\ref{fig:1200PEsplittiming} left plot). This provides strong evidence that the excess at high PE is from neutron-related signals not modeled in the simulation.

\begin{figure}[h!]
  \centering  
  \includegraphics[width=0.45\columnwidth]{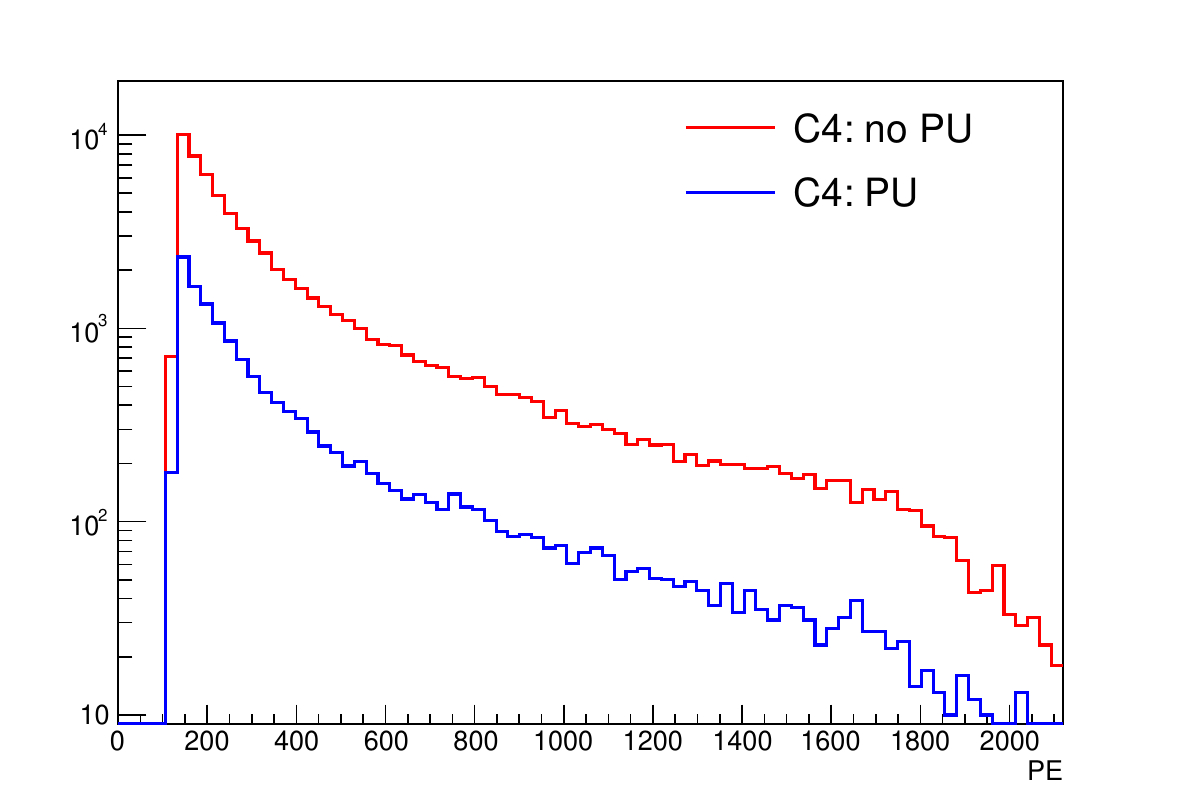}
  \includegraphics[width=0.45\columnwidth]{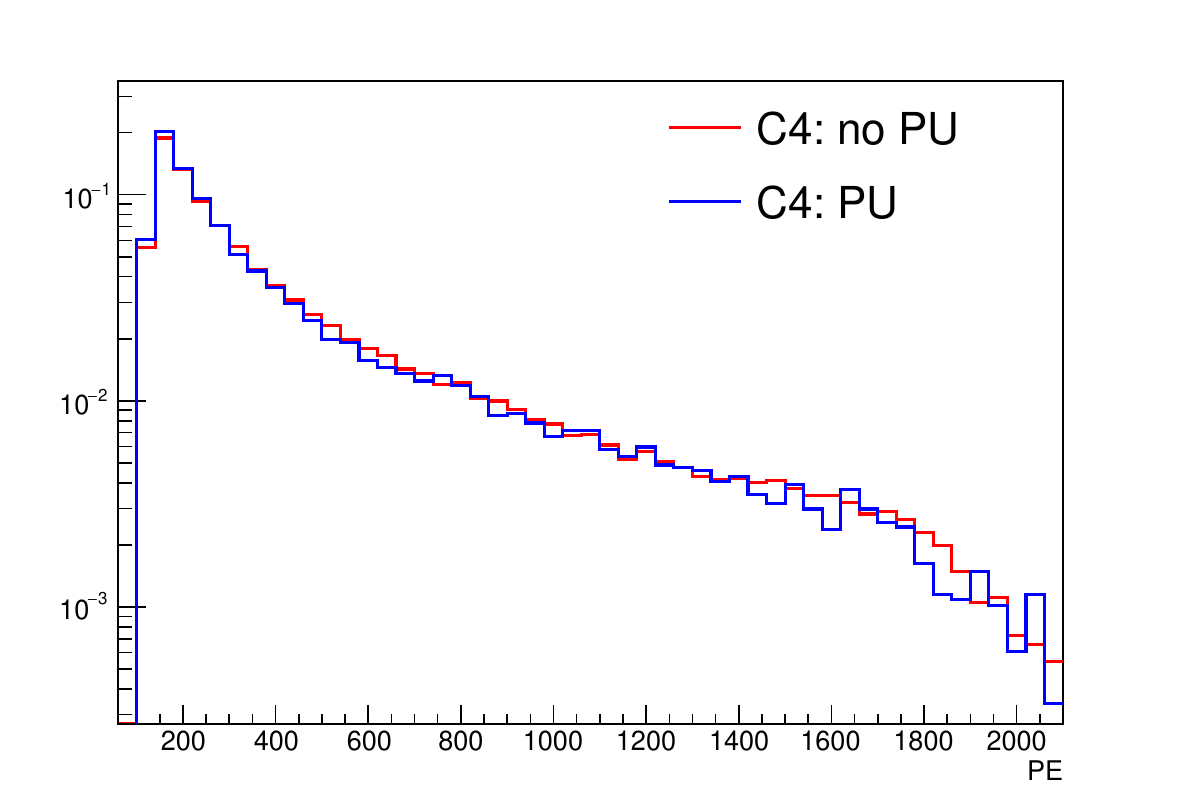}
  \caption{Distribution of light signal amplitude for pulses with (without) a preceding pulse within a 10 $\mu$s-long moving window, PU (no PU). Left: absolute number of light signals. Right: the number of light signals is normalized to unity.}
  \label{fig:PUpulses}
\end{figure}

\textit{Pile up light signals}. Pile-up (PU) light signals occur when two events happen close in time. In this analysis, the signal amplitude could be overestimated when the signal overlaps with the long tail of an earlier pulse. This could happen, for example, when a cosmic muon precedes a neutron-induced signal. We performed a data-driven estimation by tagging and separating the PU light pulses. Because of the intrinsic slow scintillation light component (1.5 $\mu$s) in LAr, a 10 $\mu$s-long moving window is used to identify PU light pulses that are contained in this timing window for each channel on the XA. All pulses except the earliest pulse in this moving time window are tagged as PU pulses. The same data selection in Sec.~\ref{sec:lightsel} is applied. Overall, we found PU pulses constitute about 20\% of all light signals. An overlay of the signal amplitude of the PU tagged pulses and the rest pulses on the C4 XA module is shown in Fig.~\ref{fig:PUpulses}. The similar shapes for PU pulses and the rest indicate PU is not causing the excess observed in high PE.

\textit{Cherenkov light from charged particles crossing the XA detector}.
Charged particles entering the plastic layer of the XA detector produce Cherenkov light. The collection and detection efficiency of this light is completely unknown. A test simulation has been performed to evaluate the relevance of this effect. Even assuming a perfect detection efficiency, those events would contribute to about 2\% \ of the signal above 100 PE, and the events producing more than 1200 PE would be about 0.2\% of the integrated signal, much smaller than the observed high PE tail.

\textit{Electric field outside active LAr}. The uncertainty of the electric field in the buffer LAr region is one of the systematics discussed in Sec.~\ref{sec:syst}. In the nominal case, the simulation takes the same 454 V/cm for the buffer LAr region as the active LAr region. In Fig.~\ref{fig:syst_pe_log}, the change to zero field in the buffer LAr region introduces a high PE tail extending to 2000 PE. This effect could partially explain the high PE excess in the data, as the actual E field in the buffer region is typically smaller than 454 V/cm and is often characterized by complicated field lines that are difficult to model in simulation.

\textit{XA Absolute PDE}. In the analysis (Sec.~\ref{sec:pdecali}), a relative PDE calibration was performed to the 3\% absolute PDE used in the simulation. However, the actual XA absolute PDE can vary by up to 50\% when SiPMs are biased at higher overvoltages with a non-negligible cross-talk probability. The absolute PDE is also affected by the detailed XA configuration. A smaller absolute PDE in simulation would underestimate the number of PE and could partially explain the high PE excess in data. This is further discussed in Sec.~\ref{sec:syst} as a systematic effect. The precise measurement of absolute PDE for XA modules requires a dedicated laboratory setup and analysis. More details can be found in ref.~\cite{VD_XA_PDE}.

\section{Systematic Errors}
\label{sec:syst}

Several important systematic effects are studied using the Fluka simulation. The uncertainty of the electric field in the buffer LAr region affects the scintillation photon yield. A simulation with zero electric field ("0 field") in this volume is evaluated, where the light yield reaches a maximum compared to the nominal electric field of 454 V/cm. The uncertainty of the initial production of scintillation light in buffer LAr and propagation to the active LAr readout volume is evaluated by setting the photon production in the buffer LAr volume to zero ("act only"). The uncertainty of the PNS position relative to the active volume inside the CB is modeled by shifting the PNS in the \textit{x} (drift) and \textit{y} (horizontal) directions by 5 cm ("dispx" and "dispy").  The uncertainty on the photon detector efficiency is studied by increasing it to 4.44\% ("heff"), a value that corresponds to the product of efficiency and cross-talk probability for one of the device configurations described in ref.~\cite{VD_XA_PDE}.  

\begin{figure}[h!]
  \centering  
  \includegraphics[width=0.9\columnwidth]{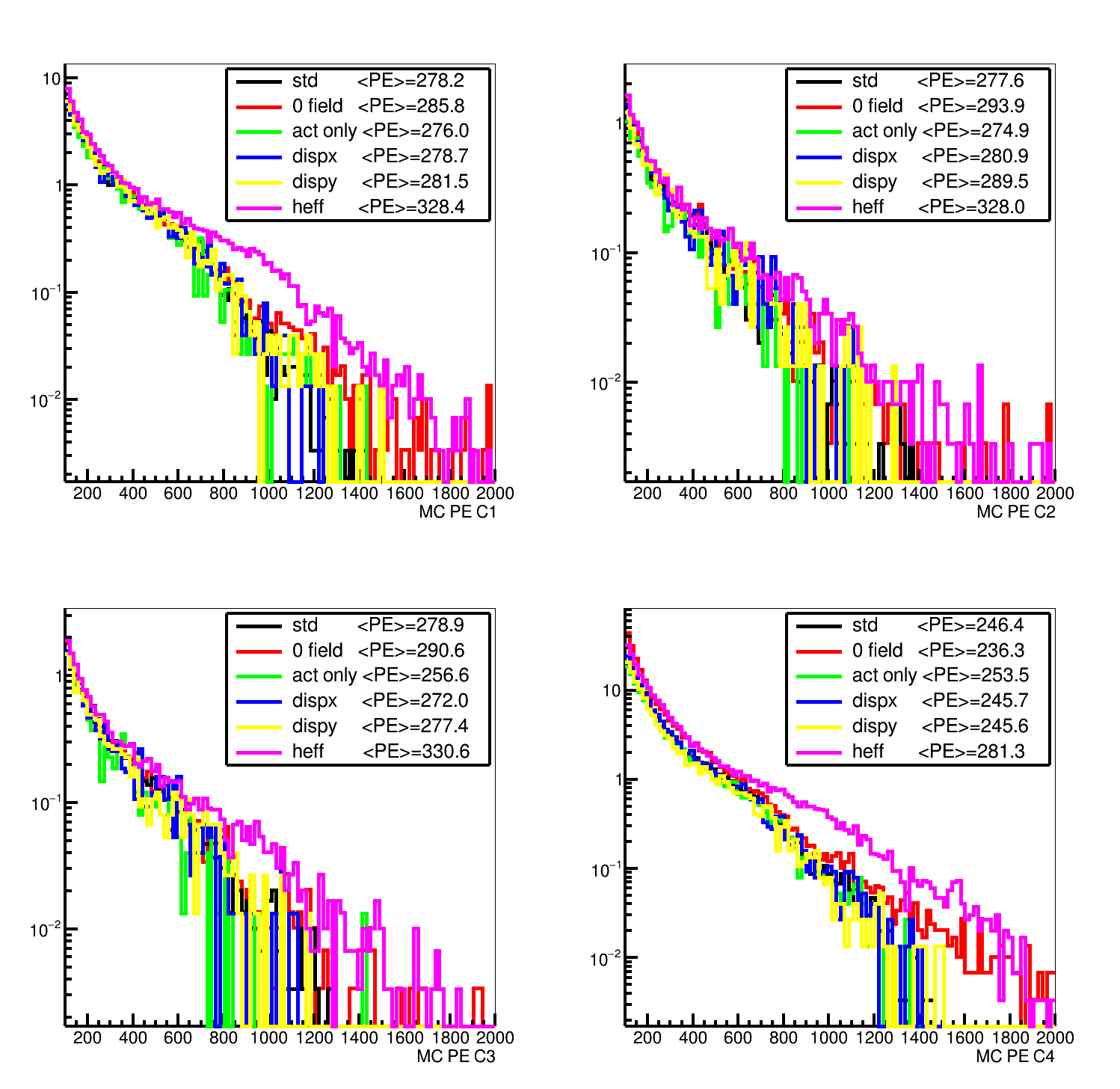}
  \caption{Comparison of the total number of photoelectrons detected by each XA on the cathode under different scenarios in the Fluka simulation.}
  \label{fig:syst_pe_log}
\end{figure}

\begin{figure}[h!]
  \centering  
\includegraphics[width=0.9\columnwidth]{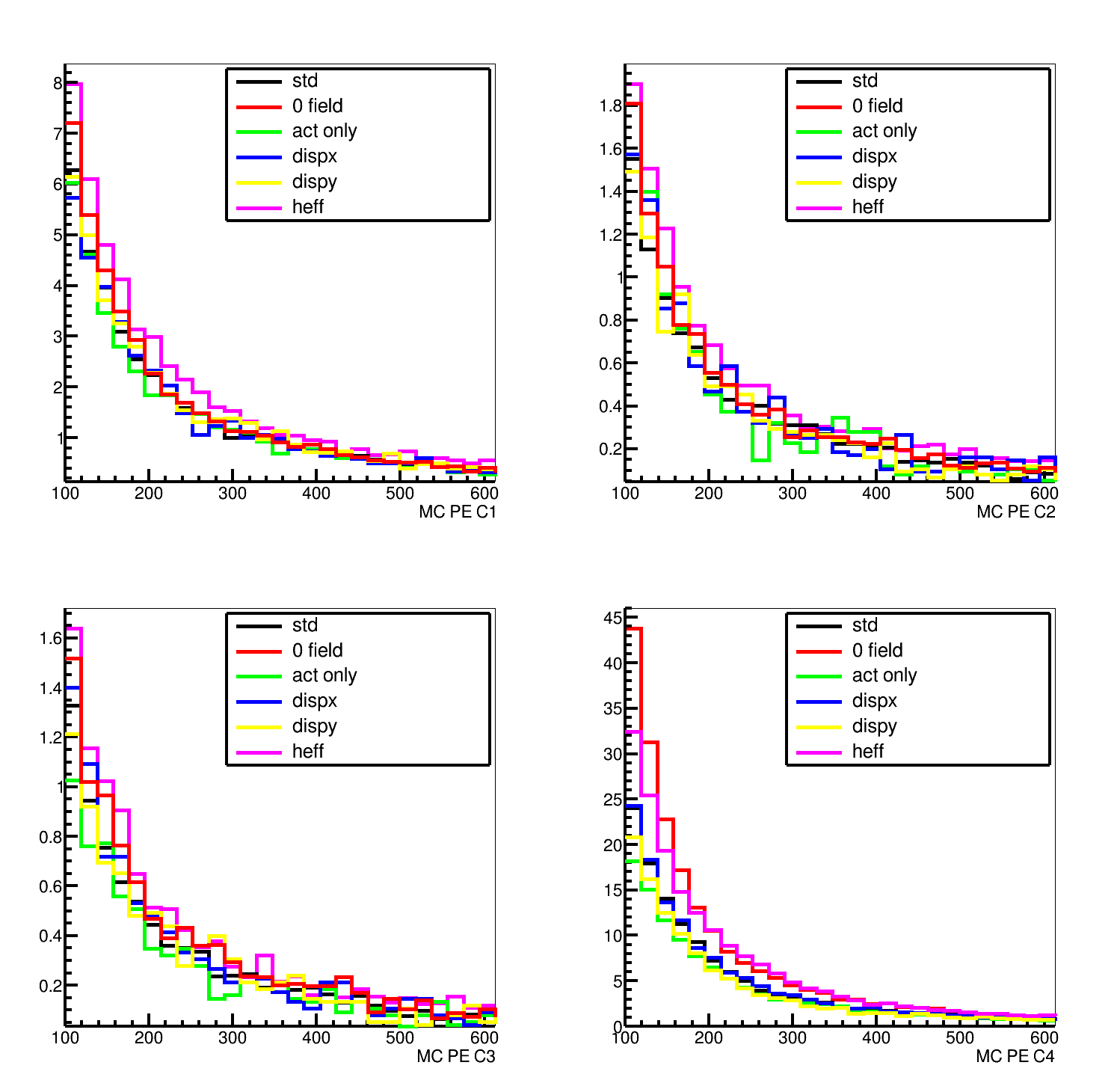}
  \caption{Comparison of the total number of photoelectrons detected by each XA on the cathode under different scenarios in the Fluka simulation below 600 PE.}
  \label{fig:syst_pe_zoom}
\end{figure}

The resulting distribution of the total number of photoelectrons detected by each XA on the cathode under these scenarios is shown in Figure~\ref{fig:syst_pe_log}. The largest systematic effect is the electric field in the buffer LAr region, which produces almost a factor of two more total PE on the C4 XA module, the closest module to the source, as also shown in the zoom-in plot in Fig.~\ref{fig:syst_pe_zoom}. A high detection efficiency obviously increases the high PE tail of the spectra, and the average number of detected PE,  especially for the most exposed detectors.

\begin{figure}[h!]
  \centering  \includegraphics[width=0.49\columnwidth]{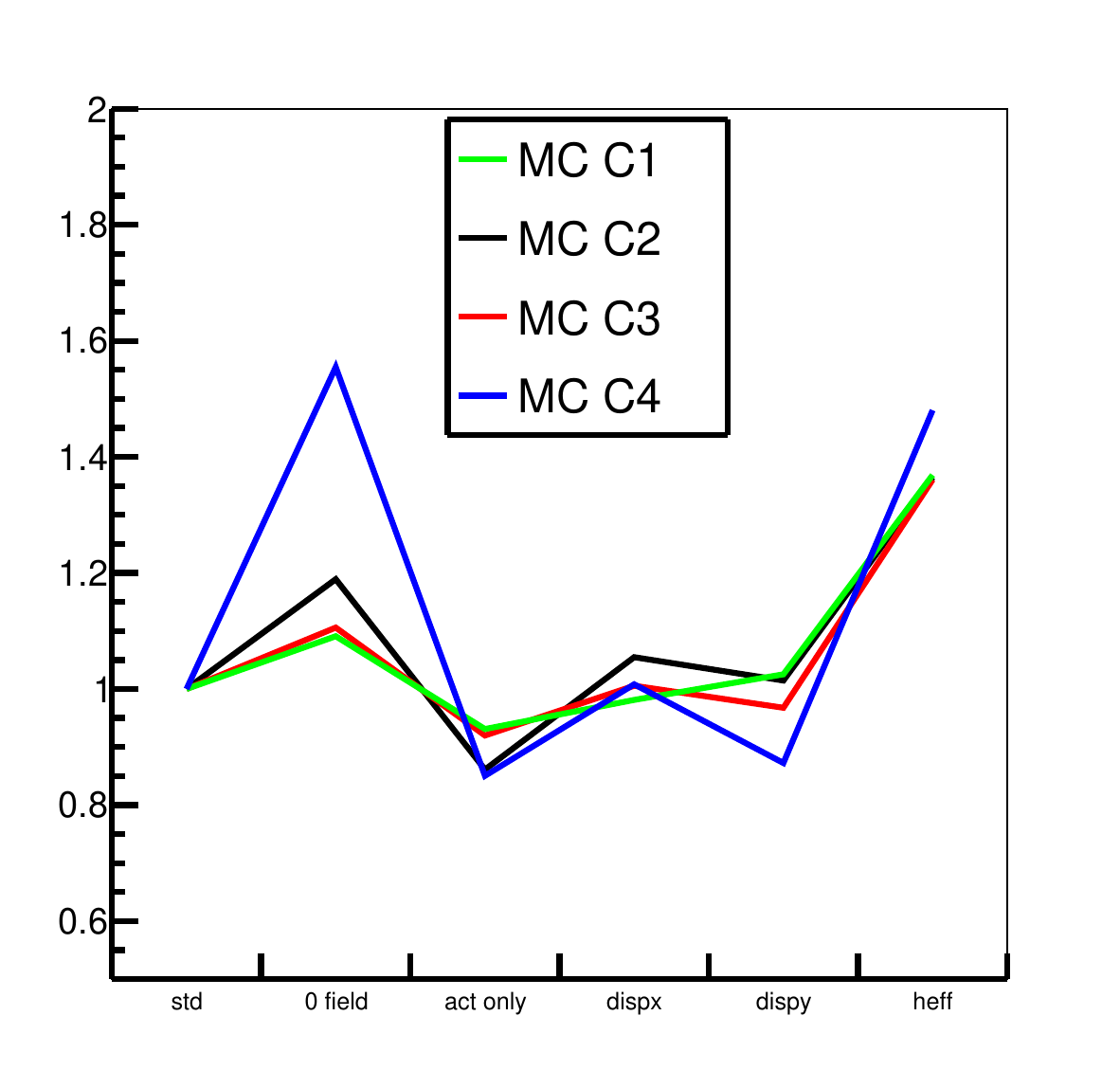}
  \includegraphics[width=0.49\columnwidth]{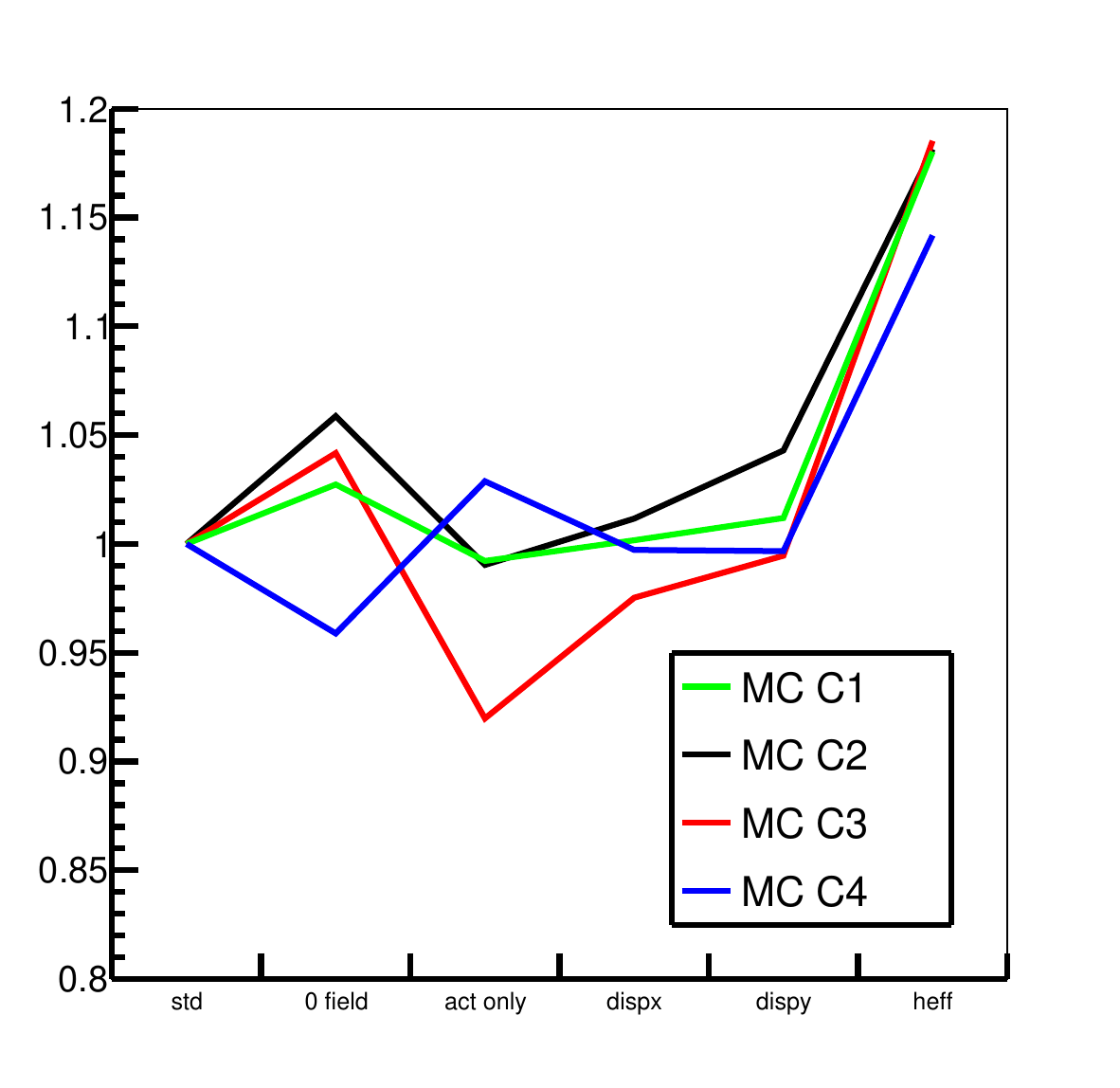}
  \caption{Integrated (left) and average (right) of the optical photons detected by XA under systematic shifts.}
  \label{fig:syst_ratio_int_pe}
\end{figure}

The integrated and average of all photons normalized to the standard simulation condition for each systematic effect is shown in Fig.~\ref{fig:syst_ratio_int_pe}. Overall, the displacement of the neutron generator gives small variations in the arriving optical photons. On the other hand, the field conditions in the non-readout LAr volume have a large impact on XA modules nearest to the source, especially for small signals (left plot in Fig.~\ref{fig:syst_ratio_int_pe}). At a higher number of photons, the electric field impact is smaller, and the modeling of interactions outside active LAr becomes important (right plot in Fig.~\ref{fig:syst_ratio_int_pe} ).

The peak ADC amplitude to photoelectron calibration constant used in the data discussed in Sec.~\ref{sec:adc2pe} affects the observed amount of light on each XA. The systematic shift of this constant up and down by one ADC unit is applied to the calculation in data, and the resulting counts are compared to the nominal case. The non-cosmic component in data has a max shift of 8.3\% to the nominal scenario from the C4 XA module.


\section{Summary and Outlook}
\label{sec:outlook}

A pulsed neutron source is deployed for the first time at the VD CB facility at the CERN neutrino platform to study MeV neutron interactions with LAr. In this paper, an analysis of light signals is presented. A good modeling of the light signals up to 650 photoelectrons per XA module from the PNS run is demonstrated. A good agreement is found in the fitted decay time constant from neutron interactions after the neutron beam stops.

The modeling of signals from neutron interactions will benefit the DUNE low energy physics program in many ways. MeV-scale physics measurements using scintillation light in neutrino LArTPC are scarce. Efficient tagging of the neutron captures could improve energy resolution~\cite{APEX_lowE, MeVNueArCC}. Meanwhile, understanding the neutron capture signals in charge and light will help reject cavern neutron backgrounds. Developing such a neutron capture tagging technique at small LArTPC prototypes before DUNE FD operation is crucial for advancing the DUNE physics program~\cite{DUNE_lowE_1,DUNE_lowE_2,DUNE_lowE_3}.

Several constraints exist at the CB facility for the physics analysis. The limited drift distance and active volume make it inefficient to contain and tag neutron capture events. However, these constraints are foreseen to be gone in future runs at larger prototypes such as ProtoDUNE-VD. The developed analysis method in this study will be applied to future PNS physics runs in larger LArTPCs. This study opens new possibilities to achieve tagging of capture events for light calorimetry calibration and measurement of the characteristic time of neutron capture on argon.

\acknowledgments

The vertical drift ColdBox test facility was constructed and operated on the CERN Neutrino Platform. We thank the CERN management for providing the infrastructure for this experiment and gratefully acknowledge the support of CERN. We thank LANL for the availability of the DD generator for this physics program. 


\bibliographystyle{JHEP}
\bibliography{biblio.bib}

@article{DUNE_SP_TDR,
    author = "Abi, Babak and others",
    collaboration = "DUNE",
    title = "{Deep Underground Neutrino Experiment (DUNE), Far Detector Technical Design Report, Volume IV: Far Detector Single-phase Technology}",
    eprint = "2002.03010",
    archivePrefix = "arXiv",
    primaryClass = "physics.ins-det",
    reportNumber = "FERMILAB-PUB-20-027-ND, FERMILAB-DESIGN-2020-04",
    doi = "10.1088/1748-0221/15/08/T08010",
    journal = "JINST",
    volume = "15",
    number = "08",
    pages = "T08010",
    year = "2020"
}

@article{VD_XA_PDE,
    author = "Botogoske, G. and others",
    title = "{Laboratory Measurement of the X-ARAPUCA's Absolute Photon Detection Efficiency for the Deep Underground Neutrino Experiment's Vertical Drift Far Detector}",
    eprint = "2511.12328",
    archivePrefix = "arXiv",
    primaryClass = "physics.ins-det",
    month = "11",
    year = "2025"
}

@techreport{ProtoDUNE3,
      author        = "Gollapinni, Sowjanya and Bertolucci, Sergio and Fields,
                       Laura and Soldner-Rembold, Stefan and Sorel, Michel",
      collaboration = "DUNE",
      title         = "{Proposal Addendum for a ProtoDUNE-III Run at NP02}",
      institution   = "CERN",
      reportNumber  = "CERN-SPSC-2025-037, SPSC-P-358-ADD-1",
      address       = "Geneva",
      year          = "2025",
      url           = "https://cds.cern.ch/record/2947730",
}

@article{DUNE_PoF,
doi = {10.1088/1748-0221/19/10/P10019},
url = {https://dx.doi.org/10.1088/1748-0221/19/10/P10019},
year = {2024},
month = {oct},
publisher = {IOP Publishing},
volume = {19},
number = {10},
pages = {P10019},
author = {Arroyave, M.A. and Behera, B. and Cavanna, F. and Feld, A. and Guo, F. and Heindel, A. and Jung, C.K. and Koch, K. and Leon Silverio, D. and Martinez Caicedo, D.A. and McGrew, C. and Paudel, A. and Pellico, W. and Rivera, R. and Rodríguez Rondon, J. and Sacerdoti, S. and Shanahan, P. and Shi, W. and Torres Muñoz, D. and Totani, D. and Uy, C. and Vermeulen, C. and Vieira de Souza, H.},
title = {Characterization and novel application of power over fiber for electronics in a harsh environment},
journal = {Journal of Instrumentation}
}

@article{DUNE_Physics_TDR,
    author = "Abi, Babak and others",
    collaboration = "DUNE",
    title = "{Deep Underground Neutrino Experiment (DUNE), Far Detector Technical Design Report, Volume II: DUNE Physics}",
    eprint = "2002.03005",
    archivePrefix = "arXiv",
    primaryClass = "hep-ex",
    reportNumber = "FERMILAB-PUB-20-025-ND, FERMILAB-DESIGN-2020-02",
    month = "2",
    year = "2020"
}

@article{DUNE_VD_TDR,
    author = "Abed Abud, Adam and others",
    collaboration = "DUNE",
    title = "{The DUNE Far Detector Vertical Drift Technology. Technical Design Report}",
    eprint = "2312.03130",
    archivePrefix = "arXiv",
    primaryClass = "hep-ex",
    reportNumber = "FERMILAB-TM-2813-LBNF",
    doi = "10.1088/1748-0221/19/08/T08004",
    journal = "JINST",
    volume = "19",
    number = "08",
    pages = "T08004",
    year = "2024"
}

@article{Arapuca_2016,
doi = {10.1088/1748-0221/11/02/C02004},
url = {https://dx.doi.org/10.1088/1748-0221/11/02/C02004},
year = {2016},
month = {feb},
publisher = {},
volume = {11},
number = {02},
pages = {C02004},
author = {Machado, A.A. and Segreto, E.},
title = {ARAPUCA a new device for liquid argon scintillation light detection},
journal = {Journal of Instrumentation}
}

@article{XA_2018,
doi = {10.1088/1748-0221/13/04/C04026},
url = {https://dx.doi.org/10.1088/1748-0221/13/04/C04026},
year = {2018},
month = {apr},
publisher = {},
volume = {13},
number = {04},
pages = {C04026},
author = {Machado, A.A. and Segreto, E. and Warner, D. and Fauth, A. and Gelli, B. and Máximo, R. and Pissolatti, A. and Paulucci, L. and Marinho, F.},
title = {The X-ARAPUCA: an improvement of the ARAPUCA device},
journal = {Journal of Instrumentation}
}

@article{APEX_lowE,
    author = "Shi, Wei and Ning, Xuyang and Pershey, Daniel and Marinho, Franciole and Fleuri, Anjarazafy and Riccio, Ciro and Jo, Jay Hyun and Zhang, Chao and Cavanna, Flavio",
    title = "{Physics prospects with MeV neutrino-argon charged current interactions using enhanced photon detection in future LArTPCs}",
    eprint = "2502.18498",
    archivePrefix = "arXiv",
    primaryClass = "physics.ins-det",
    reportNumber = "FERMILAB-PUB-25-0124-PPD",
    doi = "10.1103/f3t7-5vmd",
    journal = "Phys. Rev. D",
    volume = "112",
    number = "1",
    pages = "012019",
    year = "2025"
}

@article{MeVNueArCC,
  title = {{Benefits of MeV-scale reconstruction capabilities in large liquid argon time projection chambers}},
  author = {Castiglioni, W. and Foreman, W. and Littlejohn, B. R. and Malaker, M. and Lepetic, I. and Mastbaum, A.},
  journal = {Phys. Rev. D},
  volume = {102},
  issue = {9},
  pages = {092010},
  numpages = {25},
  year = {2020},
  month = {Nov},
  publisher = {American Physical Society},
  doi = {10.1103/PhysRevD.102.092010},
  url = {https://link.aps.org/doi/10.1103/PhysRevD.102.092010}
}

@article{Fluka,
	author = {{The FLUKA Collaboration} and {Ballarini, Francesca} and {Batkov, Konstantin} and {Battistoni, Giuseppe} and {Bisogni, Maria Giuseppina} and {Böhlen, Till T.} and {Campanella, Mauro} and {Carante, Mario P.} and {Chen, Daiyuan} and {De Gregorio, Angelica} and {Degtiarenko, Pavel V.} and {De la Torre Luque, Pedro} and {dos Santos Augusto, Ricardo} and {Engel, Ralph} and {Fassò, Alberto} and {Fedynitch, Anatoli} and {Ferrari, Alfredo} and {Ferrari, Anna} and {Franciosini, Gaia} and {Kraan, Aafke Christine} and {Lascaud, Julie} and {Li, Wenxin} and {Liu, Juntao} and {Liu, Zhiyi} and {Magro, Giuseppe} and {Mairani, Andrea} and {Mattei, Ilaria} and {Mazziotta, Mario N.} and {Morone, Maria C.} and {Müller, Stefan E.} and {Muraro, Silvia} and {Ortega, Pablo G.} and {Parodi, Katia} and {Patera, Vincenzo} and {Pinsky, Lawrence S.} and {Ramos, Ricardo L.} and {Ranft, Johannes} and {Rosso, Valeria} and {Sala, Paola R.} and {Santana Leitner, Mario} and {Sportelli, Giancarlo} and {Tessonnier, Thomas} and {Ytre-Hauge, Kristian S.} and {Zana, Lorenzo}},
	title = {The FLUKA code: Overview and new developments},
	DOI= "10.1051/epjn/2024015",
	url= "https://doi.org/10.1051/epjn/2024015",
	journal = {EPJ Nuclear Sci. Technol.},
	year = 2024,
	volume = 10,
	pages = "16",
}

@misc{lardon,
  title        = {{LARDON}},
  author       = {Laura Zambell and others},
  howpublished = {\url{https://github.com/dune-lardon/lardon}}
}

@article{Fluka_validation,
doi = {10.1088/1748-0221/15/09/P09009},
url = {https://dx.doi.org/10.1088/1748-0221/15/09/P09009},
year = {2020},
month = {sep},
publisher = {},
volume = {15},
number = {09},
pages = {P09009},
author = {Babicz, M. and Bordoni, S. and Fava, A. and Kose, U. and Nessi, M. and Pietropaolo, F. and Raselli, G.L. and Resnati, F. and Rossella, M. and Sala, P. and Stocker, F. and Zani, A.},
title = {A measurement of the group velocity of scintillation light in liquid argon},
journal = {Journal of Instrumentation}
}

@article{DUNE_lowE_1,
    author = "Capozzi, Francesco and Li, Shirley Weishi and Zhu, Guanying and Beacom, John F.",
    title = "{DUNE as the Next-Generation Solar Neutrino Experiment}",
    eprint = "1808.08232",
    archivePrefix = "arXiv",
    primaryClass = "hep-ph",
    reportNumber = "MPP-2018-214, SLAC-PUB-17199",
    doi = "10.1103/PhysRevLett.123.131803",
    journal = "Phys. Rev. Lett.",
    volume = "123",
    number = "13",
    pages = "131803",
    year = "2019"
}

@article{DUNE_lowE_2,
    author = "Zhu, Guanying and Li, Shirley Weishi and Beacom, John F.",
    title = "{Developing the MeV potential of DUNE: Detailed considerations of muon-induced spallation and other backgrounds}",
    eprint = "1811.07912",
    archivePrefix = "arXiv",
    primaryClass = "hep-ph",
    reportNumber = "SLAC-PUB-17355",
    doi = "10.1103/PhysRevC.99.055810",
    journal = "Phys. Rev. C",
    volume = "99",
    number = "5",
    pages = "055810",
    year = "2019"
}

@article{DUNE_lowE_3,
    author = "Meighen-Berger, Stephan A. and Newstead, Jayden L. and Beacom, John F. and Bell, Nicole F. and Dolan, Matthew J.",
    title = "{Enhancing DUNE{\textquoteright}s Solar Neutrino Capabilities with Neutral-Current Detection}",
    eprint = "2410.00330",
    archivePrefix = "arXiv",
    primaryClass = "hep-ph",
    doi = "10.1103/htfm-tbdq",
    journal = "Phys. Rev. Lett.",
    volume = "135",
    number = "1",
    pages = "011803",
    year = "2025"
}

@article{LArQL,
doi = {10.1088/1748-0221/17/07/C07009},
url = {https://dx.doi.org/10.1088/1748-0221/17/07/C07009},
year = {2022},
month = {jul},
publisher = {IOP Publishing},
volume = {17},
number = {07},
pages = {C07009},
author = {Marinho, F. and Paulucci, L. and Totani, D. and Cavanna, F.},
title = {LArQL: a phenomenological model for treating light and charge generation in liquid argon},
journal = {Journal of Instrumentation}
}






\end{document}